\documentstyle[11pt,amssymb]{article}

\makeatletter
\@addtoreset{equation}{section}
\makeatother

\def\dalemb#1#2{{\vbox{\hrule height .#2pt
        \hbox{\vrule width.#2pt height#1pt \kern#1pt
                \vrule width.#2pt}
        \hrule height.#2pt}}}

\def\cA{{\cal A}}

\def\cao{{\cal O}}

\def\0{{\sst{(0)}}}
\def\1{{\sst{(1)}}}
\def\2{{\sst{(2)}}}
\def\3{{\sst{(3)}}}
\def\4{{\sst{(4)}}}
\def\5{{\sst{(5)}}}
\def\6{{\sst{(6)}}}
\def\7{{\sst{(7)}}}
\def\8{{\sst{(8)}}}
\def\n{{\sst{(n)}}}

\def\ep{\epsilon}
\def\td{\tilde}

\def\half{{\textstyle{1\over2}}}

\def\qu{{\textstyle{1\over 4}}}

\let\a=\alpha \let\b=\beta \let\g=\gamma \let\d=\delta \let\e=\epsilon
  \let\q=\theta  \let\k=\kappa
\let\l=\lambda \let\m=\mu \let\n=\nu  \let\r=\rho
\let\s=\sigma \let\t=\tau  \let\f=\phi  
\let\w=\omega  \let\D=\Delta  \let\L=\Lambda
    
 \let\W=\Omega   \let\G=\Gamma
\let\la=\label  \let\re=\ref
  
\def\nn{\nonumber} \def\bd{\begin{document}} \def\ed{\end{document}}
\def\ds{\documentstyle} \let\fr=\frac \let\bl=\bigl \let\br=\bigr
\let\Br=\Bigr \let\Bl=\Bigl
\let\bm=\bibitem
\let\na=\nabla
\let\pa=\partial \let\ov=\overline
\newcommand{\be}{\begin{equation}}
\newcommand{\ee}{\end{equation}}
\def\ba{\begin{array}}
\def\ea{\end{array}}
\def\ft#1#2{{\textstyle{{\scriptstyle #1}\over {\scriptstyle #2}}}}
\def\fft#1#2{{#1 \over #2}}
\def\del{\partial}
\def\sst#1{{\scriptscriptstyle #1}}
 \def\oneone{\rlap 1\mkern4mu{\rm l}}
\def\ie{{\it i.e.\ }}
\def\via{{\it via}}
\def\semi{{\ltimes}}
\def\str{{\rm str}}
\def\Dm{{{D_{\sst{max}}}}}
\def\vac{ \left | 0 \right \rangle }
\def\kvac{ \left | k \right \rangle }

\def\sp{\; \; \;}

\def\bol{ \left | B (p^+) \right \rangle}
\def\bo1{ \left | B^0 (p^+) \right \rangle}

\def\bolt{ \left | B (p^+) \right \rangle_{\t}}

\def\boxl{ \left | B (x^-) \right \rangle}

\def\<{ \langle }
\def\>{ \rangle }

\def\S{\Sigma}

\newcommand{\hsp}{\hspace{0.5cm}}

\newcommand{\ho}[1]{$\, ^{#1}$}
\newcommand{\hoch}[1]{$\, ^{#1}$}
\newcommand{\bea}{\begin{eqnarray}}
\newcommand{\eea}{\end{eqnarray}}
\newcommand{\ra}{\rightarrow}
\newcommand{\lra}{\longrightarrow}
\newcommand{\Lra}{\Leftrightarrow}
\newcommand{\ap}{\alpha^\prime}
\newcommand{\bp}{\tilde \beta^\prime}
\newcommand{\tr}{{\rm tr} }
\newcommand{\Tr}{{\rm Tr} }
\newcommand{\NP}{Nucl. Phys. }

\newcommand{\ams}{{\it Institute for Theoretical Physics,
University of Amsterdam, \\
Valckenierstraat 65, 1018XE Amsterdam, The Netherlands} \\
{\tt ingkanit, skenderi, taylor@science.uva.nl}}

\newcommand{\auth}{\large Ingmar Kanitscheider, Kostas Skenderis and Marika Taylor}

\begin{document}
\begin{flushright}
\hfill{\bf hep-th/0611171}\\
\hfill{ITFA-2006-28}
\end{flushright}

\vspace{15pt}

\begin{center}

{\Large \bf Holographic anatomy of fuzzballs}

\vspace{20pt}

\auth

\vspace{15pt}

\vspace{8pt}

{\ams}

\vspace{15pt}

\underline{ABSTRACT}
\end{center}
We present a comprehensive analysis of 2-charge fuzzball solutions,
that is, horizon-free non-singular solutions of IIB supergravity
characterized by a curve on $R^4$.
We propose a precise map that relates any given curve to a specific
superposition of R ground states of the D1-D5 system.
To test this proposal we compute the
holographic 1-point functions associated with these solutions,
namely the conserved charges and the vacuum expectation values of chiral
primary operators of the boundary theory, and find perfect agreement
within the approximations used. In particular, all kinematical
constraints are satisfied and the proposal is compatible
with dynamical constraints although detailed quantitative tests
would require going beyond the leading supergravity approximation.
We also discuss which geometries may be dual to a given R ground state.
We present the general asymptotic form that such solutions must
have and present exact solutions which have such asymptotics
and therefore pass all kinematical constraints. Dynamical constraints
would again require going beyond the leading supergravity approximation.

\noindent

\pagebreak
\setcounter{page}{1}

\tableofcontents

\pagebreak

\section{Introduction, summary of results and conclusions}

An interesting proposal for the gravitational nature of black hole microstates
has emerged over the last few years
\cite{Lunin:2001jy,Lunin:2002bj,Lunin:2002iz}; see also \cite{Lunin:2001fv,
Balasubramanian:2000rt, Maldacena:2000dr}, and \cite{Mathur:2005zp}
for a review. According to this proposal there should exist a
horizon free geometry associated with each black hole
microstate\footnote{ Note however
that in general only a subset of the
solutions will have small enough curvatures to be accurately described
by supergravity.}. These solutions should approach the original black hole
geometry asymptotically and they should generically differ from
each other up to the horizon scale; in a sense the interior of the
horizon is replaced by a fuzzball.
In this picture the black hole
provides only the average statistical description; the true description is
in terms of the horizon free microstate geometries.

Such a picture would resolve long standing puzzles
associated with black hole physics such as the information
loss paradox since the underlying
physics of the black hole would not be conceptually different from
that of a distant star, with the temperature and entropy being
of a statistical origin. It is thus important to
scrutinize the current evidence and further develop
this proposal.

The fuzzball proposal requires the existence of 
exponential numbers of horizon
free solutions. So the most basic question is
whether one can find such a number of solutions with the required
properties. A crucial issue here is what precisely are
the ``required properties''. Furthermore, to utilize
this proposal and to address questions such as
how the Hawking temperature and other black hole
properties emerge one would like to understand the precise
relation between solutions and microstates.

A test case for the fuzzball proposal has been the 2-charge D1-D5 system.
This is a 1/4 supersymmetric system and
the ``naive'' black hole geometry has a near-horizon geometry of
the form $AdS_3 \times S^3 \times M$, where $M$ is either $T^4$ or
$K3$. The naive geometry has a naked singularity
but one expects that a horizon would emerge from
$\alpha'$ corrections. At any rate, the description
in terms of D-branes (at weak coupling) is well defined
and one can obtain a statistical entropy in much the same way
as for the 3 charge geometry which has a finite radius horizon.
Indeed, the D1-D5 system can be mapped by dualities
to a system of a fundamental string carrying momentum modes
and the degeneracy of the system can be computed by standard methods.
To be more specific, let us take $M=T^4$; then
the degeneracy is the same as that of 8 bosonic
and 8 fermionic oscillators at level $N=n_1 n_5$, where $n_1$ and
$n_5$ are the number of D1 and D5 branes, respectively.
The fuzzball proposal in this context is that there should
exist an exponential number of horizon free solutions, one for each
microstate, each carrying these two D-brane charges.

An exponential number of solutions
was constructed by Lunin and Mathur in \cite{Lunin:2001jy} and proposed to
correspond to microstates.  These were found by dualizing
a subset of the FP solutions \cite{Callan:1995hn,Dabholkar:1995nc}, namely
those that are associated with excitations of four bosonic oscillators.
These provide enough solutions to account for
a finite fraction of the entropy but one still needs an
exponential number of solutions (associated with the additional
four bosonic and eight fermionic oscillators in the example
of $T^4$) to account for the total entropy.
Such solutions, related to the odd cohomology
of $T^4$ and the middle cohomology of the internal manifold
have been discussed in \cite{Taylor:2005db} and
\cite{Lunin:2002iz}, respectively, and we will
complete this program in a forthcoming publication
\cite{KST}. We thus indeed find that there are an
appropriate number of solutions to account for
all of the D1-D5 entropy\footnote{Note however that this
is a continuous family of supergravity solutions. To properly count them
one needs to appropriately quantize them. Such a
quantization has been discussed in \cite{Rychkov:2005ji}, see
also \cite{Palmer:2004gu,Bak:2004rj} for a counting using supertubes.}.

Do these solutions, however, have the right properties to be
associated with D1-D5 microstates, and if yes, what is the
precise relation? The aim of this paper is to address this
question for the solutions corresponding to the universal
sector of the $T^4$ and $K3$ compactifications.

As mentioned above the solutions of interest were obtained by
dualizing FP solutions so let us briefly review these solutions
and their relation to string perturbative states. A more detailed
discussion will be given in section \ref{FP}.
The FP solutions (which are general chiral null models)
involve the metric, B-field and the dilaton
and are characterized by a null curve $F^I(x^+)$ with $ {I=1,\ldots, 8}$ 
in $R^8$.
The solution describes the long range fields sourced by a string wrapping
one compact direction and having a transverse profile
given by the null curve $F^I(x^+)$. The ADM conserved charges,
i.e. the mass, momentum and angular momentum,
associated with this solution are given precisely
by the energy, momentum  and angular momentum
of the classical string that sources the solution.

On general grounds, one would expect that this classical string
should be produced by a coherent state of string oscillators.
Indeed, we show in section \ref{FP} that associated to a classical curve $F^I(x^+)$,
\be \label{curve}
F^I(x^+) = \sum_{n>0} \frac{1}{\sqrt{n}} \left(
\a_n^I e^{-i n \left(\frac{x^+}{w R_9}\right)} + (\a_n^I)^*
e^{i n \left(\frac{x^+}{w R_9}\right)}\right),
\ee
where $x^+=x^0+x^9$, $x^9$ is the compact direction of radius $R_9$,
$w$ is the winding number and
$\a_n^I$ are (complex) numerical coefficients, there is a
coherent state $|F^I)$ of the first quantized string in an
unconventional lightcone gauge with $x^+= w R_9 \s^+$, where $\s^+$
is a worldsheet lightcone coordinate, 
such that the expectation
value of all conserved charges match the conserved charges
associated with the solution. More precisely, let
\be
X^I = \sum_{n>0} \frac{1}{\sqrt{n}} \left(
\hat{a}_n^I e^{-i n \s^+} + (\hat{a}_n^I)^\dagger e^{i n \s^+}\right)
\ee
be the 8 transverse left moving coordinates with $\hat{a}_n^I$
the quantum oscillators normalized such that
$[\hat{a}_n^I, (\hat{a}_m^J)^\dagger]=\d^{IJ} \d_{mn}$.
The corresponding coherent state is given by
\be
|F^I)=\prod_{n,I}| \a_n^I )
\ee
where $| \a_n^I )$ is a coherent state of the left-moving
oscillator $\hat{a}_n^I$,
i.e. it satisfies $\hat{a}_n^I | \a_n^I ) = \a_n^I | \a_n^I )$,
and the eigenvalues $\a_n^I$ are the coefficients appearing in (\ref{curve}).
By construction
\be
(F^I| X^I |F^I) = F^I
\ee
with root mean deviation of order $1/\sqrt{m}$, where $m \equiv
(F^I| \hat{m} |F^I)$
the expectation value of the occupation operator\footnote{Usually the
occupation operator is called $N$ but we reserve this letter for the
level of the Fock states, $N = \sum n \hat{a}_{-n}^I \hat{a}_{n}^I$.
Note also that after the duality to the D1-D5 system the
occupation number becomes the eigenvalue of $j_3$ which is usually
called $m$.}
$\hat{m} = \sum \hat{a}_{-n}^I \hat{a}_{n}^I$. In other words, the
expectation value is given by the classical string that sources the
solution,
and this is an
accurate description as long as the excitation numbers are high.
For low excitation numbers the state produced is fuzzy and
the supergravity solution would require quantum corrections
(as one would indeed expect). Note that the
right-movers are in their ground state throughout this discussion.

Given winding $w$ and momentum $p_9$ quantum numbers there
are also corresponding Fock states
\be \label{fock}
\prod (\hat{a}_{-n^I}^{I})^{m_I} |0\>, \qquad N_L = \sum n^I m^I = - w p_9
\ee
where $N_L$ is the total left-moving excitation level ($m_I$ are integers).
It is sometimes stated in the literature that the solutions of
\cite{Callan:1995hn,Dabholkar:1995nc}
represent these states. This cannot be exactly correct
as the string coordinates have zero expectation on these
states, so semiclassically they do not produce the required
source. The statement is however approximately correct since
these states strongly overlap with the corresponding coherent state
for high excitation numbers. So in the regime where
supergravity is valid the coherent state can be approximated
by Fock states.  Notice that one can organize the
Fock states (\ref{fock}) into eigenstates of the angular
momentum operator by using as building blocks
linear combination of oscillators that themselves
are eigenstates (e.g. $(\hat{a}_{-n}^I \pm i \hat{a}_{-n}^{I+1})$).
The coherent states are however (infinite) superpositions of
states with different angular momenta and are thus not
eigenstates of the angular momentum operator.

We now return to the discussion of the dual D1-D5 system. The solutions
of \cite{Lunin:2001jy} were obtained by dualizing the FP solutions
we just discussed but with a curve that is restricted to
lie on $R^4$. The corresponding underlying states are
now R ground states of the CFT associated with the D1-D5 system.
This CFT is a deformation of a sigma model with target space
the symmetric product of the compactification manifold $X$, $S^N(X)$
($N=n_1 n_5$ and $n_1, n_5$ are the number of D1 and D5 branes).
The R ground states can be obtained by spectral flow of the chiral
primaries of the NS sector. Recall that the chiral primaries are
associated with the cohomology of the internal space. For the
discussion at hand only the universal cohomology is relevant
and this leads (after spectral flow) to the following R ground states
\be \label{d1d5}
\prod
({\cal {O}}^{R(\pm,\pm)}_{n_l})^{m_l} |0\>
\qquad \hsp \sum_{} n_l m_l = N=n_1 n_5 \, ,
\ee
where $n_l$ is the twist, $m_l$ are integers 
and the superscripts denote (twice) the R-charges of the
operator. Here the ground states are described in the language of the
orbifold CFT; each ground state of the latter will map to a ground
state of the deformed CFT. Notice that there is 1-1 correspondence
between these states and the Fock states in (\ref{fock}).
Namely one can map the operators ${\cal {O}}^{R(\pm,\pm)}$
to the harmonic oscillators\footnote{This correspondence
straightforwardly extends to the general case where all
R ground states are considered and all bosonic and fermionic
oscillators are used in (\ref{fock}).},
\be \label{map}
\hat{a}_{-n}^{\pm 12} \quad \leftrightarrow \quad
{\cal {O}}_n^{R(\mp,\mp)},
\qquad \hat{a}_{-n}^{\pm 34} \quad \leftrightarrow
\quad {\cal {O}}_n^{R(\pm,\mp)}.
\ee
where $\hat{a}_{-n}^{\pm 12} \equiv (\hat{a}_{-n}^1 \pm i \hat{a}_{-n}^2)/\sqrt{2}$
and
$\hat{a}_{-n}^{\pm 34} \equiv (\hat{a}_{-n}^3 \pm i \hat{a}_{-n}^4)/\sqrt{2}$.
In particular, the frequency $n$ is mapped to the twist of the operator
and the R-charge to the angular momentum in the 1-2 and 3-4 plane.
However, the underlying 
algebra of these operators is different from the algebra of the harmonic
oscillators.

Motivated by this correspondence it was proposed in \cite{Lunin:2001jy}
that each of the solutions obtained via dualities
from the FP solution corresponds to a R ground state
and via spectral flow to a chiral primary \cite{Lunin:2002bj}.
One of the original motivations for this work was to understand
how such a map might work. Whilst it was clear from these
works that the frequencies
involved in the Fourier decomposition of the curve should map
to twists of operators, it was unclear what the meaning of the
amplitudes is in general and moreover a generic curve has far more parameters
than an operator of the form (\ref{d1d5}). In our
discussion of the FP system we have seen that the geometry is
more properly viewed as dual to a coherent state rather than a
single Fock state.  The coherent state however viewed as
linear superposition of Fock states (see (\ref{cohe})) contains states
that do not satisfy the constraint $N_L=-p_9 w$ and
therefore do not map to R ground states after the dualities.
This then leads to the following proposal 
for the map between geometries and states \cite{Skenderis:2006ah}
\footnote{A map between density matrices of the CFT states built
from 4 bosonic oscillators and modified fuzzball solutions has been
recently discussed in \cite{Alday:2006nd}. Here we provide a map
between the original fuzzball solutions and superpositions
of R ground states of the D1-D5 system.}:

{\it Given a curve $F^i(v)$
we construct the corresponding coherent state in the FP system and then
find which Fock states in this coherent state satisfy $N_L=- p_9 w$.
Applying the map (\ref{map}) then yields the superposition of R ground
states that is proposed to be dual to the D1-D5 geometry.}

Let us see how this works in some simple examples. The simplest case
is that of a circular planar curve that we may take to lie in the 1-2
plane:
\be \la{circ-00}
F^1(v) = \frac{\sqrt{2 N}}{n} \cos 2 \pi n \frac{ v}{L}, \qquad
F^2(v) = \frac{\sqrt{2 N}}{n} \sin 2 \pi n \frac{ v}{L},
\qquad F^3=F^4=0,
\ee
where $L$ is the length of the curve and the overall factors
are fixed by requiring that the solution has the correct
charges (this will be explained in the main text).
The corresponding coherent state can immediately be read off from the
curve
\be \la{circ-01}
|a_n^{-12}; a_n^{+12}; a_n^{-34}; a_n^{+34})=
|\sqrt{N/{n}};0;0;0).
\ee
In this case there is a single state with $N_L = N = - w p_9$
contained in this coherent state,
namely
\be
|N/n\> = (\hat{a}_{-n}^{-12})^{N/n} |0\>.
\ee
Using the map (\ref{map}) we get that the D1-D5 solution
based on the circle is dual to the R ground state
\be
|circle)=\left({\cal {O}}_n^{R(+,+)}\right)^{N/n}
\ee
which was the proposal in \cite{Lunin:2001jy}.

As soon as
one moves to more complicated curves, however, the correspondence
becomes more complex, as there is more than one Fock state
with $N_L=-w p_9$. For example
the next simplest case is the solution based on an ellipse
\be \la{ellp-00}
F^1(v) = \frac{\sqrt{2 N}}{n} a \cos 2 \pi n \frac{ v}{L}, \qquad
F^2(v) = \frac{\sqrt{2 N}}{n} b \sin 2 \pi n \frac{ v}{L},
\qquad F^3=F^4=0,
\ee
with $a^2 + b^2=2$. Following our prescription
we obtain the following superposition
\be \la{ellp-01}
|ellipse)= \sum_{k=0}^{N/n}
%2^{-N/2n} \sqrt{\frac{(\frac{N}{n})!}{(N/n-k)! k!}} \a^{N/n-k} \b^k
\frac{1}{2^{\frac{N}{n}}} \sqrt{\frac{(\frac{N}{n}   )!}{(\frac{N}{n}-k)! k!}}
(a+b)^{\frac{N}{n}-k}
(a-b)^k
\left({\cal {O}}_n^{R(+,+)}\right)^{\frac{N}{n}-k}
\left({\cal {O}}_n^{R(-,-)}\right)^{k},
\ee
as is explained in section 2.3.
The superposition for a general curve will involve a large number of
Fock states.

Given such a map from curves to superpositions of states
the question is whether the correspondence
can be checked quantitatively. The D1-D5 solutions
approach $AdS_3 \times S^3$ (times $T^4$ or $K3$)
in the decoupling limit so one can use the AdS/CFT correspondence to
make detailed quantitative tests.
Recall that the deviations of the solution from $AdS_3 \times S^3$
encode vacuum expectation values of chiral primary operators
(and possible deformations of the CFT by such operators),
so by analyzing the asymptotics one can in principle
completely characterize the ground state of the boundary theory.

Before proceeding to explain this, let us contrast the somewhat different
meanings that one attaches to the statement
``a geometry is dual to a state $|S\>$''. In the context
of the FP system, the state $|S\>$ is meant to provide
the source for the supergravity solution and because
of that we argued it should be a coherent state. In the
context of the D1-D5 system however the same statement means that
the ground state of the dual field theory is the state $|S\>$
(so $|S\>$ need not be approximated by a classical solution)
and the vevs of gauge invariant operators on this state,
$\<S|{\cal O}|S\>$, are encoded in the asymptotics of the
solution.

The D1-D5 system is governed by a 1+1 dimensional theory with
${\cal N}=(4,4)$ supersymmetry. This theory has
Coulomb and Higgs branches (which are distinct even quantum mechanically)
\cite{Witten:1997yu,Dijkgraaf:1998gf,Seiberg:1999xz}.
The boundary CFT is the IR limit of the theory on the Higgs branch. Thus
the fuzzball solutions should be in correspondence with
the Higgs branch. Note that due to strong infrared fluctuations 
in 1+1 dimensions one usually encounters wavefunctions 
%that spread over the whole of Higgs branch 
rather than continuous moduli spaces of the quantum states. 
So more properly one should view the fuzzball solutions as dual to
wavefunctions on the Higgs branch.
These wavefunctions, however, may be localized
around specific regions in the large $N$ limit and one should
view our proposed correspondence in this way.

The vevs of gauge invariant operators in this 1+1 dimensional theory can be
computed from the asymptotics of the solution.
The existence of such a relationship follows from the basic AdS/CFT dictionary
that relates bulk fields
to boundary operators and the bulk partition function
to boundary correlation functions. The implementation
of this program is however very subtle and precise formulae
for the 1-point functions for solutions with asymptotics to $AdS \times S$
were only recently obtained \cite{Skenderis:2006uy}.

Naively the vev of an operator of dimension $k$ is linearly related
to coefficients of order $z^k$ in the asymptotic expansion of the 
solution, where $z$ is a radial coordinate
(with the boundary of AdS located at $z=0$.) The actual map however 
is more complicated and involves
in addition a variety of non-linear contributions from terms of lower
order $z^l$, $l<k$. There are four sources of such non-linear
contributions, as we now discuss. 

Recall that the holographic 1-point functions
are derived by functionally differentiating the {\it renormalized}
on-shell action w.r.t. the corresponding sources
(see, for example, the review \cite{Skenderis:2002wp}). 
The most transparent way to describe the outcome of this
computation is to use a radial Hamiltonian 
language where the radial coordinate plays the role of time. 
Ignoring for the moment the compact part of the geometry,
so analyzing only the $(p+1)$ dimensional theory,
one finds that the renormalized 1-point function
of an operator of dimension
$k$ is exactly equal to the part of the corresponding 
radial canonical momentum that has
dimension $k$ \cite{Papadimitriou:2004ap,Papadimitriou:2004rz}.
This coefficient is related to the asymptotic coefficients 
but the map is in general non-linear (due to the non-linear
nature of gravitational field equations). This is the
first source of non-linearities and is essentially due
to general covariance, since it is the canonical momentum
that transforms properly under diffeomorphisms, not each coefficient
of the asymptotic expansion. So although the relation between
the 1-point function and momentum is linear, the relation
with the asymptotic coefficients is non-linear \cite{deHaro:2000xn}.

Taking into account the compact part of the geometry
leads to additional subtleties \cite{Skenderis:2006uy}.
Firstly, one needs to understand the
map between bulk fields and boundary operators beyond the linearized
approximation, i.e. we need the non-linear Kaluza-Klein reduction
map. Furthermore, this map should be gauge invariant
i.e. independent of the parametrization of the compact space.
The latter is dealt with by constructing gauge invariant variables,
which are non-linear in terms of the original fields. Finally,
1-point functions may receive contributions from boundary terms
in higher dimensions. Such contributions are responsible for the
extremal correlators \cite{D'Hoker:1999ea} and induce terms 
non-linear in momenta in the 1-point functions.

For the case at hand, the first step is to reduce the 10 dimensional
solution over $T^4$ or $K3$. We show that the fuzzball solutions 
reduce to solutions of 6-dimensional supergravity coupled
to tensor multiplets.
% with $n_t=5\ (n_t=21)$ for $T^4$ ($K3$),
%that use an $SO(1,1)$ subgroup of $SO(5, n_t)$ group that 
%appears in the supergravity. 
These solutions (in the decoupling limit) 
are asymptotic to $AdS_3 \times S^3$. The next step is to find the 
non-linear gauge invariant KK map from 6 to 3 dimensions. 
Following \cite{Skenderis:2006uy}, this is done to second order 
in the fluctuations using 
(and extending) the results of \cite{Arutyunov:2000by,Pank}. 
The results up to this order are sufficient to derive (after taking 
into account the subtle issue of extremal correlators) the 
vevs of all 1/2 BPS operators up to dimension 2. 
This includes in particular the conserved charges and the stress 
energy tensor. We emphasize that the non-linear terms are crucial 
in getting the right physics. We also discuss the 
vevs of higher dimension operators but these results are only 
qualitative as we did not compute the non-linear contributions;
these could be computed along the lines
described above, but the computation becomes very tedious.
One point functions for this system have also 
been discussed in the context of black rings 
\cite{Alday:2005xj}, although the non-linear 
terms (which play a crucial role) were not included there.

The final results for the vevs of the fuzzball solution are 
given in section \ref{v-fuzz}. In particular, the vevs of the stress energy 
is (non-trivially) zero for all solutions, consistent with the 
fact that the solutions are supersymmetric. The vevs of the other 
operators are 
\bea
\left < {\cal O}_{S^1_i} \right > &=& \frac{n_1 n_5}{4 \pi}
(- 4 \sqrt{2} f^{5}_{1i}); \qquad (i{=}1,{\ldots}, 4) \\
\left < {\cal O}_{S^2_I} \right > 
&=& \frac{n_1 n_5}{4 \pi} ( \sqrt{6} (f^1_{2I} -
f^5_{2I}) ); \qquad (I{=}1,{\ldots}, 9) \nn \\
\left < {\cal O}_{\S^2_I} \right > &=& \frac{n_1 n_5}{4 \pi} \sqrt{2}
( -  (f^1_{2I} +f^5_{2I}) + 8 a^{\a -} a^{\b +} f_{I\a\b} ); 
\qquad (\a{=}1,{\ldots},3) \nn \\
\left < J^{+\a} \right >  &=& \frac{n_1 n_5}{2 \pi} a^{\a+} (dy -dt); \hsp
\left < J^{-\a} \right > = - \frac{n_1 n_5}{2 \pi} a^{\a-} (dy + dt), 
 \nn
\eea
where ${\cal O}_{S^1_i}$ are dimension 1 operators,
${\cal O}_{S^2_I}, {\cal O}_{\S^2_I}$ are dimension 2 operators,and 
$J^{\pm\a}$ are R-symmetry currents. 
These operators correspond
to the lowest lying KK states, the KK spectrum consisting of
two towers of spin 1 supermultiplets, the $S$ and $\S$ towers,
and a tower of spin 2 supermultiplets, which contain the gauge field
that is dual to the R-symmetry current. The coefficients 
$f^{5}_{1i}, f^1_{2I}, f^5_{2I}, a^{\pm \a} $ appear in the
asymptotic expansion of the harmonic functions that specify the solution, see  
(\ref{p0})-(\ref{a1}), and  $f_{I\a\b}$ is a certain triple overlap
of spherical harmonics.  Expressed in terms of the defining curve $F^i$, 
the degree $k$ coefficients
involve symmetric rank $k$ polynomials of $F^i$, see (\ref{p00}).
In general, the vev of an operator of dimension 
$k$ depends linearly on degree $k$ coefficients and non-linearly
on lower degree coefficients but such that the sum of degrees is
$k$.

Any proposal for the field theory dual of these geometries should 
reproduce these vevs. Now, except when the 
curve is circular, operators charged wrt the R-symmetry acquire a vev.  
This implies immediately that the ground state of the field theory dual 
cannot be an eigenstate of R-symmetry
since if this were the case only neutral operators would acquire a
vev \cite{Skenderis:2006ah}. 
So none of the fuzzball solutions, except the circular ones, can 
correspond to a single R-ground state. Indeed, we have argued 
above (and in \cite{Skenderis:2006ah}) that
these solutions should instead be dual to particular superpositions
of R-ground states. 

To test this proposal we discuss in some detail the case of the
ellipse, comparing the vevs extracted from the supergravity solution
with those implicit from the corresponding superposition of states in
the field theory. We find complete matching for all kinematical 
properties of these vevs, thus demonstrating the consistency of our
proposal. Moreover, the first dynamical test - matching of the R
charges - is passed. To match the other vevs would require a knowledge
of certain multiparticle three point functions at strong coupling,
and is thus not currently possible. However, approximating the
required three point functions using free harmonic oscillators leads
to vevs which are in remarkable agreement with those extracted from
the supergravity solution. This agreement suggests that certain 
three point functions in the dual CFT may be well approximated by free
field computations, a result which in itself merits further investigation. 
Our proposal therefore passes all kinematical and all accessible 
dynamical tests, with other dynamical tests requiring going beyond the
supergravity approximation.

Given that the original 
fuzzball solutions do not correspond to single R ground states,
one may wonder whether there are other supergravity solutions that 
do correspond to a given R ground state. 
A necessary condition for this would be that the vevs of all charged
operators are all zero, and this will only be the case if the solution
preserves an $SO(2) \times SO(2)$ symmetry (among the original solutions
only the circular one had this symmetry). We give the most 
general asymptotic supergravity solution consistent with these requirements. 
Different solutions with such asymptotics are parametrized
by the vevs of the neutral operators, and to obtain these vevs
one needs the complete solutions.

One way to produce solutions 
with an $SO(2) \times SO(2)$ symmetry is to take appropriate 
superpositions of the non-symmetric solutions. We discuss 
how to do such an averaging in general and we work out the 
details for the ellipse and for a curve that is a straight line
followed by a semi-circle. This latter case yields the 
Aichelburg-Sexl metric namely the metric describing a massless
particle moving along a greater circle on $S^3$ and sitting at 
the center of $AdS_3$. Solutions with the same $SO(2) \times SO(2)$
symmetry can also be produced using disconnected circular curves; one
would expect that such solutions are related to Coulomb rather than
Higgs branch physics. 

We then discuss the relationship between 
such symmetric geometries and R ground states. We argue that 
the vevs for neutral operators in a particular ground state 
can be related to three point functions at the conformal point. Thus
with knowledge of the latter one can distinguish 
whether a given geometry corresponds to a
particular R ground state. However, we find that implementing
this procedure generically requires going beyond the
leading supergravity approximation: one would need to know three point
functions of multi particle operators, not captured by supergravity, as
well as $1/N$ corrections. Thus we cannot currently determine
which geometries are indeed dual to R ground states; indeed even the 
solutions based on disconnected curves (which should be Coulomb
branch) could not be ruled out. 

So what do our results imply for the fuzzball program?  
Firstly, our results support the overall picture; the fuzzball
solutions can be in correspondence with the black hole microstates
in a way that is compatible with the AdS/CFT correspondence and
our computations provide the most stringent test to date.
The detailed correspondence however is more complicated than 
anticipated. In particular a generic fuzzball solution 
corresponds to a superposition of many R ground states,
and in general one would need to go beyond the 
leading supergravity to properly describe the system, even 
in this simplest 2-charge system. It has long been appreciated that
most of the fuzzball solutions, despite being regular, have regions 
of high curvature so are at best extrapolations of the actual 
solutions describing the microstates. Here we see that even for
solutions with low curvature everywhere, such as the ones 
based on large ellipses, one needs to go beyond the 
leading supergravity to test any proposed correspondence.

There has been a lot of interest in finding and analyzing fuzzball 
geometries in systems with more charges which have 
classical horizons \cite{Bena} but a precise matching between these 
geometries and black hole microstates has not been established. 
Such a matching is clearly necessary, both to demonstrate that the correct
geometries have been identified and to find for what fraction of the total
entropy these account. A precise correspondence would also be
important in understanding the quantization of the geometries and, most
importantly of all, how the black hole properties emerge.

A key result of our work is that the vevs encoded by a given geometry
give significant information about the field theory dual, and
distinguish between geometries with the same charges (mass, angular
momentum). In particular, dipole and higher multipole moments are
related to the vevs of operators with dimension two or greater.
Vevs determined by kinematics can by themselves rule out
proposed correspondences, as shown in \cite{Skenderis:2006ah} and here, 
and vevs determined by dynamics are strong
tests of a given proposal, when they can be computed on both sides. 
In particular, whilst our solutions based on disconnected curves 
pass all kinematical tests to correspond to R ground states on 
the Higgs branch, they should be ruled out by dynamical tests. 

Previous work has often focused on computing two point functions and
relating them to those in the dual field theory, and vice versa,
see for example \cite{Balasubramanian:2005qu}, but extracting vevs is much
easier, since one needs only the geometry itself, rather than 
solving fluctuation equations in the geometry. Thus one can easily 
extract vevs from geometries with few symmetries, where the corresponding
fluctuation equations are intractable. It hence seems worthwhile to
explore whether the techniques developed here can give useful information
in the context of other fuzzball geometries. One can analyze any 
fuzzball geometry which has a throat region using AdS/CFT
techniques, with the formalism developed here being directly
applicable to three charge black strings in six dimensions. Black
rings in six dimensions could also be explored using the same
formalism; indeed the extracted data should uniquely identify 
the field theory dual.

\bigskip

The plan of the paper is as follows. In section \ref{FP} we discuss the
relationship between solitonic string supergravity solutions and
coherent states of the fundamental string. In section \ref{fuzzball}
we introduce the dual solutions in the D1-D5 system, and discuss the
embedding of their decoupling limit into 6-dimensional supergravity.
In section \ref{harm-ex} we discuss the asymptotic expansion of these
six dimensional solutions near the $AdS_3 \times S^3$ boundary. In
section \ref{vev-proc} we explain how the vevs of field theory operators can
be extracted from these asymptotics. In section \ref{v-fuzz} we give the
explicit values of these vevs for the fuzzball solutions in full
generality, and in section \ref{examples} we specialize to the examples of
solutions sourced by circular and elliptical curves. In section \ref{fti} 
we recall relevant features of the dual field theory, and discuss how
the vevs can be related to three point functions at the conformal point.
In section \ref{corresp} we move on to the correspondence between fuzzball
geometries and superpositions of chiral primaries, giving evidence for
our proposed correspondence in terms of the matching of the vevs for
the ellipsoidal case. In section \ref{symmetric-s} we discuss the asymptotics of
a geometry dual to a single chiral primary, and give some examples of
solutions which have such asymptotics. In section \ref{dyn-2} we discuss
the correspondence between symmetric geometries and chiral primaries,
emphasizing that dynamical tests require going beyond the leading
supergravity approximation. In section \ref{flat-reg} we discuss how the
asymptotically flat part of the geometry can be included in the field
theory description.

Throughout the paper we use a number of technical results which are
contained in appendices. Appendix \ref{apa} contains various properties
of $S^3$ spherical harmonics whilst appendix \ref{apa1} proves an addition
theorem for harmonic functions on $R^4$. Appendix \ref{apb} discusses the
perturbative expansion of six-dimensional field equations about the
$AdS_3 \times S^3$ background. Appendix \ref{apd} discusses the
supergravity computation of certain three point functions, whilst 
appendix \ref{apc} contains a derivation of the one point function 
for the energy momentum tensor in this system. Appendix \ref{apf}
concerns the three point functions in the orbifold CFT; we argue
that these differ from those computed in supergravity and that they
are therefore not protected by any non-renormalization theorem.

\section{FP system and perturbative states} \label{FP}

We begin by discussing solitonic string supergravity solutions
and their relation to perturbative string states. The FP solutions
are characterized by a curve $F^I(x^+)$ describing the transverse
displacement of the string. For later purposes only 4 transverse
directions will be excited so the curve is confined to $R^4$
but for now we keep the discussion general. The supergravity solution
describing an oscillating string is given by  \cite{Callan:1995hn,Dabholkar:1995nc}
\bea
&&ds^2=H^{-1}(-dx^- dx^+ + K (dx^+)^2 
- 2 A_I dx^I dx^+) + d x_I d x_I \nonumber \\
&& H=1 + \frac{Q_f}{|\vec{x} - \vec{F}(x^+)|^6}, \quad
K = \frac{Q_f |\dot{F}|^2}{|\vec{x} - \vec{F}(x^+)|^6}, \quad
A_I = \frac{Q_f \dot{F}_I}{|\vec{x} - \vec{F}(x^+)|^6}
\eea
with suitable $B$ field and dilaton. Here $x^{\pm} = x^0 \pm x^9$
are lightcone coordinates, $\vec{x}$ are 8 transverse coordinates and
$x^9 \equiv x^9 + 2 \pi R_9$. $\dot{F}_I$ denotes the derivative with
respect to $x^+$. The fundamental string charge
$Q_f$ is proportional to the number of fundamental strings. The ADM
mass and momentum along the compact direction are respectively
\cite{Callan:1995hn,Dabholkar:1995nc}
\be \la{adm}
M = k Q_f (1 + |\dot{F}|_0^2); \hsp
P^9 = - k Q_f |\dot{F}|_0^2,
\ee
where the subscript denotes the zero mode and $k = 3 \w_7/2 \k^2$ with
$\w_7$ the volume of the $S^7$. The angular momenta in the transverse
directions are similarly given by
\be \la{ang-mom2}
J^{IJ} = k Q_f (F^J \dot{F}^I - F^I \dot{F}^J)_0.
\ee
As we will review below, these are
exactly the conserved quantities of
a string which wraps around the compact direction
$w$ times and whose transverse profile is given by $F^I$.

\subsection{String quantization}

To relate the supergravity solutions to perturbative string states,
let us consider quantizing a string propagating in a flat background;
we discuss this in some detail since the preferred gauge choice is
a non standard light cone gauge. The relevant part of the worldsheet action is
\be
S = \frac{1}{4 \pi \a'} \int d^2 \s (\pa_{+} X^M \pa_{-} X_M + \cdots),
\ee
where the worldsheet metric is gauge fixed to $- g_{\t \t} = g_{\s \s}
= 1$. Fermions will not play any role in the discussion here and will
be suppressed. We will also set $\a' = 2$ to simplify formulae.
Null worldsheet coordinates are introduced by
setting $\s^{\pm} = (\t \pm \s)$ and a lightcone gauge can be chosen
for $V$ such that
\be
X^+ = (w^+ \s^+ + \td{w}^+ \s^-).
\ee
A similar choice of lightcone gauge for open strings has been 
discussed in \cite{Skenderis:2003da}.
The other fields are then expanded in harmonics as
\bea
X^- &=& x^- + (w^- \s^+ + \td{w}^- \s^-) + \sum_{n} \frac{1}{\sqrt{|n|} }
(a^{-}_n e^{- i n \s^+} +
\td{a}^-_n e^{- i n \s^-}); \\
X^I &=& x^I + p^I (\s^+ + \s^-) + \sum_{n}
\frac{1}{\sqrt{|n|}} (a^{I}_n e^{- i n \s^+} + \td{a}^I_n e^{- i n \s^-}). \nn
\eea
Reality of $X^M$ demands that $a^M_{-n} = (a^M_n)^{\dagger}$.
The Virasoro constraints are
\be
T_{++} = \pa_{+} X^M \pa_{+} X_M = 0; \hsp
T_{--} = \pa_{-} X^M \pa_{-} X_M = 0.
\ee
At the classical level this enforces
\bea
&& (- w^+ w^- + (p^I)^2) \d_{m,0} + i \frac{m }{\sqrt{|m|}} ( w^+ a^{-}_m
- 2 p^I a^{I}_m) 
%\\
%&& \hsp \hsp \hsp \hsp
+ \sum_{n} \frac{n(n-m)}{\sqrt{| n (n-m)|}}  a^{I}_n a^{I}_{m-n} = 0;
\nn 
\\
&& (- \td{w}^+ \td{w}^- + (p^I)^2) \d_{m,0} + i \frac{m
  }{\sqrt{|m|}}  ( \td{w}^+ \td{a}^{-}_m - 2 p^I \td{a}^{I}_m) 
+ \sum_{n} \frac{n(n-m)}{\sqrt{ | n (n-m) |}}
\td{a}^{I}_n \td{a}^{I}_{m-n} = 0, \nn
\eea
thereby determining the non-dynamical field $X^-$ in terms of the
dynamical transverse fields $X^I$, as in standard lightcone gauge.
The conserved momentum and winding charges are given by
\be
P^{M} = \frac{1}{4 \pi } \int^{2\pi}_{0} d\s (\pa_{\t} X^M); \hsp
W^M = \frac{1}{2 \pi } \int^{2\pi}_{0} d\s (\pa_{\s} X^M),
\ee
which take the values
\bea
P^M &=& \left ( \frac{1}{4} 
(w^- + w^+  + \td{w}^- + \td{w}^+), 
 \frac{1}{4} (w^+ - w^- + \td{w}^+ - \td{w}^-), p^I \right ); \\
W^M &=& \left ( \frac{1}{2} (w^- + w^+  - \td{w}^- - \td{w}^+), 
\frac{1}{2} (w^+ - w^- - \td{w}^+ + \td{w}^-), 0 \right ). \nn
\eea
In order for the string not to wind the time direction, one thus needs
\be
W^0 = \frac{1}{2} (w^- + w^+  - \td{w}^- - \td{w}^+) = 0.
\ee
We are interested in states with only left moving excitations and no
transverse momentum, namely the $\td{w}^+ = 0$ sector. For these the
momentum and winding charges are
\bea
P^M &=& (\half w R_9 - \frac{p_9}{R_9}, \frac{p_9}{R_9},0);
\hsp W^M = (0,w R_9, 0); \la{mom-win} \\
w^+ & \equiv & w R_9; \hsp
w^-  \equiv  - 2 \frac{p_9}{R_9}. \nn
\eea
Restricting to such states the $L_0$ constraint becomes
\be \la{lo-con}
p_9 w + \sum_{n > 0} n a^I_{-n} a^{I}_{n} \equiv
p_9 w + N_L = 0.
\ee
The angular momenta in the transverse directions are given by the
usual expressions
\be \la{ang-mom}
J^{IJ} = \frac{1}{4 \pi } \int_0^{2\pi} d\s (X^{J} \pa_{\t} X^I -
X^{I} \pa_{\t} X^J) = -i \sum_{n > 0} ( a^I_{-n} a_n^J - a_{-n}^J a_n^I).
\ee
Quantization proceeds in the standard way, with the oscillators
satisfying the commutation relations
\be
\left [\hat{a}^I_{n}, (\hat{a}^J_m)^{\dagger} \right ] =
\d_{m,n} \d^{IJ},
\ee
and states being built out of creation operators
$(\hat{a}^I_m)^{\dagger}$ acting on the vacuum. The classical
expressions continue to hold, replacing $a_{m}^I$ by operators
$\hat{a}^I_m$, with appropriate shift in $L_0$ (which is negligible
in the large charge limit).

\subsection{Relation to classical curves}

On rather general grounds, one expects that the supergravity solution
characterized by a null curve corresponds to a coherent state
of string oscillators. To be more precise, let us Fourier
expand the classical curve
\be
F^I(x^+) = \sum_{n>0} \frac{1}{\sqrt{n}} \left(
\a_n^I e^{-i n \s^+} + (\a_n^I)^* e^{i n \s^+}\right)
\ee
where $\a_n^I$ are (complex) numerical coefficients and
$x^+ = w R_9 \s^+$. Then the coherent state $|F^I)$ of string oscillators
that corresponds to this curve is given by
\be
|F^I)=\prod_{n,I}| \a_n^I )
\ee
where $| \a_n^I )$ is a coherent state of the oscillator $\hat{a}_n^I$,
i.e. it satisfies,
\be
\hat{a} | \a ) = \a | \a )
\ee
where we suppress the super and subscripts for clarity.
Recall the coherent states are related to the Fock states by
\be \label{cohe}
|\a)=e^{-|\a|^2/2} \sum_k \frac{\a^k}{\sqrt{k!}} |k\>
\ee
and 
\be \la{harm-osc}
|k\> = \frac{1}{\sqrt{k!}} (\hat{a}^\dagger)^k |0\>
\ee
is the standard $k$th excited state. By construction
\be \label{exc_n}
(F^I | \hat{N}_L |F^I) \equiv N_L =  \sum_{n > 0} n | \a^I_n |^2.
\ee
From (\ref{lo-con}) and (\ref{mom-win}) we find that
\be \la{aver-charg}
(F^I | \hat{P}^0 |F^I) = (\half w R_9 + \frac{1}{w R_9} N_L); \hsp
(F^I | \hat{P}^9 |F^I) = - \frac{1}{w R_9} N_L.
\ee
Now note that the zero mode of $(\dot{F}^I)^2$ is given by
$2 N_L/(w R_9)^2$. This means that the mass and momentum of the
supergravity solution associated with this curve are,
using (\ref{adm}),
\be
M = k Q_f (1 + \frac{2 N_L}{(w R_9)^2}); \hsp
P^9 = - k Q_f \frac{2 N_L}{(w R_9)^2},
\ee
which agree with the expressions (\ref{aver-charg}) provided that
\be
k Q_f = \half w R_9,
\ee
which is the relationship found in
\cite{Callan:1995hn,Dabholkar:1995nc}. Moreover,
\be
(F^I | \hat{J}^{IJ} |F^I) 
= \frac{1}{2} w R_9 ( F^J \dot{F}^I - F^I  \dot{F}^J)_0,
\ee
which manifestly agrees with the expression (\ref{ang-mom2}).

\subsection{Examples}

Consider an elliptical curve in the 1-2 plane, such that
\be
F^1 = \frac{\sqrt{2N}}{n} a \cos (n \s^+); \hsp
F^2 = \frac{\sqrt{2N}}{n} b \sin (n \s^+),
\ee
with $(a^2 + b^2) = 2$; this case was discussed in the
introduction around (\ref{circ-00}) and (\ref{ellp-00}).
The amplitude of the curve is fixed
such that 
the angular momentum in the 1-2 plane is
\be
J^{12} = - \frac{N}{n} a b,
\ee
and the total excitation number defined in (\ref{exc_n})
is $N_L = N = - w p_9$. This ensures that the mass and momenta 
match that of the supergravity solution, as described in the previous
subsection.

Introducing the usual combinations of oscillators with definite
angular momenta in the 1-2 plane
\be
\hat{a}_n^{\pm 12} \equiv \frac{1}{\sqrt{2}} (\hat{a}_n^1 \pm i
  \hat{a}^2_n),
\ee
the coherent state corresponding to the curve is
\be
| a_{n}^{-12}; a_n^{+ 12} ) = | \frac{\sqrt{N}}{2 \sqrt{n}} (a + b);
 \frac{\sqrt{N}}{2 \sqrt{n}} (a - b) ),
\ee
which in the case of the circle ($\a = \b$) reduces to (\ref{circ-01}).
Extracting from this coherent state those states which satisfy
$N_L = N$ gives
\be \la{ellp-02}
|ellipse)= \sum_{k=0}^{N/n}
%2^{-N/2n} \sqrt{\frac{(\frac{N}{n})!}{(N/n-k)! k!}} \a^{N/n-k} \b^k
\frac{1}{2^{\frac{N}{n}}} \sqrt{\frac{(\frac{N}{n}   )!}{(\frac{N}{n}-k)! k!}}
(a+b)^{\frac{N}{n}-k}
(a-b)^k |k_{-12} = (\frac{N}{n}-k); k_{+12} = k \>,
\ee
which leads to the corresponding superposition (\ref{ellp-01}) in the
dual D1-D5 system.

\section{The fuzzball solutions} \la{fuzzball}

We now consider the two charge fuzzball solutions in the D1-D5
system, obtained from the FP chiral null models by a chain of
dualities. These fuzzball solutions were constructed by Lunin and
Mathur \cite{Lunin:2001fv,Lunin:2001jy} and are given by
\bea
ds^2 &=& f_{1}^{- 1/2} f_{5}^{- 1/2} \left ( - (dt - A)^2 + (dy +
B)^2 \right ) + f_{1}^{1/2} f_{5}^{1/2} dx \cdot dx +
f_{1}^{1/2} f_{5}^{- 1/2} dz \cdot dz; \nn \\
e^{2 \Phi} &=& f_1 f_5^{-1}; \la{ez1} \\
C_{ti} &=& f_{1}^{-1} B_{i}; \hsp
C_{ty} = f_{1}^{-1}; \nn \\
C_{yi} &=& f_{1}^{-1} A_{i}; \hsp
C_{ij} = c_{ij} - f_{1}^{-1} (A_i B_j - A_j B_i), \nn
\eea
where $i,j$ are vector indices in the transverse $R^4$ and 
the metric is in the string frame. These fields solve the
equations of motion following from the type IIB action
\be
S = \frac{1}{2 \k_{10}^2} \int d^{10}x \sqrt{-g_{10}} \left(e^{-2 \Phi}
(R_{10} + 4 (\pa \Phi)^2) - \frac{1}{12} F_3^2 + \cdots\right),
\ee
where $F_3$ is the curvature of the two form $C$ and $2 \k_{10}^2 = (2
\pi)^7 (\a')^4$ (we set $g_s=1$ since it plays no role in our discussion),
provided the following equations hold
\bea
&&dc = \ast_{4} df_5, \hsp d B = \ast_{4} dA, \nonumber \\
&& \Box_{4} f_1 = \Box_{4} f_5 = \Box_{4} A_i =0, \qquad
\partial^i A_i = 0. \label{cond}
\eea
where the Hodge dual $\ast_4$ and $\Box_{4}$ are defined on the four
(flat) non-compact overall transverse directions $x^i$. The compact part
of the geometry does not play a role; it could be either $T^4$ or $K3$.

A solution to the conditions (\ref{cond}) based on an arbitrary closed
curve $F^i(v)$ of length $L$ in $R^4$ is given by
\be \la{mathu1}
f_{5} = 1 + \frac{Q_5}{L} \int^{L}_{0} \frac{dv}{\left | x - F \right
  |^2}; \hsp
f_1 = 1 + \frac{Q_5}{L} \int_{0}^L
\frac{dv | \dot{F} |^2}{\left | x - F \right |^2}; \hsp
A_i = \frac{Q_5}{L} \int_{0}^{L} \frac{\dot{F}_i dv}{\left | x - F
  \right |^2}.
\ee
It was argued in \cite{Lunin:2001jy} that
these solutions are related to the R ground states (and via
spectral flow to  chiral primaries \cite{Lunin:2002bj})
common to both the $T^4$ and $K3$ CFTs.
%Other supergravity solutions related to R ground states/chiral primaries
%specific to the compact manifold such as those discussed in
%\cite{Taylor:2005db,Lunin:2002iz} are not considered here.
The physical interpretation of these solutions is that the D1 and D5
brane sources are distributed on a curve in the transverse $R^4$. The
D5-branes are uniformly distributed along this curve, but
the D1-brane density at any point on the curve depends on the tangent
to the curve. The total one brane charge is given by
\be \la{one-brane}
Q_1 = \frac{Q_{5}}{L} \int_{0}^L | \dot{F} |^2 dv.
\ee
Both the $Q_i$ have dimensions of length squared and are related to
the integral charges by
\be \la{int-charg}
Q_1 = \frac{(\a')^3 n_1}{V}; \hsp
Q_5 = \a' n_5,
\ee
where $(2 \pi)^4 V$ is the volume of the compact manifold.
Furthermore, the length of the curve
is given by
\be \la{cur-leng}
L = 2 \pi Q_5/R,
\ee
where $R$ is the radius of the $y$ circle.

The holographic analysis in this paper will be done for the
general class of solutions (\ref{ez1}) satisfying (\ref{cond}).
Results appropriate for the solutions determined by
(\ref{mathu1}) will be obtained by specializing the general
results to this case and we will indicate how this is done
at each step of the analysis.

\subsection{Compactification to six dimensions}

Since only the breathing mode of the compact manifold is excited,
it is convenient to compactify and work with solutions of six-dimensional
supergravity. The effective six-dimensional (Einstein) metric
coincides with the six-dimensional part of the (string frame) metric
above (because the would be 
six-dimensional dilaton  $\phi_6 = \Phi - \qu \ln {\rm det} g_{M^4}$,
where $g_{M^4}$ is the metric on the compact space, 
is constant). Thus the
six-dimensional metric
\be \la{six-met}
ds^2 = f_{1}^{- 1/2} f_{5}^{- 1/2} \left ( - (dt - A)^2 + (dy +
B)^2 \right ) + f_{1}^{1/2} f_{5}^{1/2} dx \cdot dx
\ee
along with the scalar field and tensor field of (\ref{ez1})
satisfy the equations of motion following
from the reduced action
\be \la{eqb}
S = \frac{1}{2 \k_6^2} \int d^6 x \sqrt{-g} \left (R - (\pa \Phi)^2 -
\frac{1}{12} e^{2 \Phi} F_3^2 \right ),
\ee
where $R$ is the six-dimensional curvature and $F_3$ is the curvature of
the antisymmetric tensor field $C$. These equations of motion are
\bea
R_{MN} &=& \qu e^{2 \Phi} (F_{MPQ} F_{N}^{\sp PQ} - \frac{1}{6} F^2
g_{MN}) + \pa_{M} \Phi \pa_{N} \Phi; \nn \\
D_{M} (e^{2 \Phi} F^{MNP} ) &=& 0; \hsp
\Box \Phi = \frac{1}{12} e^{2 \Phi} F^2. \la{six-eq}
\eea
Note that the six-dimensional scalar field
originates from the breathing mode of the compactification manifold.

The equations of motion which follow from the action (\ref{eqb}) can be embedded
into those of $d=6$, $N=4b$ supergravity coupled to $n_t$ tensor
multiplets, the covariant field equations for which were constructed
in \cite{Romans}. The bosonic field content of this theory is the graviton
and five self-dual tensor fields from the supergravity multiplet,
along with $n_t$ anti-self dual tensor fields and $5n_t$ scalars from the
tensor multiplets.

Following the notation of \cite{Sez,Arutyunov:2000by}
the bosonic field equations may be written as
\bea \la{x2}
R_{MN} &=& H^{m}_{MPQ} H_{N}^{m \sp PQ} + H^{r}_{MPQ} H^{r \sp PQ}_{N}
+ 2 P^{mr}_{M} P_{N}^{mr}; \\
D^{M}P_{M}^{mr} &=& \frac{\sqrt{2}}{3} H^{m MNP} H^{r}_{MNP},
\eea
along with Hodge duality conditions on the 3-forms
\be
H^{m}_{MNP} = \frac{1}{6} \ep_{MNPQRS} H^{m QRS}; \hsp
H^{r}_{MNP} = - \frac{1}{6} \ep_{MNPQRS} H^{r QRS}. \la{x1}
\ee
In these equations $m,n$ are $SO(5)$ vector indices running from
1 to 5 whilst $r,s$ are
$SO(n_t)$ vector indices running from 6 to $5+n_t$.
The three form field strengths are given by
\be
H^{m} = G^{A} V_{A}^m; \hsp
H^{r} = G^{A} V_{A}^r,
\ee
where $A \equiv \{n,r\} = 1, \cdots, 5+n_t$; $dG^{A} = 0$ and the vielbein on the
coset space $SO(5,n_t)/(SO(5) \times SO(n_t)$ satisfies
\be
V_{A}^m V_{B}^m - V_{A}^r V_{B}^r = \eta_{AB},
\ee
with $\eta_{AB} = \rm{diag} (+++++--- \cdots -)$. The associated
connection is
\be
d V V^{-1} =  \left ( \begin{array} {c c} Q^{mn} & \sqrt{2} P^{ms} \\
\sqrt{2} P^{nr} & Q^{rs} \end{array} \right ).
\ee
The equations of motion (\ref{six-eq}) can be
embedded into this theory using an $SO(1,1)$ subgroup as follows.
Let
\be
V^{m=5}_{5} = \cosh (\Phi); \hsp
V^{m=5}_{6} = \sinh (\Phi); \hsp
V^{r=6}_{5} = \sinh (\Phi); \hsp
V^{r=6}_{6} = \cosh (\Phi),
\ee
so that the connection is $\sqrt{2} P^{56} = d\Phi$. Now let\footnote{
The field strengths 
$G^5$ and $G^6$ were called $G^{\pm}$ in \cite{Skenderis:2006ah}.}
\be
G^{5} = \qu (F + e^{2 \Phi} \ast_6 F); \hsp
G^{6} = \qu (F - e^{2 \Phi} \ast_{6} F),
\ee
which are both closed using the three form equation in (\ref{six-eq}).
This implies that
\be
H^{m=5} = \qu e^{\Phi} (F + \ast_{6} F); \hsp
H^{r=6} = \qu e^{\Phi} (F - \ast_{6} F),
\ee
which manifestly have the correct Hodge duality properties to satisfy
(\ref{x1}). Substituting $H$ and $P$ into (\ref{x2}) also correctly
reproduces the Einstein and scalar field equations of (\ref{six-eq}).

Since this embedding uses only an $SO(1,1)$ subgroup it does not depend on the
details of the compactification manifold.  Thus one can use this
six-dimensional supergravity to
analyze the fuzzball geometries in both $T^4$ and $K3$ systems.
More generally,
the (anomaly free) case of $n_t=21$ gives the complete six dimensional
theory obtained by K3 compactification of type IIB supergravity.
For $T^4$ compactification of type IIB one obtains the maximally
supersymmetric non-chiral six-dimensional theory, whose field content
is a graviton, eight gravitinos, 5 self-dual and 5 anti-self dual
three forms, 16 gauge fields, 40 fermions and 25 scalars. (Bosonic) solutions of
this supergravity which do not have gauge fields switched on are
solutions of the chiral supergravity given above, with $n_t=5$.

\subsection{Asymptotically AdS limit}

In the appropriate decoupling limit, the solutions (\ref{ez1})
become asymptotically AdS. This corresponds to harmonic functions
with leading behavior $r^{-2}$. In terms of the harmonic
functions in (\ref{mathu1})
the decoupling limit amounts to
removing the constant terms in the harmonic functions $f_1$ and $f_5$.
(Later on in section \ref{flat-reg} we will discuss the
interpretation of these constant terms in
the dual CFT.) The solutions are then manifestly asymptotic to $AdS_3
\times S^3$ as $r \rightarrow \infty$. Firstly the metric asymptotes
to
\be \la{me}
ds_6^2 = \frac{r^2}{\sqrt{Q_1 Q_5}} (-dt^2 + dy^2) + \sqrt{Q_1 Q_5}
\left ( \frac{dr^2}{r^2} + d \Omega_3^2 \right );
\ee
whilst the three-forms and scalar field from (\ref{ez1}) asymptote to
\bea
F_{rty} &=& \frac{2r}{Q_1}; \hsp
F_{\Omega_3} = 2 Q_5; \hsp e^{2 \Phi_0} = \frac{Q_1}{Q_5}.
\eea
It is convenient to shift the scalar field so that $\Phi \to \Phi
- \Phi_{0}$ and rescale 
$G^{5} \to e^{\Phi_0} G^{5}$ and same for $G^6$. 
%, G^{6} \to e^{\Phi_0} G^{6}$.
Then the relevant background fields of the six-dimensional supergravity are
\bea
g^{o (m=5)}&=&H^{o(m=5)} = \frac{r}{ \sqrt{ Q_1 Q_5}} dr \wedge dt \wedge dy +
\sqrt{Q_1 Q_5} d \Omega_3; \la{g3} \\
V^{o(m=5)}_{5} &=& 1; \hsp V^{o(r=6)}_{6} = 1, \nn
\eea
with the off-diagonal components of the vielbein vanishing; the
anti-self dual field $g^{o (r=6)}=H^{r=6}$ 
vanishing and $\Phi$ being zero also.
Note that with the coordinate rescalings
$t \rightarrow t \sqrt{Q_1
  Q_5} $  and $y \rightarrow y \sqrt{Q_1 Q_5} $, the curvature radius
appears only as an overall scaling factor in both the metric
(\ref{me}) and the three form (\ref{g3}). When one rescales the
coordinates in this way, the new $y$ coordinate will have periodicity
$\td{R}  = R/\sqrt{Q_1 Q_5}$. 

The goal is to extract from the subleading asymptotics around the AdS
boundary the vevs of chiral primaries in the dual theory, and thus
investigate the matching with R vacua.
The strategy is as follows. First one expands the solution
systematically near the AdS boundary. Then one extracts from the
asymptotic solution the values of 6-dimensional gauge invariant
fields. These must then be reduced to three dimensional fields using
the KK map, and then the vevs can be extracted using holographic
renormalization.

\section{Harmonic expansion of fluctuations} \la{harm-ex}

Let us consider the asymptotic expansion of the
solution. The perturbations of the six-dimensional supergravity fields
relative to the $AdS_3 \times S^3$ background can be
expressed as
\be
g_{MN} = g^{o}_{MN} + h_{MN};
\hsp G^{A} = g^{o A} + g^{A}; \hsp
\phi^{mr}. \la{dq2}
%\hsp g^{A} = db^{A};
%V^{i}_{A} &=& \d^{i}_{A} + \phi^{ir} \d_{A}^r + \half \phi^{Ar}
%\phi^{jr} \d^{j}_{A}, \hsp
%V^{r}_{A} = \d^{r}_{A} + \phi^{ir} \d^{i}_{A} +
%\half \phi^{ir} \phi^{is} \d^{s}_{A}. \nn
\ee
%where $M=\{\m,a\}$ and $\mu$ ($a$) is  an AdS ($S^3$) index.
%Note that the latter quadratic expansion of the hyperbolic functions
%is sufficient for an analysis of the field equations to quadratic
%order.
These fluctuations can then be expanded in spherical harmonics as follows:
\bea
h_{\m \n} &=& \sum h_{\m\n}^I (x) Y^{I} (y), \la{flc1} \\
h_{\m a} &=& \sum (h_{\m}^{I_v} (x) Y_a^{I_v} (y) +
h_{(s)\m}^{I} (x) D_a Y^I (y) ), \nn \\
h_{(a b)} &=& \sum (\rho^{I_t} (x) Y_{(ab)}^{I_t} (y) +
\rho_{(v)}^{I_v} (x) D_a Y_b^{I_v} (y) +
\rho_{(s)}^{I} (x) D_{(a} D_{b)} Y^{I} (y) ), \nn \\
h^{a}_{a} &=& \sum \pi^{I} (x) Y^{I} (y), \nn \\
g^{A}_{\m\n \r} &=& \sum 3 D_{[\m} b_{\n \r]}^{(A)I} (x) Y^{I} (y), \nn
\\
g^{A}_{\m \n a} &=& \sum ( b_{\m \n}^{(A)I} (x) D_{a} Y^{I} (y) + 2
D_{[\m} Z_{\n]}^{(A) I_{v}} (x) Y_{a}^{I_v} (y)); \nn \\
g^{A}_{\m a b} &=& \sum (D_{\m} U^{(A)I}(x) \ep_{abc} D^{c} Y^I(y) +
2 Z_{\m}^{(A) I_v} D_{[b} Y^{I_v}_{a]}); \nn \\
g^{A}_{a b c} &=& \sum (- \ep_{abc} \L^{I} U^{(A)I}(x) Y^{I} (y)) ; \nn \\
%b^{A}_{\m a} &=& \sum (Z_{\m}^{A I_v } (x) Y_{a}^{I_v }(y) +
%Z_{(s)\m}^{A I} (x) D_{a} Y^I (y) ); \nn \\
%b^{A}_{ab} &=& \sum \ep_{abc} (U^{A I} (x) D^{c} Y^I (y) + U_{(v)}^{A
%  I_{v} } (x)  Y^{c I_v } (y) ); \nn \\
\phi^{mr}  &=& \sum \phi^{(mr) I} (x) Y^{I} (y), \nn
\eea
Here $(\mu, \nu)$ are AdS indices and $(a,b)$ are $S^3$ indices,
with $x$ denoting AdS coordinates and $y$ denoting sphere coordinates. The
subscript $(ab)$ denotes symmetrization of indices $a$ and $b$ with
the trace removed. Relevant properties of the spherical harmonics are
reviewed in appendix \ref{apa}. We will often use a notation
where we replace the index $I$ by the degree of the harmonic $k$
or by a pair of indices
$(k,I)$ where $k$ is the degree of the harmonic and $I$ now parametrizes
their degeneracy, and similarly for $I_v, I_t$.

Imposing the de Donder gauge condition $D^{A} h_{aM} = 0$
on the metric fluctuations removes the fields with subscripts $(s,v)$.
In deriving the spectrum and computing correlation functions, this is
therefore a convenient choice. The de Donder gauge choice is however not
always a convenient choice for the asymptotic expansion of solutions;
indeed the natural coordinate choice in our application takes us
outside de Donder gauge. As discussed in \cite{Skenderis:2006uy}
this issue is straightforwardly dealt with by
working with gauge invariant combinations of the fluctuations; we will
present the relevant gauge invariant combinations later.

\subsection{Asymptotic expansion of the fuzzball solutions}

Now consider the asymptotic expansion at large radius of the
fuzzball solutions. The natural radial coordinate in which to
expand the solutions is the radial coordinate $r$ of the transverse $R^4$,
even though with this choice it will turn out that the metric is not in
de Donder gauge.

The harmonic functions appearing in the solution (\ref{ez1}) can be
expanded as
\bea
f_5 &=&
\frac{Q_5}{r^2} \sum_{k,I} \frac{f^{5}_{kI} Y^I_k(\q_3)}{r^k}; \nn \\
f_1 &=&
\frac{Q_1}{r^2} \sum_{k,I} \frac{f^{1}_{kI} Y^I_k(\q_3)}{r^k}; \la{p0}
\\
A_i &=& \frac{Q_5}{r^2} \sum_{k,I} \frac{(A_{kI})_i
Y^I_k(\q_3)}{r^k}, \nn
\eea
for some coefficients $f^{5}_{kI},
f^{1}_{kI}$ and $(A_{kI})_i$. There are restrictions on the
coefficients $(A_{kI})_i$ because $\pa^i A_i=0$ which will be given below.

In the case of the (near-horizon) harmonic functions
of (\ref{mathu1}), the coefficients
$f^{5}_{kI}, f^{1}_{kI}, (A_{kI})_i$ are given in terms of the
curve $F^i(v)$. To obtain these coefficients we make use
of the following addition theorem for harmonic functions on $R^4$:
\be \la{add-th}
\frac{1}{(x^i - y^i)^2}  = \sum_{k \ge 0} \frac{y^k}{
  (k+1) r^{2+k}} Y^{I}_{k}(\q^x_3) Y^{I}_{k}(\q_3^y).
\ee
In this expression $x^{i}$ and $y^i$ are Cartesian coordinates
on $R^4$, with the corresponding polar coordinates being
$(r,\q_3^x)$ and $(y,\q_3^y)$ respectively.
$Y^{I}_k (\q_3)$ are (normalized) spherical harmonics of degree $k$
on $S^3$ with $I$ labeling their degeneracy; the degeneracy of
degree $k$ harmonics is $(k+1)^2$. For the $k=1$ harmonics of
degeneracy four, it is convenient to use the label $i$, $Y^{i}_1$.
The addition theorem can also be expressed as
\be
\frac{1}{\left | x - y \right |^{2}} = \sum_{k \ge 0}
\frac{1}{(k+1) r^{2+k}}
Y^{I}_{k}(\q^x_3) (C^{I}_{i_1 \cdots i_k} y^{i_1} \cdots y^{i_k} ),
\ee
where $C^I_{i_1 \cdots i_k}$ are the orthogonal symmetric
traceless rank $k$ tensors on $R^4$ which are in one-to-one
correspondence with the (normalized) spherical harmonics $Y^{I}_k
(\q_3)$ of degree $k$ on the $S^3$. This formula is the exact
analogue of the well-known addition theorem for electromagnetism
(see \cite{Jackson}) and also of the addition theorem for harmonic
functions on $R^6$ discussed in the appendix of
\cite{Skenderis:2006di}, and it can be proved in the same way, as we
show in appendix \ref{apa1}.

Using the addition theorem we obtain
\bea
f^5_{kI} &=& \frac{1}{ (k+1) L}
\int_{0}^{L} dv C^{I}_{i_1 \cdots i_k} F^{i_1} \cdots F^{i_k}; \nn \\
f^1_{kI} &=&  \frac{Q_5}{ Q_1 (k+1) L}
\int_{0}^{L} dv \left | \dot{F} \right |^2
C^{I}_{i_1 \cdots i_k} F^{i_1} \cdots F^{i_k}; \la{p00} \\
(A_{kI})_i &=& \frac{1}{ (k+1) L}
\int_{0}^{L} dv \dot{F}_{i}
C^{I}_{i_1 \cdots i_k} F^{i_1} \cdots F^{i_k}. \nn
\eea
Furthermore, in the final equality of (\ref{p0})
the summation is restricted to $k \ge 1$ because
of  the closure of the curve $F^{i}$ ($\int dv \dot{F}_i =0$).
Note that we will often suppress implicit summations over the index $I$
in later expressions for compactness.

Before substituting these expressions into the supergravity fields, we
need to consider which fluctuations are physical. Suppose we use
translational invariance to impose the condition
\be \la{c1}
\int_{0}^{L} dv F_i = 0,
\ee
which was the choice made in previous literature, for example, in \cite{Lunin:2001jy}.
This corresponds to choosing the origin of the coordinate
system to be at the center of mass of the D5-branes.
However, the center of mass of the D1-branes
does not coincide with that of the D5-branes in general; thus this
condition does {\it not} take one to the center of mass of the whole
system.

Indeed with this choice the leading correction to the AdS
background derives from the $k=1$ terms in the harmonic function
$f_1$. The choice (\ref{c1}) gives a leading metric deviation
\be
h_{\m \n} = D_{\m} D_{\n} \l; \hsp
h_{ab} = g_{ab} \l,
\ee
with
\be
\l = \sum_{i} \frac{f^1_{1 i} Y^i_1}{2 r},
\ee
which satisfies $\Box \l = - \l$. Such a perturbation is unphysical
because it can be removed by a superconformal transformation (with
parameter $- \l$). The physical
origin of the term is that with the choice (\ref{c1}) we are not working in
the centre of mass of the system. Instead of imposing that the $k=1$
term in the D5-brane harmonic function vanishes, we should impose that
the $k=1$ term in $\sqrt{f_1 f_5}$ vanishes, namely
\be \la{k1-con}
f^{5}_{1i} + f^{1}_{1 i} = 0.
\ee
When the solution is related to a closed curve
this reduces to
\be \la{c2}
\int_{0}^{L} dv F^{i} (1 + \frac{Q_5}{Q_1} | \dot F |^2 ) = 0.
\ee
Then all unphysical $k=1$ terms in the metric vanish automatically.

Now consider the asymptotic expansion of $A_i$. The restriction on the coefficients
in the asymptotic expansion imposed by the condition $\pa_i A^i = 0$ is most
easily understood as follows. The form $A$ may be written as
\be
A = Q_5 \sum_{k,I,i} \frac{(A_{kI})_i}{r^{2+k}}
Y^I_k \left (Y^i_1 dr + r d Y^i_1 \right ),
\ee
using
\be
dx^i = dr Y^{i}_1 + r dY^{i}_{1}.
\ee
Projecting the products of spherical harmonics onto the basis of
spherical harmonics gives
\bea
A &=& Q_5 \sum_{l,L,k,I,i} \frac{(A_{kI})_i}{r^{2+k}}
(a_{iIL} Y^L_l dr + \frac{b_{IiL}}{ \Lambda^L} r dY^L_l) \\
 && + \hsp Q_5 \sum_{k_v,I_v,k,I,i} \frac{(A_{kI})_i}{r^{1+k}}
E^{\pm}_{I_v I i} Y^{I_v \pm}_{k_v}, \nn
\eea
where the spherical harmonic overlaps $(a_{iIJ},b_{IiJ},E^{\pm}_{I_v I i})$
are defined in (\ref{ap-ov0}), (\ref{ap-ov-1}) and (\ref{gt1}) respectively.
The term in $A$ proportional to the vector harmonic is already
divergenceless on its own. The first two combine into divergenceless
combination iff scalar harmonics with degree $l =(k-1)$ appear in this
asymptotic expansion:
\bea
A &=& Q_5 \sum_{L,k,I,i} \frac{(A_{kI})_i}{r^{2+k}} a_{IiL}
( Y^L_{k-1} dr - \frac{r}{(1+k)} dY^L_{k-1}) \\
&&+ Q_5 \sum_{k_v,I_v,k,I,i} \frac{(A_{kI})_i}{r^{1+k}}
E^{\pm}_{I_v I i} Y^{I_v \pm}_{k_v}. \nn
\eea
Vanishing of the other terms requires
\be
\sum_{I,i} (A_{kI})_i a_{i I L} = 0 \hsp \hsp l \neq (k-1).
\ee
In particular this means that $(A_{1j})_i$ must be antisymmetric (since
$a_{ijL}$ is symmetric in $i,j$). Note that this condition is
clearly satisfied for the $(A_{1j})_i$ defined in (\ref{p00}).

The leading term in the asymptotic expansion is given in terms 
of degree one vector harmonics as
%\be
%A = \frac{Q_5}{r^3} (A_{1 j})_i Y^{j}_{1} dx^i,
%\ee
%where by its definition in (\ref{p00}) $(A_{1 j})_{i} = - (A_{1 i})_{j}$. Noting that
%
%the gauge field becomes
\be
A =  \frac{Q_5}{r^2} (A_{1 j})_i Y^{j}_{1} dY^{i}_{1}
\equiv \frac{\sqrt{Q_5 Q_1} }{r^2} (a^{\a -} Y_{1}^{\a -} +
a^{\a+} Y_{1}^{\a+}),
\ee
where $(Y_{1}^{\a -},Y_{1}^{\a +})$ with $\a =1,2,3$
form a basis for the $k=1$ vector harmonics, which coincide with
the Killing one forms of $SU(2)_L$ and $SU(2)_R$ respectively.
%From the
%antisymmetry of $(A_{1 j})_{i}$, we need only the antisymmetric part
%of the product of the representations, which is precisely the $k=1$
%vector harmonic representation.
Here we define
\be \label{a1}
a^{\a \pm} = \frac{\sqrt{Q_5}}{\sqrt{Q_1}} \sum_{i > j} e^{\pm}_{\a ij } (A_{1j})_i
%=
%\frac{\sqrt{Q_5}}{\sqrt{Q_1} L} \sum_{i > j} e^{\pm}_{\a ij }
%\int_{0}^{L} dv \dot{F}_{i} F_j,
\ee
where the spherical harmonic triple overlap $e^{\pm}_{\a ij }$ is defined in
(\ref{ap-ov3}) and explicit values in a particular basis are given in
(\ref{ecof}). For solutions defined
by a curve $F^i(v)$, the coefficients $(A_{1j})_i$ are given in (\ref{p00}).
The dual field to leading order is
\be
B =  \frac{\sqrt{Q_5 Q_1} }{r^2} (a^{\a -} Y_{1}^{\a -} -
a^{\a +} Y_{1}^{\a +}),
\ee
where we use the Hodge duality property of the vector harmonics given
in (\ref{vec-dual}).

\bigskip

Putting these results together the leading perturbations of the metric are
\bea
- h_{tt} &=& h_{yy} = \frac{1}{2}
\left ( - (f^1_{2I} + f^5_{2I}) Y^I_{2} + (f^5_{1i} Y^i_1)^2  \right );
\nn \\
h_{rr} &=& \frac{1}{2
  r^4} \left (  (f^1_{2I} + f^5_{2I}) Y^I_{2} - (f^5_{1i} Y_1^i)^2  \right ); \nn
\\
h_{ta} &=& \left ( a^{\a -} Y_{1}^{\a-} +
a^{\a +} Y_{1}^{\a +}
\right ); \la{p2} \\
h_{ya} &=& \left (
a^{\a -} Y_{1}^{\a-} - a^{\a +} Y_{1}^{\a +}
\right ); \nn \\
h_{ab} &=& g^{o}_{ab} \frac{1}{2r^2}
\left (  (f^1_{2I} + f^5_{2I}) Y^I_{2} - (f^5_{1i} Y^i_1)^2  \right ) -
\frac{2}{r^2} a^{\a -} a^{\b +}
(  (Y_{1}^{\a -})_a (Y_{1}^{\b +})_b
+  (Y_{1}^{\a -})_b (Y_{1}^{\b +})_a). \nn
\eea
Note that the condition (\ref{k1-con}) has been used to eliminate
$f^{1}_{1i}$. Terms quadratic in spherical harmonics will need to
be projected back onto the
basis of spherical harmonics in order to determine the contributions
to each perturbation component in (\ref{flc1}).

In these expressions we have suppressed the scale factor
$\sqrt{Q_1 Q_5}$. As mentioned previously, after rescaling $t
\rightarrow t \sqrt{Q_1 Q_5}$ and $y \rightarrow y \sqrt{Q_1 Q_5}$,
the metric has an overall scale factor $\sqrt{Q_1 Q_5}$. Scale
factors will similarly be suppressed in the other fields. The overall
scaling will be taken into account via the normalization of the
three-dimensional action.

Now consider the other supergravity fields; from (\ref{g3}) and
(\ref{dq2}) one finds the following three form fluctuations are
\bea
g^{5}_{tya} &=& \frac{1}{4} D_{a} \left (2(f_{1i}^5 Y^i_1)^2 -
(f_{2I}^5+f_{2I}^1)Y_2^I \right ); \nn \\
g^{5}_{ta b} &=& - (a^{\a -} D_{[a} (Y_{1}^{\a -})_{b]}
- a^{\a +} D_{[a} (Y_{1}^{\a +})_{b]});  \la{pm2} \\
g^{5}_{y a b}
&=& - (a^{\a -} D_{[a} (Y_{1}^{\a -})_{b]} +
a^{\a +} D_{[a} (Y_{1}^{\a +})_{b]}); \nn \\
g^{5}_{r a b} &=&
\frac{1}{r^3} \left (\frac{1}{4} {\e_{ab}}^c
(f_{2I}^1+f_{2I}^5) D_c Y_2^I + 4
a^{\a -} a^{\b +} (Y^{\a -}_{1})_{[a} (Y^{\b +}_{1})_{b]} \right ); \nn \\
g^{5}_{abc} &=& \frac{1}{r^2} {\e_{abc}}
(f_{2I}^1+f_{2I}^5) Y_2^I - \frac{6}{r^2}
a^{\a -} a^{\b +} D_{[a} (Y^{\a -}_{1})_{b} (Y^{\b +}_{1})_{c]}). \nn
\eea
%\bea
%b^{5}_{ty} &=& \frac{1}{4\sqrt{Q_1Q_5}}(2(f_1^5Y^1)^2 -
%(f_2^5+f_2^1)Y^2), \nn \\
%b^{5}_{ta} &=& \hp\sqrt{\frac{Q_5}{Q_1}}(A_L^{\a} (e_L^{\a})_a
%- A_R^{\a} (e_R^{\a})_a), \\
%b^{5}_{ya} &=& - \hp\sqrt{\frac{Q_5}{Q_1}}(A_L^{\a} (e_L^{\a})_a
%+ A_R^{\a} (e_R^{\a})_a), \nn \\
% b^{5}_{ab} &=& - \frac{\sqrt{Q_1Q_5}}{8r^2}{\e_{ab}}^c
%(f_2^1+f_2^5) D_c Y^2 - \frac{2Q_5^{3/2}}{r^2\sqrt{Q_1}}
%A_L^{\a} A_R^{\b} (e^{\a}_L)_{[a} (e^{\b}_R)_{b]}, \nn
%\eea
and
\bea
g^{6}_{t y r } &=& \frac{1}{2} f_{1i}^5 Y_1^i; \la{pm3} \\
g^{6}_{t y a} &=& \frac{1}{4} D_{a} \left (2f_{1i}^5 Y_1^{i} r +
(f_{2I}^5-f_{2I}^1) Y_2^I \right ); \nn \\
g^{6}_{r a b} &=& \frac{1}{2r^2} {\e_{ab}}^c f_{1i}^5 D_c Y^i_1
+ \frac{1}{4 r^3}{\e_{ab}}^c (f_{2I}^5-f_{2I}^1) D_c Y^I_2; \nn \\
g^{6}_{a b c } &=& \frac{3}{2r} \e_{abc} f_{1i}^5 Y^i_1
+ \frac{1}{r^2}\e_{abc} (f_{2I}^5-f_{2I}^1) Y^I_2. \nn
\eea
%
%\bea
%b^{6}_{ty} &=& \frac{1}{4\sqrt{Q_1Q_5}}(2f_1^5Y^1 r +
%(f_2^5-f_2^1)Y^2), \\
%b^{6}_{ab} &=& -\frac{\sqrt{Q_1Q_5}}{2r} {\e_{ab}}^c f_1^5 D_c Y^1
%- \frac{\sqrt{Q_1Q_5}}{8r^2}{\e_{ab}}^c (f_2^5-f_2^1) D_c Y^2. \nn
%\eea
Finally the scalar field is expanded as
\be \la{p1}
\phi^{(56)} \equiv \Phi
= - \frac{f^{5}_{1i}}{r} Y^i_1 + \half \frac{f^1_{2I} - f^5_{2I}}{r^2} Y^I_2.
\ee
All other fluctuations, $g^{A}$ with $A \neq 5,6$ and $\phi^{mr}$ with
$m \neq 5$, $r \neq 6$ vanish.

\subsection{Gauge invariant fluctuations}

We now wish to extract gauge invariant combinations of these
fluctuations. Gauge invariant means that the fluctuations do not
transform under coordinate transformations $\d x^{M} = \xi^M$, or,
in the case of the three dimensional metric and gauge fields, they
have the correct transformation properties. Using the fact that
the metric and three forms transform (up to linear order in fluctuations) as
\bea
\d h_{MN} &=& D_{M} \xi_{N} + D_{N} \xi_{M} + D_{M} \xi^P h_{PN} +
D_{N} \xi^{P} h_{PM} - \xi^{P} D_{P} h_{MN}; \la{trans1} \\
\d g^{A}_{MNP} &=& 3 D_{[M} \xi^{S} g^{oA}_{NP] S} + 3 D_{[M} \xi^{S}
  g^{A}_{NP] S} + \xi^{S} D_{S} g_{MNP}^A, \nn
\eea
one can systematically compute combinations which are gauge
invariant to quadratic order in fluctuations. That is, the gauge
invariant fluctuations $\hat{\psi}^{Q}$ are given by the following
schematic expression
\be \la{schem}
\hat{\psi}^{Q} = \sum_{R} a_{QR} {\psi}^{R} + \sum_{R,S} a_{QRS}
    {\psi}^{R} \psi^{S},
\ee
where $\psi^Q$ collectively denotes all fields and
the quadratic contributions are rather complicated in
general. Note that each gauge invariant field at linearized level
should reduce to the corresponding
field in de Donder gauge on setting the fields with subscripts $(s,v)$
to zero in (\ref{flc1}). Clearly by retaining higher order terms in (\ref{trans1})
one could compute the invariants to arbitrarily high order in the fluctuations.

For the discussion at hand, however, we do
not need the most general expressions. Since we are working
perturbatively in the radial coordinate, we need only retain terms in (\ref{schem})
with the same radial behavior. In particular, as we discuss only
leading order and next to leading order perturbations, we will need at most
quadratic invariants. In fact the only combinations which will be needed here are
\bea
\hat{\pi_2}^{I} &=& \pi_2^{I} + \Lambda^{2} \rho_{2(s)}^{I}; \\
\hat{U}^{(5)I}_2 &=& U^{(5)I}_2 - \half \rho^{I}_{2(s)}; \nn \\
\hat{h}^{0}_{\m\n} &=& h^{0}_{\m\n} - \sum_{\a,\pm} h_{\m}^{1 \pm \a}
h_{\n}^{1 \pm \a}. \nn
\eea
In addition the fluctuations $(\Phi_1^{i}, \Phi_2^{I}, U^{(6)i}_1,
U^{(6)I}_2)$ are by themselves gauge invariant up to the necessary order
and the fields $(Z_{\m}^{(5) 1 \pm \a}, Z_{\m}^{(6) 1 \pm \a},
h_{\m}^{1 \pm \a})$
by themselves transform
correctly as gauge fields. Thus only in the metric do we need to take
into account a quadratic contribution.

\section{Extracting the vevs systematically} \la{vev-proc}

In this section we will compute the vevs following the systematic
procedure of \cite{Skenderis:2006uy}.
First one should identify the six-dimensional
equations of
motion that these fields satisfy to appropriate order, in this case
quadratic. Secondly one should remove derivative terms in these
equations of motion by a field redefinition: this defines the
  Kaluza-Klein reduction map between six-dimensional and
three-dimensional fields. Finally, once one has the three dimensional
fields and their equations of motion, one extracts vevs using the by now
familiar methods of holographic renormalization.

\subsection{Linearized field equations}

Let us first consider the linearized field equations.
As discussed in \cite{Skenderis:2006uy}, the equations of motion for
the gauge invariant fields at linear order are precisely the same as those in de
Donder gauge, provided one replaces all fields with the corresponding
gauge invariant field. So now let us briefly review the linearized spectrum in
de Donder gauge derived in
\cite{Sez}. Consider first the scalars. It is
useful to introduce the following combinations of these
fields which diagonalize the linearized equations of motion:
\bea
s^{(r) k}_{I} &=& \frac{1}{4(k+1)} ({\phi}^{(5r) k}_{I} +2 (k+2)
{U}^{(r)k}_{I}), \la{diageqm} \\
t^{(r)k}_{I} &=& \frac{1}{4} ({\phi}^{(5r)k}_{I} - 2 k {U}^{(r)k}_{I}),
\nn \\
\s^k_{I} &=& \frac{1}{12 (k+1)} (6 (k+2) \hat{U}^{(5)k}_{I} -
\hat{\pi}_I^{k}), \nn \\
\t^k_{I} &=& \frac{1}{12 (k+1)} ( \hat{\pi}^k_I + 6k \hat{U}^{(5) k}_{I}). \nn
\eea
Note that these combinations are applicable when the
background $AdS_3 \times S^3$ has unit radius.
Here the fields $s^{(r)k}$ and $\s^k$ correspond to scalar chiral
primaries. In what follows we will need only the $r=6$ fields
and will thus drop the $r$ superscript. The masses of the scalar fields are
\be \la{masses}
m_{s^k}^2 = m_{\s^k}^2 = k (k-2), \hsp
m_{t^k}^2 = m_{\t^k}^2 = (k+2) (k+4), \hsp
m_{\rho^k}^2 = k(k+2).
\ee
Note also that $k \ge 0$ for $(\t^k,t^{(r)k})$; $k \ge 1$ for
$s^{(r)k}$; $k \ge
2$ for $(\s^k, \rho^k)$.

\bigskip

Next consider the vector fields. It is useful to introduce
the following combinations which diagonalize the equations of motion:
\be
h^{\pm}_{\m I_v} = \half (C_{\m I_v}^{\pm} - A^{\pm}_{\m
  I_v}), \hsp
Z_{\m I_v}^{(5) \pm} = \pm \qu (C_{\m I_v}^{\pm} + A_{\m I_v}^{\pm}).
\ee
For general $k$ the equations of motion are Proca-Chern-Simons
equations which couple $(A_{\m}^{\pm}, C^{\pm}_{\m})$ via a first
order constraint \cite{Sez}. The three dynamical fields at each degree $k$
have masses $(k-1, k+1, k+3)$, corresponding to dual operators of
dimensions $(k,k+2,k+4)$ respectively. The lowest dimension operators
are the R symmetry currents, which couple to the $k=1$ $A^{\pm \a}_{\m }$
bulk fields. The latter satisfy the Chern-Simons equation
\be \la{cs11}
F_{\m \n}(A^{\pm \a}) = 0,
\ee
where $F_{\m\n}(A^{\pm \a})$ is the curvature of the connection and
the index $\a = 1,2,3$ is an $SU(2)$ adjoint index. Only
these bulk vector fields will be needed in what follows, and
therefore the equations of motion for general $k$ discussed in \cite{Sez}
are not given here. There are also the massive vectors $Z_{\m I_v}^{(6) \pm}$
but their mass is sufficiently high that they are irrelevant for
our discussion.

\bigskip

Finally there is a tower of KK gravitons with $m^2 = k (k+2)$ but
again only the massless graviton will play a role here. Note that it is the
combination $\hat{H}_{\m \n} = h^{o}_{\m \n} + \pi^{0} g^{o}_{\m \n}$ which
satisfies the linearized massless Einstein equation
\be \la{Ein-eq}
( {\cal{L}_{E}} + 2) \hat{H}_{\m \n}
\equiv \half ( - \Box \hat{H}_{\m \n} + D^{\r} D_{\m} \hat{H}_{\r \n} + D^{\r}
D_{\n} \hat{H}_{\r \m} - D_{\m} D_{\n} \hat{H}^{\r}_{\r} + 4 \hat{H}_{\m \n}) = 0.
\ee
That this is the appropriate combination follows from the reduction of
the six-dimensional Einstein term in the action over the sphere; keeping terms linear
in fluctuations the three dimensional action is
\be
S_{3} \sim \int d^3x \sqrt{-g} (( 1+ \half \pi^0) R + \cdots),
\ee
and the Weyl transformation
$\hat{H}_{\m \n} = h^{0}_{\m \n} + \pi^{0} g^{o}_{\m \n}$ is required to
bring the action to Einstein frame.

\subsection{Field equations to quadratic order}

{}From the asymptotic expansion we now identify the fields of
(\ref{diageqm}). In the asymptotic expansion we have retained only terms to quadratic
order, that is of order $1/r$ and $1/r^2$ relative to the
background. These terms are sufficient to determine vevs for the scalar
chiral primaries of dimension one and two; the R symmetry currents and
the energy momentum tensor. Using the tables in \cite{Sez}, one finds
that the corresponding supergravity fields are $(s^1, s^2, \s^2,
A_{\m}^{1 \pm}, {H}_{\m \n})$ respectively. Terms in other
supergravity fields at the same order do not capture field theory
data: they are simply induced by the non-linearity of the supergravity
equations. Therefore we need only consider the above fields.

The next step is to derive the six-dimensional equations
satisfied by the fluctuations, at non-linear order.  The generic field
equation for each field $\psi^Q$ expanded in the number of fields is (schematically)
\be
{\cal {L}}_{Q} \psi^Q = {\cal L}_{Q R S } \psi^R \psi^S +
{\cal L}_{Q R S T } \psi^R \psi^S \psi^T + \cdots,
\ee
where ${\cal L}_{Q_1 \cdots Q_n}$ is generically a non-linear differential
operator. (Note that each field $\psi^Q$ should be the appropriate
diffeomorphism invariant combination.) The complete set of
corrections to the field equations involves many terms even to
quadratic order.

Fortunately what is
required for extracting field theory data is the equations of motion
expanded perturbatively near the conformal boundary, where the radial
coordinate acts as the perturbation parameter. This means that we need
only retain terms on the right hand side which affect the radial
expansion at sufficiently low order to impact on the vevs. In practice
for our discussion, the relevant quadratic corrections are
those involving two $s^1$ fields or two gauge fields,
since all other quadratic terms do not contribute at the required
order. (Note that there are no corrections involving one $s^1$ field
and one gauge field.)
That all other terms can be neglected will be justified when one
carries out the holographic renormalization procedure and
considers the perturbative solution of the field equations.

The scalar field corrections to the
field equations were computed in \cite{Arutyunov:2000by,
  Pank}\footnote{We thank Gleb Arutyunov for making the latter
  available to us.}. These computations along with the
corrections quadratic in the gauge field are discussed in detail in
appendix \ref{apb}.
 Consider first the scalar field equations.
There are no quadratic corrections to the $(s^{1},s^2)$
equations from either $s^1$ fields or gauge fields, and thus the
relevant equations remain the linearized equations.
The $\s^2$ field equation does however get corrected by terms
quadratic in scalars:
\be \la{cor-s2}
\Box \s^2_I =  \frac{11}{3} (s^1_i s^{1}_j - (D_{\m} s^1_i) (D^{\m} s^{1}_j))
  a_{I ij}.
%- \frac{1}{4} (F_{\m \n} (A^{+\a}) F^{\m \n} (A^{-\b}) ) f_{2 \a \b},
\ee
The coefficient $a_{Iij}$ is the triple overlap of the corresponding
spherical harmonics (see appendix \ref{apa}).
As discussed in the appendix \ref{apb}, there are
also corrections to this equation quadratic in the gauge fields which involve
the field strengths $F_{\m \n} (A^{\pm \a})$ associated with the
connections $A^{\pm \a}_{\m}$ respectively. However, according to the
linearized field equations (\ref{cs11}) these field strengths vanish and thus these
corrections do not play a role.

\bigskip

Next consider the corrections to the Einstein equation, which are also
discussed in more detail in \ref{apb}. Note that
these corrections were not computed in \cite{Arutyunov:2000by, Pank}.
The appropriate three dimensional metric to quadratic order is
\be \label{3dmet}
H_{\m \n} = h^{0}_{\m \n} - \sum_{\a, \pm} h_{\m}^{1 \pm \a} h_{\n}^{1
  \pm \a} + \pi^0 g^{o}_{\m \n}.
\ee
As discussed previously the quadratic term is necessary in order for
the metric to transform correctly under diffeomorphisms. Then the
equation satisfied by the metric, up to quadratic order in the scalar
fields $s^1_i$ and the gauge fields is
\be \la{cor-Ein}
({\cal{L}}_{E} + 2) H_{\m \n} = 16 (D_{\m} s^1_i D_{\n} s^{1}_i -
  g^{o}_{\m \n} s^{1}_i s^{1}_i ),
\ee
where the linearized Einstein operator was defined in (\ref{Ein-eq}).
This equation can be rewritten as
\be
{\cal G}_{\m \n} - g^{o}_{\m \n}=
16 \left(D_{\m} s^{1}_i D_{\n} s^{1}_i -\half
g^o_{\m \n} ((D s^1_i )^2 - (s^{1}_i)^2)\right),
\ee
where ${\cal G}_{\m \n}$ is the linearized Einstein tensor. The
rhs of this equation is the stress energy tensor of $s^1$.
Note that the gauge field contributions to the energy momentum tensor involve
the field strengths, and thus are zero when one imposes the lowest
order field equation (\ref{cs11}).

\bigskip

Finally, let us consider the equations for the gauge field. As
  discussed in \cite{Arutyunov:2000by,Pank} the corrections
quadratic in the gauge field correct the linearized equation to the
non-Abelian Chern-Simons equation. That is, the six-dimensional
  equation is
\be \la{cs9}
\ep^{\m \n \r} (\pa_{\n} A^{\pm \a}_{\r} +  \half A^{\pm \b}_{\n}
  A^{\pm \g}_{\r} \ep_{\a \b \g}) = 0,
\ee
where the $\ep_{\a \b \g}$ arises from the triple overlap of vector
harmonics defined in (\ref{vector}).
Note that the $SU(2)_L$ and $SU(2)_R$
  gauge fields are decoupled from each other.
There are also corrections quadratic in the scalars $s^1$, which
provide a source for the field strength:
\be \la{cs99}
\ep^{\m \n \r} (\pa_{\n} A^{\pm \a}_{\r} + \cdots) = \pm 4 s^{1}_{i} D^{
\m} s^{1}_{j} e^{\pm}_{\a ij},
\ee
where the ellipses denote the non-linear Chern-Simons terms and the triple overlap is
defined in (\ref{ap-ov3}).

\subsection{Reduction to three dimensions}

Given the corrected six-dimensional field equations (\ref{cor-s2}), (\ref{cor-Ein}) and
(\ref{cs9}), we now need to determine the corresponding
three-dimensional field equations.
As discussed in \cite{Skenderis:2006uy}, the KK map between six and
three dimensional fields is in general non-linear.
The non-linear corrections arise from field redefinitions used to
remove derivative couplings.
{}From the form of the corrected field equations, it is apparent that
only the scalar fields $\s^2$ are affected (at this order) by such field
redefinitions. That is, the derivative couplings in (\ref{cor-s2}) can be removed
by the field redefinition
\be \la{def1}
{\Sigma}^2_I = \sqrt{32} (\s^2_I + \frac{11}{6} s^{1}_i s^{1}_j a_{I
  ij} + \cdots),
\ee
where $\Sigma^2_I$ is the three dimensional field.
(The prefactor ensures canonical normalization of
the three dimensional field, as we will shortly discuss.) This field
redefinition defines the KK reduction map between six and three
dimensional fields.

The resulting set of three dimensional field equations can then
be integrated to the following three-dimensional bulk action
\bea \la{pp1}
&& \frac{n_1 n_5}{4 \pi} \int d^{3}x \sqrt{-G} (R_G + 2 - \half (DS^1_i)^2
+ \half (S^1_i)^2 - \half (DS^2_I)^2 - \half (D\S^2_I)^2 ) \\
&& + \frac{n_1 n_5}{8 \pi} \int (A^{+}_{\a} dA^{+ \a} + \frac{1}{3}
\ep_{\a \b \g} A^{+\a} A^{+\b} A^{+\g} - A^{-}_{\a} dA^{- \a} - \frac{1}{3}
\ep_{\a \b \g} A^{-\a} A^{-\b} A^{-\g}) + \cdots \nn.
\eea
The ellipses denote fields dual to operators of higher dimension not
being considered here, along with higher order interactions.
The boundary terms in this action will be discussed later in the
context of holographic renormalization.

An overall rescaling of the scalar fields arises from demanding that the
three-dimensional scalar fields are canonically normalized, up to the
overall scaling of the action;
it follows from the quadratic actions given in \cite{Arutyunov:2000by}.
Thus the three dimensional fields $S_{I}^k$ and $\S_{I}^k$ are related
to the six-dimensional fields $s^k_I$ and $\s^k_I$ via
\be \la{def2}
S^{k}_I = 4 \sqrt{k (k+1)} (s^k_I + \cdots), \hsp
\S^{k}_{I} = 4 \sqrt{k (k-1)} (\s^k_I + \cdots).
\ee
The ellipses denote non-linear terms in the KK map of which only
(\ref{def1}) will be relevant here; other terms do not contribute to the order
we need. The normalization of the gauge field terms also follows from the
actions given in \cite{Arutyunov:2000by}.
Note that the leading scalar field corrections to the gauge field
equation (\ref{cs99}) are also implicitly contained in the action
(\ref{pp1}), recalling that $D$ is a covariant derivative and the
scalar fields are charged under the $SO(4)$ gauge group.

The overall prefactor in the action (\ref{pp1}) follows from the
chain of dimensional reductions
\be
\frac{1}{2 \k_{10}^2} \int d^{10}x \sqrt{-g_{10}} e^{-2 \Phi} (R_{10} +
\cdots) \rightarrow
\frac{1}{2 \k_{6}^2} \int d^6x \sqrt{-g} (R + \cdots) \rightarrow
\frac{1}{2 \k_{3}^2} \int d^3 x \sqrt{-G} (R_{G} + 2 \cdots).
\ee
Implicitly in the latter expression the curvature scale is contained
in the prefactor, so that the background $AdS_3$ metric $G$ has unit
radius. Then
\be
2 \k_{10}^2 = (2 \pi)^7 (\a')^4; \hsp
2 \k_{6}^2 = \frac{1}{(2 \pi)^4 V} 2 \k_{10}^2; \hsp
2 \k_3^2 = \frac{1}{2 \pi^2 Q_1 Q_5} 2 \k_6^2,
\ee
which using (\ref{int-charg}) implies that
\be
\frac{1}{2 \k_3^2} = \frac{n_1 n_5}{4 \pi},
\ee
as in (\ref{pp1}).

\subsection{Holographic renormalization and extremal couplings}

Having determined the three-dimensional fields and the equations of
motion which they satisfy we are now ready to determine vevs using the
procedure of holographic renormalization.
We will first briefly review this procedure, using the Hamiltonian
formalism developed in \cite{Papadimitriou:2004ap,Papadimitriou:2004rz}.
Let ${\cal{O}}_{\Psi^{k}}$ be the dimension $k$ operator
dual to the three dimensional supergravity field $\Psi^k$, the latter
being related to the six dimensional fields $\psi^Q$ by non-linear KK maps.
Then its vev can be expressed
as
\be \la{holrel}
\left < {\cal{O}}_{\Psi^k} \right > = \frac{n_1 n_5}{4 \pi} \left ( (\pi_{\Psi^k})_{(k)}
+ \cdots \right );
\ee
where we will explain the meaning of the ellipses below.
Now $\pi_{\Psi^k}$ is the radial canonical momentum for the field
$\Psi^k$ and
$(\pi_{\Psi^k})_{(k)}$ is the $k$th component in its
expansion in terms of eigenfunctions of the dilatation operator.
The results of \cite{Papadimitriou:2004ap,Papadimitriou:2004rz} show that
there is a one to one correspondence between momentum coefficients and
terms in the asymptotic expansion of the fields.

That is, the near boundary expansion of the metric and scalar fields
is
\bea
ds_{3}^{2} &=& \frac{dz^2}{z^2} + \frac{1}{z^2} \left(g_{(0)uv} + z^2
\left(g_{(2)uv} + {\rm{log}}(z^2) h_{(2) uv} + ({\rm{log}}(z^2))^2
\tilde{h}_{(2) uv}\right) + \cdots\right) dx^{u} dx^v; \nn \\
\Psi^1 &=& z ({\rm{log}}(z^2) \Psi^1_{(0)}(x) + \td{\Psi}^1_{(0)}(x) +
\cdots ); \label{as_exp} \\
\Psi^{k} &=& z^{2-k} \Psi^{k}_{(0)}(x) + \cdots + z^{k}
\Psi^{k}_{(2k-2)}(x) + \cdots, \hsp k \neq 1. \nn
\eea
In these expressions $(G_{(0)uv},  \Psi^1_{(0)}(x), \Psi^{k}_{(0)}(x))$
are sources for the stress energy tensor and scalar operators of dimension
one and $k$ respectively; as usual one must treat separately the operators of
dimension $\Delta = d/2$, where $d$ is the dimension of the boundary. Note that
the 2-dimensional boundary coordinates are labeled by $(u,v)$.

The correspondence between the momentum
coefficients and these expansion coefficients for the scalar fields is then
\bea \label{mome}
(\pi_{\Psi^k})_{(k)} &=& ( (2k-2) \Psi^{k}_{(2k-2)}(x) +
\cdots ); \\
(\pi_{\Psi^1})_{(1)} &=&  (2 \td{\Psi}^1_{(0)} + \cdots). \nn
\eea
The ellipses denote non-linear terms in the relations that involve
the sources and do not play a role here.

The ellipses in (\ref{holrel}) denote terms non-linear in momenta.
Such terms are
related to extremal correlators and play a crucial role which we
will discuss in detail. Before doing so, however, it is convenient to
first discuss the gauge fields.

\subsubsection{R symmetry currents}

Let us now consider the vevs for R symmetry currents; these were previously discussed in
\cite{Ban,Kraus} and we will briefly summarize their results
here. Given the asymptotic form of the metric
(\re{as_exp}) the Chern-Simons gauge fields have corresponding
asymptotic
field expansions
\be
A^{\pm \a} = {\cal A}^{\pm \a} + z^2 A_{(2)}^{\pm \a} + \cdots.
\ee
Here ${\cal A}^{\pm \a}$ are fixed boundary values which are respectively holomorphic and
anti-holomorphic. A key point is that the vev will be obtained from
the leading order term in this expansion which is not affected by the
other supergravity fields. Supergravity couplings affect only the
subleading behavior of the gauge field, and thus we can neglect
them. Put differently, the vev for the R symmetry current involves
only the gauge field and there are no non-linear contributions.

The following boundary action
\be \la{b-cs}
S_{B} = \frac{n_1 n_5}{16 \pi} \int d^2x \sqrt{-\g} \g^{uv} ({\cal A}^{+
  \a}_{u} {\cal A}^{+ \a}_v + {\cal A}^{-\a}_{u} {\cal A}^{-\a}_{v})
\ee
ensures that the variational problem for the gauge fields is well-defined with these
boundary conditions; $\g_{u v}$ is the induced boundary metric.
\footnote{In \cite{Ban} the additional boundary term
%\be \la{b-cs2}
$\Delta S_{{\cal A}} = - \frac{n_1 n_5}{16 \pi} \int d^2x \sqrt{-\g}
(\g^{u  v}+ \ep^{uv}) {\cal A}^{+ \a}_u {\cal A}^{-\a}_v$
%\ee
was added to the action. The variational problem is still consistent,
but this term couples
left and right movers so it is not appropriate for our purposes.}.
With these boundary terms the on-shell variation of the action yields
the currents
\be \la{cur1}
\left < J^{\pm \a}_{u} \right > = \frac{1}{\sqrt{-\g}} \left (\frac{\delta S}{\delta
{\cal A}^{\pm \a u}_{\a}} \right ) = \frac{n_1 n_5}{8 \pi}
(g_{(0)uv} \mp \ep_{uv}) {\cal A}^{\pm \a v}.
\ee
As discussed recently in \cite{Kraus} the resulting currents have the
desired properties. In particular, momentarily switching to the
Euclidean signature and using conformal gauge for the
boundary metric so that $g_{(0) uv} dx^{u} dx^v = dw d\bar{w}$, the
currents are
\bea
J^{+ \a}_w &=& \frac{n_1 n_5}{4 \pi} {\cal A}^{+ \a}_w; \hsp
J^{+ \a}_{\bar{w}} = 0; \\
J^{- \a}_w &=& 0; \hsp
J^{- \a}_{\bar{w}} = \frac{n_1 n_5}{4 \pi} {\cal A}^{- \a}_{\bar{w}}. \nn
\eea
Thus the $SU(2)_L$ and $SU(2)_R$ right currents are holomorphic and
anti-holomorphic respectively, as expected for the boundary
CFT. Moreover the current modes defined by
\be
J^{+\a}_{n} = \frac{1}{2 \pi i} \oint dw  w^n J^{+\a}_w; \hsp
J^{-\a}_n = \frac{1}{2 \pi i} \oint  d \bar{w} \bar{w}^n J^{-\a}_{\bar{w}},
\ee
obey the correct $SU(2)$ current algebras.
%(Here the coordinates $w,\bar{w}$ are taken to have periodicity $2 \pi$.)

\subsubsection{Scalar operators} \la{scal-disc}

Consider next the scalar operators; here the non-linear terms in
(\ref{holrel}) play a crucial role.
Just as in \cite{Skenderis:2006uy} we need to take into account the
rather subtle issue of extremal couplings.
Recall that an extremal correlation function is one for which the
dimension of one operator is
equal to the sum of the other operator dimensions. The corresponding
bulk couplings in supergravity vanish: this is physically necessary,
because such couplings would induce conformal anomalies which are
known to be zero (and non-renormalized). In \cite{D'Hoker:1999ea} it
was appreciated that extremal correlators are obtained not from bulk couplings,
but instead from certain finite
boundary terms. These would arise from demanding a well posed variational
problem in the higher dimensional theory, and then keeping track of
all boundary terms when carrying out the KK reduction.

These same extremal couplings play a key role in determining the
vevs. Suppose the operator ${\cal {O}}_{\Psi^k}$ has a non-vanishing
extremal $n$-point function with operators $\{ {\cal {O}}_{\Psi^{k_{a}}}
\}$, with $a=1,\cdots (n-1)$. Then this implies an additional term in
the holographic renormalization relation
\be \la{hol-rel2}
\left < {\cal{O}}_{\Psi^k} \right > = \frac{n_1 n_5}{ 4 \pi}
\left (\left (\pi_{\Psi^k} \right )_{(k)} +
A_{k k_1 \cdots k_{(n-1)}} \prod_{k_a} \left (\pi_{\Psi^{k_a}} \right
)_{(k_a)} + \cdots \right )
\ee
The coupling $A_{k k_1 \cdots k_{(n-1)}}$ must be such that one obtains the
correct $n$-point function upon functional differentiation.

\bigskip

Now consider how this issue affects the vevs being determined here:
there are potentially contributions to vevs of dimension two operators
from their couplings to two dimension one operators. The latter include
both the operators dual to the scalars $S^1_i$ and the R-symmetry
currents dual to the gauge fields $A^{\pm \a}_{\m}$. Let us consider first
the following extremal three point functions between scalar operators
\be
\Sigma^{2} : \left < {\cal{O}}_{\Sigma_I^2} {\cal{O}}_{S^1_i}
{\cal{O}}_{S_{j}^1} \right >;
\hsp
S^{2} : \left < {\cal{O}}_{S^2_I} {\cal{O}}_{S^1_i} {\cal{O}}_{S^{1}_j}
\right >.
\ee
If these three point functions are non-zero, there will necessarily be
additional quadratic contributions to the vevs of the dimension two
operators.

In the discussions of \cite{Skenderis:2006uy} one could use
the known free field extremal correlators of ${\cal N} = 4$ SYM along with
non-renormalization theorems to fix the additional terms in
(\ref{hol-rel2}). As we will discuss momentarily comparing with field theory is in this
case rather more subtle. From the supergravity side there are two methods to compute
these quadratic terms. The first would be to
start with the six-dimensional action, demand that the
variational problem is well-defined (which fixes boundary terms), and
then dimensionally reduce
to three dimensions. This is straightforward in principle, but to
extract the required coefficient we
need boundary terms cubic in the fields, which in turn requires
expanding the field equations to
cubic order. Thus we choose to use a second method: we compute the extremal
correlator in supergravity by computing the corresponding non extremal correlator and
then using a careful limiting procedure. This computation of
the extremal correlators and hence the
non-linear terms (\ref{hol-rel2}) is presented in appendix \ref{apd}.

Since all non-extremal three point functions
between three ${\cal O}_{S^{I}}$ operators vanish
\cite{Mihailescu:1999cj,Arutyunov:2000by}, one also obtains no
extremal three point function and therefore no extra contributions to
$\left < S^2_I \right >$ beyond the standard term given in
(\ref{holrel}). The cubic coupling between one $\Sigma$ field and two $S$
fields is however generically non-vanishing
\cite{Mihailescu:1999cj,Arutyunov:2000by} and therefore we do obtain
an extremal three point function which leads to the following result
for the scalar contributions to the one point function
(\ref{cop1}), (\ref{cop2})
\be \label{fq1}
\<{\cal O}_{\S^2_I}\> =
 \left(\frac{n_1 n_5}{4 \pi} \right)\left(\pi^{\S^2_I}_{(2)}
- \frac{1}{4 \sqrt{2}} a_{Iij} \pi^{S^1_i}_{(1)}
\pi_{(1)}^{S^{1}_j} \right).
\ee
An extremal coupling between the dimension two
scalar operators and two R symmetry currents would require a
term in the rhs of (\ref{fq1}) proportional to ${\cal A}_u {\cal A}^u$.
However such term is gauge dependent and thus forbidden. We conclude
that there are no additional contributions to (\ref{fq1}).

Before leaving this section we should note why the extremal
correlators were fixed via a limit of the non-extremal supergravity
correlators and other indirect arguments rather than from a dual field
theory computation.
The relevant three point functions of scalar operators in the orbifold CFT
were computed in \cite{Jevicki:1998bm} and \cite{Lunin:2001pw}.
There is no known non-renormalization
theorem to protect them and thus no justification for extrapolating
them to the strong coupling regime. Indeed, as we discuss in
appendix \ref{apf}, certain correlation functions seem to disagree between
supergravity and the orbifold CFT.

\subsubsection{Stress energy tensor}

Finally we discuss the vev for the stress energy tensor. This being a dimension
two operator, we again need to take into account terms quadratic in two
dimension one operators.
Terms quadratic in the scalar fields $S^1_i$ and in the gauge fields
$A_{\m}^{\pm \a}$ both contribute.
Let us momentarily suppress the gauge field contributions. Then as discussed in
the previous section, the three dimensional metric couples at leading
order to the scalar field $S^1_i$ in the three dimensional equations of
motion and thus we need to derive the one
point functions for this coupled system. This computation is
very similar to the Coulomb branch analysis given in \cite{BFS1,BFS2}
and is summarized in appendix \ref{apc}.

Next consider the additional contributions to
the stress energy tensor quadratic in the gauge field.
These immediately follow from the variation of the boundary terms
(\ref{b-cs}), since the bulk Chern-Simons
terms cannot contribute. Thus the total result for the stress energy
tensor follows from (\ref{1pts}) plus gauge field terms giving:
\be \la{set}
\< T_{uv} \> = \frac{n_1 n_5}{2 \pi} \left ( g_{(2)uv} + \half R g_{(0) uv}
+ \frac{1}{4} (\td{S}^1_{(0)})^2 g_{(0)uv} + \qu ({\cal A}^{+ \a}_{(u}
{\cal A}^{+ \a}_{v)} + {\cal A}^{- \a}_{(u} 
{\cal A}^{- \a}_{v)}) + \cdots \right ),
\ee
where the terms in ellipses (source terms for the scalars) are given in
(\ref{1pts}) but do not contribute in our solutions. (Recall that parentheses
 denote the symmetrised traceless combination of indices.)

Now consider the effect of a large gauge transformation of the form
${\cal A}^{+ 3}_w \rightarrow {\cal A}^{+3}_w + \eta$. As discussed in
\cite{Kraus} (see also \cite{Balasubramanian:2000rt}) this induces the shifts
\be
L_0 \rightarrow L_0 + \eta J_0^{+3} + \qu k \eta^2; \hsp
J_0^{+3} \rightarrow J_0^{+3} + \half k \eta,
\ee
where the Virasoro generator is
defined as $L_0 = \frac{c}{24} + \oint dw T_{ww}$ and
the level of the $SU(2)$ algebra is $k \equiv n_1 n_5$. This is
clearly a spectral flow transformation, and shows the
relationship between bulk coordinate transformations on the $S^3$ and
spectral flow in the boundary theory.

\section{Vevs for the fuzzball solutions} \la{v-fuzz}

We are now ready to extract the vevs from the asymptotic expansions of
the fields in the fuzzball solutions given in (\ref{p2}), (\ref{pm2}),
(\ref{pm3}) and (\ref{p1}). The appropriate (gauge invariant)
combinations of six-dimensional scalar and gauge fields are
\bea \la{field1}
{\bf s^1_i} &=& {\bf \frac{1}{4 r} (f^1_{1i} - f^5_{1i}) + \cdots; \hsp
s^2_I = \frac{1}{8 r^2} (f^1_{2I} - f^5_{2I}) + \cdots;} \\
{\bf \s^2_I} &=& {\bf - \frac{1}{8 r^2} (f_{2I}^1 + f_{2I}^5)}
+ \frac{1}{24 r^2} (f_{1i}^5) (f_{1j}^5)  a_{Iij} + \frac{1}{r^2}
a^{\a -} a^{\b +} f_{I \a \b} + \cdots. \nn \\
{\bf A^{+ \a}_{t}} &=& {\bf - 2 a^{\a +}} + \cdots; \hsp
{\bf A^{+ \a}_{y} = 2 a^{\a +}} + \cdots , \nn \\
{\bf A^{- \a}_{t}} &=& {\bf - 2 a^{\a -}} + \cdots; \hsp
{\bf A^{- \a}_{y} = -2 a^{\a -}} + \cdots. \nn
\eea
The graviton is given by
\bea
H_{tt} &=& f_{1i}^5 f^{5}_{1i} - a^{\a +} a^{\a +} - a^{\a -}
a^{\a -} + \cdots; \label{H} \\
H_{yy} &=& -  f_{1i}^5 f^{5}_{1i} - a^{\a +} a^{\a +} - a^{\a -}
a^{\a -} + \cdots; \nn \\
H_{ty} &=& a^{\a +} a^{\a +} - a^{\a -} a^{\a -} + \cdots; \nn \\
H_{rr} &=& - \frac{2}{r^4} f_{1i}^5 f^{5}_{1i} + \cdots. \nn
\eea
Next we extract the three-dimensional fields, which involves
rescaling and shifting the scalar fields as defined in (\ref{def1})
and (\ref{def2}):
\bea
S^{1}_{i} &=& -\frac{2 \sqrt{2} }{r} f^5_{1i} +\cdots ; \hsp
S^{2}_{I} = \frac{\sqrt{3}}{\sqrt{2} r^2} (f^1_{2I} - f^5_{2I}) + \cdots ; \\
\Sigma_2^I &=& \sqrt{32} (- \frac{1}{8 r^2} (f_{2I}^1 + f_{2I}^5)
+ \frac{1}{2 r^2} (f_{1i}^5) (f_{1j}^5)  a_{Iij} + \frac{1}{r^2}
a^{\a -} a^{\b +} f_{I \a \b} +\cdots). \nn
\eea
where we used (\ref{k1-con}) in $S^{1}_{i}$.
Note that the gauge fields and the metric are not rescaled or shifted upon the
dimensional reduction to this order.

Thus for the scalar operators we obtain using (\ref{holrel}) and
(\ref{fq1}) the vevs
\bea
\left < {\cal O}_{S^1_i} \right > &=& \frac{n_1 n_5}{4 \pi}
(- 4 \sqrt{2} f^{5}_{1i}); \la{vv2} \\
\left < {\cal O}_{S^2_I} \right > &=& \frac{n_1 n_5}{4 \pi} ( \sqrt{6} (f^1_{2I} -
f^5_{2I}) ); \nn \\
\left < {\cal O}_{\S^2_I} \right > &=& \frac{n_1 n_5}{4 \pi} \sqrt{2}
( -  (f^1_{2I} +f^5_{2I}) + 8 a^{\a -} a^{\b +} f_{I\a\b} ). \nn
\eea

The currents follow from (\ref{cur1}) as
\bea
\left < J^{+\a} \right >  = \frac{n_1 n_5}{2 \pi} a^{\a+} (dy -dt); \hsp
\left < J^{-\a} \right > = - \frac{n_1 n_5}{2 \pi} a^{\a-} (dy + dt).
\eea
To evaluate the vev of the stress energy tensor using (\ref{set})
we first need to
bring the metric into the Fefferman-Graham coordinate
system. This requires the following change of radial
coordinate
\be
z = \frac{1}{r} - \frac{1}{2 r^3} (f^5_{1i})^2 + \cdots
\ee
After changing radial coordinate in this way the metric becomes
\bea
ds_3^2 &=& \frac{dz^2}{z^2} + \frac{1}{z^2} (1 - 2 (f^5_{1i})^2 z^2)
(-dt^2 + dy^2) \\
&& - a^{\a+} a^{\a +} (dt - dy)^2
- a^{\a -} a^{\a -} (dt+dy)^2 + \cdots \nn
\eea
The metric perturbation in the second line is traceless with
respect to the leading order metric. Now applying the formula
(\ref{set}) we find that
\be
\<T_{uv} \> = 0.
\ee
This is the anticipated answer, since these solutions are supposed to
be dual to R vacua. The cancellation is however very non-trivial and
needed all the machinery of holographic renormalization.

\subsection{Higher dimension operators}

Having  extracted the vevs for all operators up to dimension two
using the systematic procedure developed in \cite{Skenderis:2006uy},
it is worth considering whether any predictions can be made for vevs
of higher dimension operators. These could of course be determined by
the same systematic procedure used above, by retaining all terms to
sufficiently high order, but this would involve considerable
computation.

It is therefore useful to recall at this point the result obtained
in \cite{Skenderis:2006di} for the vevs extracted from
supergravity solutions corresponding to
the Coulomb branch of ${\cal N}=4$ SYM. When these solutions are
asymptotically expanded in the radial coordinate of the defining harmonic
function, non-linear terms in the vevs of CPOs arising from non-linear
terms in the higher dimensional fields, non-linear terms in the KK
reduction map and non-linear terms in the holographic renormalization
relations all cancel out\footnote{Strictly speaking, the cancellation
  was proven in \cite{Skenderis:2006di} for operators of dimension four
    and less for which the corresponding vevs had been extracted using
    the rigorous procedures of \cite{Skenderis:2006uy}. However, the
    linearized approach gave results which agreed with the
    (non-renormalized) weak coupling field theory results for all
    dimension operators.}! The vevs are given by the {\it linear} terms
in the higher dimensional fields. ``Non-linear'' in this context means terms
which are non-linear in spherical harmonics.

Now consider what happens here if one retains only the linear terms in the
fields, the dimensional reductions and the holographic renormalization
relations. Then from (\ref{field1}), only the terms in boldface are retained.
This means that there is no graviton
perturbation to this order, and thus that the three-dimensional mass
vanishes, in accordance with the expectation that these geometries
describe R vacua. Furthermore, these terms give precisely the same
results as before for the scalar ${\cao}_S$ and current vevs, in which
all non-linear contributions
canceled. It is an interesting question to understand why the linear terms
alone
determine the stress energy tensor and ${\cao}_S$  vevs.
Note that just as in  \cite{Skenderis:2006di}
a priori there is absolutely no justification for
neglecting the non-linear terms, given that there is no small
parameter. Presumably this question can be answered by understanding
holographic renormalization directly in the higher dimension and developing the
map between higher-dimensional fields and operators.

However, the linear terms clearly {\it fail} to give the correct answer for
the operators dual to $\S^2$.  Thus the linearized approximation
in this situation fails already at dimension two, which is the first
place where non-linear terms can play a role (but note that it still holds
for the dimension two operator ${\cao}_{S^2}$).

Nevertheless one may proceed with the  linearized procedure
in order to get a rough idea of the behavior of the vevs for higher
dimension operators. From the asymptotic expansion of the solution
we extract the following linear terms for the scalars
\bea
s^{k}_{I} &=& \frac{1}{4k r^k} (f^1_{kI} - f^{5}_{kI}) + \cdots \\
\s^{k}_{I} &=& - \frac{1}{4 k r^k} (f^{1}_{kI} + f^{5}_{kI}) + \cdots
\nn
\eea
{}From these asymptotics the vevs of the dual operators contain the
linear terms
\bea
\left < {\cal{O}}_{S^{k}_{I}} \right > &=&
\left (\frac{n_1 n_5}{4 \pi} \right ) 2 (k-1)
\frac{\sqrt{k+1}}{\sqrt{k}} (f_{kI}^1 - f^5_{kI} + \cdots); \la{h-1} \\
\left < {\cal{O}}_{\S^k_I} \right > &=&
- \left (\frac{n_1 n_5}{4 \pi} \right ) 2 (k-1) \frac{\sqrt{k-1}}{\sqrt{k}} (f^1_{kI}
+ f^{5}_{kI} + \cdots), \nn
\eea
where the ellipses denote the non-linear terms.
Recall that $(f^1_{k I}, f^{5}_{kI})$ are proportional to the
$k$th multipole moments of the D1 and D5 brane charge
distributions, respectively. We will argue in the section \ref{corresp} that
these linear terms do not give the expected answer for the vevs of operators
${\cal{O}}_{\S^k_I}$, although they seem to be sufficient to
give the expected answer for the vevs of operators ${\cal{O}}_{S^k_I}$,
at least for circular curves.

Following analogous arguments for the dimension $k_v$ vector chiral primaries
$J^{I_v \pm}_{k_v}$ dual to bulk vectors $A^{I_v \pm}_{k_v}$,
we get the following structure
\be
\left < J^{I_v \pm}_{k_v} \right > \propto \left (\frac{n_1 n_5}{4
  \pi} \right ) (A_{kI})_i E^{\pm}_{I_v I i} ( dt \mp dy) \la{h-2} + \cdots,
\ee
where the ellipses denote again the non-linear terms,
the spherical harmonic triple overlap $E^{\pm}_{I_v I i}$ is
defined in (\ref{gt1}) and $(A_{kI})_i$ is defined in terms of the curve
$F^i(v)$ in (\ref{p00}). To extract the exact coefficient relating the
asymptotics of the bulk vector fields to the current vev, one would
need to analyze the relevant Proca-Chern-Simons bulk equation and
obtain the holographic renormalization relation for this case.

In the discussions of \cite{Skenderis:2006di}, the vevs obtained by
the linearized approach gave correctly all the (non-renormalized)
field theory vevs. Here the linearized approach does not give
correctly vevs for chiral primaries. Moreover, we will also
argue that there are additional vevs which are
not captured by the linearized approximation at all. For example, when one
linearizes the solution following the above procedure
the (non primary) scalar fields $(t^k_I, \t^k_I)$ are
identically zero, but arguments given in section \ref{corresp}
suggest that the corresponding operators should in general have
non-zero expectation values. Perhaps these vevs could still be extracted by
an appropriate linearized analysis, but it is not apparent what the
prescription should be. By contrast, the systematic method of
\cite{Skenderis:2006uy} used in earlier sections will certainly give
the correct answer for these vevs.

Note also that the linearized approximation manifestly gives different answers in
different coordinate systems. For the example of the solution based
on a circular curve we discuss in the next section,
the linearized approximation in the coordinate system (\ref{nh})
actually gives the conjectured answers for scalar vevs, but
linearizing in the hatted coordinate system (flat coordinates on $R^4$)
gives different answers. Both in the fuzzball solutions considered here
and in the Coulomb branch solutions discussed in \cite{Skenderis:2006di} there
are preferred coordinate systems, those in which the harmonic
functions are naturally expressed.
For the Coulomb branch this coordinate systems was precisely that in
which linearizing gives the correct vevs, but here it does not.
In general, however, there will be no preferred coordinate system or it may not
be visible (as in (\ref{nh})), and therefore there would be no natural way to linearize;
one would have to apply the general methods of \cite{Skenderis:2006uy}.

\section{Examples} \la{examples}

We discuss in this section the application of the general results to
two specific examples: solutions based on circular and ellipsoidal
curves, respectively.

\subsection{Circular curves} \la{sp-ex}

A commonly used example of the fuzzball solutions is that in which the
curve $F^i(v)$ is a (multiwound) circle
\cite{Balasubramanian:2000rt,Maldacena:2000dr,Lunin:2001fv},
\be \label{fcirc}
F^1 = \mu_n \cos \frac{2 \pi n v}{L}, \qquad F^2 =
\mu_n \sin \frac{2 \pi n v}{L},
\qquad F^3=F^4=0.
\ee
The ten-dimensional solution in this case is conveniently written as
\bea
ds^2 &=& f_{1}^{- 1/2} f_{5}^{- 1/2} \left ( - (d\tilde{t} - \frac{\mu_n
  \sqrt{Q_1 Q_5}}{\hat{r}^2 + \mu_n^2 \cos^2 \hat{\theta}} \sin^2 \hat{\q}
d \phi)^2 + (d\tilde{y} -
\frac{\mu_n \sqrt{Q_1 Q_5}}{\hat{r}^2 + \mu_n^2 \cos^2 \hat{\theta}}
\cos^2 \hat{\q} d \psi)^2
\right ) \nn \\
&& + f_{1}^{1/2} f_{5}^{1/2} \left ( (\hat{r}^2 + \mu_n^2 \cos^2 \hat{\q}) (
\frac{d\hat{r}^2}{\hat{r}^2 + \mu_n^2} + d \hat{\q}^2) + \hat{r}^2 \cos^2
\hat{\q} d \psi^2 + (\hat{r}^2 + \mu_n^2) \sin^2 \hat{\q} d \phi^2 \right )
\nn \\
&& +
f_{1}^{1/2} f_{5}^{- 1/2} dz \cdot dz; \la{ez2} \\
e^{2 \Phi} &=& f_1 f_5^{-1}, \nn \\
f_{1,5} &=& 1 + \frac{Q_{1,5}}{\hat{r}^2 + \mu_n^2 \cos^2 \hat{\q}}, \nn
\eea
whilst the tensor field is as in (\ref{ez1}) with
\be
A = \mu_n \frac{\sqrt{Q_1 Q_5}}{(\hat{r}^2 + \mu_n^2 \cos^2 \hat{\q})} \sin^2
\hat{\q} d\phi; \hsp
B = - \mu_n \frac{\sqrt{Q_1 Q_5}}{(\hat{r}^2 + \mu_n^2 \cos^2 \hat{\q})} \cos^2
\hat{\q} d\psi.
\ee
This solution is precisely of the form (\ref{ez1}), using a
non-standard coordinate system on $R^4$. That is, the hatted
coordinates $(\hat{r}, \hat{\q},\phi,\psi)$ are related to usual coordinates
$(r,\q,\phi,\psi)$ on $R^4$ such that the metric is
\be
ds^2 = dr^2 + r^2 (d\q^2 + \sin^2 \q d\phi^2 + \cos^2 \q d \psi^2),
\ee
via
\be
\hat{r} \cos \hat{\q} = r \cos \q; \hsp
r^2 = \hat{r}^2 + \mu_n^2 \sin^2 \hat{\q}. \la{non-stand}
\ee
Note that this relation implies
\be
\frac{1}{\hat{r}^2 + \mu_n^2 \cos^2 \hat{\q}} = \frac{1}{\sqrt{
    (r^2 + \mu_n^2)^2 - 4 \mu_n^2 r^2 \sin^2 \q}},
\ee
with the latter admitting the following asymptotic expansion
\be \la{v1}
\frac{1}{\sqrt{
    (r^2+\mu_n^2)^2 - 4 \mu_n^2 r^2 \sin^2 \q}} =
\sum_{k \in 2 Z} (-1)^{k/2} \frac{\mu_n^k
  Y^{0}_k(\q_3)}{ \sqrt{k+1} r^{2+k}},
\ee
where the harmonic function is expanded in normalized spherical
harmonics $Y_{k}^{0}$ which are singlets under the $SO(2)^2$ Cartan of
$SO(4)$. These harmonics are given in (\ref{singlet});
there is precisely one such singlet at each even degree.
The asymptotic expansion in (\ref{v1}) follows from (\ref{add-th})
upon using the fact that the lhs of (\ref{v1}) is equal to
\be
\frac{1}{L} \int_{0}^{L} \frac{dv}{ \left | x - F \right |^2},
\ee
with $F^i$ given in (\ref{fcirc}), so $\theta_3^F=\pi/2$ and
$Y^{0}_{k} (\pi/2) =(-1)^{k/2} \sqrt{k+1}$ .

The parameters $(n,\mu_n)$ labeling the curve are
related to the charges via
\be \la{ch-ab}
n \mu_n = \frac{L}{2 \pi} \sqrt{\frac{Q_1}{Q_5}} = \frac{\sqrt{Q_1
    Q_5}}{R} \equiv \mu,
\ee
or equivalently $\mu_n =1/ (n\td{R})$, where
$\td{R} = R/\sqrt{Q_1 Q_5}$.
In deriving these results we
have used (\ref{one-brane}) and (\ref{cur-leng}).

\bigskip

The near horizon limit of (\ref{ez2}) gives the six-dimensional fields
\bea
ds_6^2 &=& \sqrt{Q_1 Q_5} \left ( - (\hat{r}^2 + \mu_n^2) dt^2 + \hat{r}^2
dy^2 + \frac{d\hat{r}^2}{(\hat{r}^2 + \mu_n^2)} \right )
\la{nh} \\
&& + \sqrt{Q_1 Q_5}
\left ( d\hat{\q}^2 + \sin^2 \hat{\q} (d \phi + \mu_n d \td{t})^2 +
\cos^2 \hat{\q} (d \psi - \mu_n d \td{y} )^2 \right ); \nn \\
G^5 &=& \sqrt{Q_1 Q_5} \hat{r} dt \wedge dy \wedge
d\hat{r} + \sqrt{Q_1 Q_5} \cos \hat{\q} \sin \hat{\q} d\hat{\q}
\wedge (d \phi + \mu_n dt) \wedge (d \psi - \mu_n dy).
\nn
\eea
with the scalar field ${\Phi}$ and the anti-self dual field
$G^6$ vanishing. As previously, it is convenient to use the rescaled
coordinates $\td{t} = \sqrt{Q_1 Q_5} t$ and $\td{y} = \sqrt{Q_1 Q_5}
y$ so that the overall scale factor is manifest. Note that the
coordinate $y$ has periodicity $2 \pi \td{R}$. When $n=1$
there is a coordinate transformation
($\phi \rightarrow \phi + \mu_n t$,
$\psi \rightarrow \psi + \mu_n y$)
that makes the metric exactly $AdS_3 \times S^3$.
For $n>1$ one can similarly shift the angular
coordinates, but the resulting spacetime metric has a conical defect.
As discussed in \cite{Balasubramanian:2000rt,Lunin:2002iz}, such a coordinate change
is equivalent to carrying out a spectral flow to the NS sector; in the
case of $n=1$ the flow is
to the vacuum. One way of seeing this is that under such a shift the Killing spinors
change periodicity about the circle direction $\td{y}$. In the above
coordinate system they are periodic, whilst after the coordinate
transformation they are anti-periodic \cite{Balasubramanian:2000rt}.

\bigskip

In the context of this paper, however, we are interested in R vacua of the
CFT, and thus we do not wish to flow to the NS sector. This means we should
interpret the solution in the original coordinate system, where the Killing spinors are
periodic. From (\ref{nh}) we can immediately read off the three
dimensional gauge field as
\be
A^{-3} = \mu_n (dy + dt); \hsp A^{+3} = \mu_n (dy - dt).
\ee
The superscript indicates that the relevant
Killing vectors are those given in the appendix in (\ref{kform}), such
that $A^{+3}$ and $A^{-3}$ commute. The fact that there is
a coordinate transformation where the solution is (locally)
$AdS_3 \times S^3$ means that the three dimensional scalar
fields $(S^1_i,S^2_I,\S^2_I,\cdots)$ vanish. Note that the
latter result is immediately obvious in the hatted coordinate system
but it is not manifest in the coordinate system $(r,\q,\f,\psi)$.
That the $S$ fields vanish in the latter coordinate system
follows from (\ref{field1}) since $f_{k I}^1 =
f_{k I}^5$. To see the vanishing of $\S^2_0$
one has to use in (\ref{field1}) the identity
\be
- \frac{1}{8} (f^1_{2 0} + f^5_{2 0}) + f_{0 3 3} a^{3+} a^{3-} = 0,
\ee
which follows from (\ref{v1}) and the identity (\ref{o2}).

Now given the three dimensional fields we derive the corresponding
vevs,
\bea
\left < T_{uv} \right > &=& \left < {\cal O}_{S^1_i} \right > = \left
< {\cal O}_{S^2_I} \right > =\left < {\cao}_{\S^2_0} \right > = 0;
\la{1pt-sp} \\
\left < J^{+3} \right > &=& \frac{n_1 n_5}{4 \pi} \mu_n (dy - dt); \hsp
\left < J^{-3} \right > = \frac{n_1 n_5}{4 \pi} \mu_n (dt + dy). \nn
\eea
Note that the R-symmetry charges
\bea \la{ch-sp}
j^3 \equiv \int_{0}^{2 \pi \td{R}} dy J^{+3}_{\td{y}}
= \frac{n_1 n_5}{2 n}; \\
\bar{j}^3 \equiv \int_{0}^{2 \pi \td{R}} dy J^{-3}_{\td{y}}
= \frac{n_1 n_5}{2n}, \nn
\eea
are quantized in half integral units provided that $n$
is a divisor of $n_1 n_5$.

\subsection{Ellipsoidal curves} \la{ellipsoid}

The next simplest case to consider is a solution determined by a
planar ellipsoidal curve:
\be \label{fell}
F^1 = \mu_n a \cos \frac{2 \pi n v}{L}, \qquad F^2 = \mu_n b \sin \frac{2 \pi n v}{L},
\qquad F^3=F^4=0,
\ee
with $\mu_n$ as in (\ref{ch-ab}).
The D1-brane charge constraint (\ref{one-brane}) implies that $(a^2 +
b^2) = 2$. The vevs for this solution are given by
\bea
\left < T_{uv} \right > &=& \left < {\cal O}_{S^1_i} \right > = 0; \nn \\
\left < J^{+3} \right > &=& \frac{N}{4 \pi} \mu_n a b (d{y} - d{t}); \hsp
\left < J^{-3} \right > = \frac{N}{4 \pi} \mu_n a b (d{t} + d {y}); \nn \\
\left < {\cal O}_{S^2_{m,\bar{m}}} \right > &=& \left
< {\cao}_{\S^2_{m,\bar{m}}} \right > = 0;
\hsp m \neq \bar{m}
\la{1pt-sp-ell} \\
\left < {\cal O}_{S^2_{1,1}} \right > &=& \left < {\cal
  O}_{S^2_{-1,-1}} \right > = - \frac{N}{8 \sqrt{2}
\pi} \mu_n^2 (a^2 - b^2); \nn \\
\left < {\cal O}_{S^2_{0,0}} \right > &=& \frac{N}{4 \sqrt{2} \pi}
\mu_n^2 (a^2 b^2 - 1); \nn \\
\left < {\cal O}_{\S^2_{1,1}} \right > &=& \left < {\cal
  O}_{\S^2_{-1,-1}} \right > = - \frac{\sqrt{3} N}{8 \sqrt{2} \pi}
\mu_n^2 (a^2 - b^2); \nn \\
\left < {\cal O}_{\S^2_{0,0}} \right > &=&
\frac{\sqrt{3} N}{4 \sqrt{2} \pi} \mu_n^2 (a^2 b^2 - 1). \nn
\eea
Here we denote by $(m, \bar{m})$ the $(SU(2)_L, SU(2)_R)$ charges.
The vanishing of the vevs of
operators with charges $m \neq \bar{m}$ follows from the fact that the curve
preserves rotational symmetry in the 3-4 plane. The equality of the
vevs for operators with charge $(1,1)$ and $(-1,-1)$ follows from the
orientation of the ellipse in the 1-2 plane: its axes are orientated
with the 1-2 axes. Explicit representations of the corresponding spherical
harmonics are given in (\ref{ch-sp-ha}).

One can also consider a planar ellipsoidal curve of different
orientation, described by the curve
\be \la{fell2}
F^1 = \mu_n ( a \cos \frac{2 \pi n v}{L} + c \sin \frac{2 \pi n v}{L})
, \qquad F^2 = \mu_n ( b \sin \frac{2 \pi n v}{L} + d \cos \frac{2 \pi n v}{L}),
\ee
with $F^3 = F^4 = 0$ and $\mu_n$ as in (\ref{ch-ab}). The D1-brane charge constraint
(\ref{one-brane}) in this case requires that $(a^2 + b^2 + c^2 + d^2)
= 2$. The non-vanishing vevs are
\bea
\left < J^{+3} \right > &=& \frac{N}{4 \pi} \mu_n (a b - c d ) (d{y} - d{t}); \hsp
\left < J^{-3} \right > = \frac{N}{4 \pi} \mu_n (a b - c d) (d{t} + d {y}); \nn \\
\left < {\cal O}_{S^2_{\pm 1,\pm1}} \right > &=& - \frac{N}{8 \sqrt{2}
\pi} \mu_n^2 ( (a \pm i d)^2 + (c \pm i b)^2 ); \nn \\
\left < {\cal O}_{S^2_{0,0}} \right > &=& \frac{N}{4 \sqrt{2} \pi}
\mu_n^2 ( (a b - cd)^2 - 1); \la{1pt-sp-ell2} \\
\left < {\cal O}_{\S^2_{\pm 1,\pm 1}} \right > &=& -
\frac{\sqrt{3} N}{8 \sqrt{2} \pi}
\mu_n^2 ( (a\pm id)^2 + (c \pm i b)^2); \nn \\
\left < {\cal O}_{\S^2_{0,0}} \right > &=& \frac{\sqrt{3} N}{4
\sqrt{2} \pi} \mu_n^2 ( (a b - cd)^2 - 1). \nn \eea The vevs for
operators with charge $(1,1)$ and $(-1,-1)$ are no longer equal,
since the axes of the ellipse are no longer orientated with the 1-2
axes. The vevs are however complex conjugate, as they must be since
the operators are complex conjugate to each other.

\section{Dual field theory} \la{fti}

To understand the interpretation of the holographic results it will be useful
to review certain aspects of the dual CFT and the ground states of the
R sector. The dual CFT is believed to be a deformation of
the ${\cal N}=(4,4)$ supersymmetric sigma model with target space
$S^N(X)$, where $N=n_1 n_5$ and the compact space
is either $T^4$ or $K3$. Most of the discussion below will
be for the case of $T^4$, although the results extend simply to $K3$.
The orbifold point is roughly the analogue
of the free field limit of ${\cal N}=4$ SYM in the context of $AdS_5/CFT_4$
duality.

The chiral primaries and R ground states can be precisely described
at the orbifold point. In particular, there exists a family of
chiral primaries in the NS-NS sector associated with the $(0,0)$, $(2,0)$,
$(0,2)$, $(1,1)$ and $(2,2)$ cohomology of the internal manifold
(we do not discuss the chiral primaries associated
with odd cohomology in this paper). These can be labeled as
\bea
&& {\cal O}^{(0,0)}_{n}, \hsp h = \bar{h} = \half (n-1); \la{oper-2} \\
&& {\cal O}^{(2,0)}_{n}, \hsp h = \bar{h}+1 = \half (n+1); \nn \\
&& {\cal O}^{(1,1) q}_{n}, \hsp h = \bar{h}= \half n;
\qquad q=1,\ldots, h^{1,1}
\nn \\
&& {\cal O}^{(0,2)}_n, \hsp h = \bar{h} - 1 = \half (n-1); \nn \\
&& {\cal O}^{ (2,2)}_n, \hsp h = \bar{h} = \half (n+1), \nn
\eea
where $n$ is the twist, $h^{1,1}$ in the dimension
of the $(1,1)$ cohomology  and $h=j^3$, $\bar{h} = \bar{j^3}$.
The operator ${\cal O}^{(0,0)}_{1}$ is the identity operator.
The complete set of chiral primaries associated with this cohomology is built
from products of the form
\be \la{oper-1}
\prod_{l=1}
({\cal {O}}^{(p_l+1,q_l+1)}_{n_l})^{m_l},
\qquad \hsp \sum_{l=1}^I n_l m_l = N \, ,
\ee
where $p_l, q_l$ take the values $\pm 1$ (so that one gets the
product of operators in (\ref{oper-2}); we suppress the index $q$)
and symmetrization over the $N$ copies of the CFT is implicit.

In \cite{deBoer:1998ip} the spectrum of chiral primary operators of
the orbifold CFT was matched with the KK spectrum. One should note
however that the matching is not canonical in the sense that the
operators at the orbifold point and the fields in supergravity are
characterized by additional labels not visible in the other description.
In particular,
the supergravity spectrum is also organized in
representations of an additional\footnote{$\widetilde{SO}(4)$ was called
$SO(4)_R$ in \cite{Sez}.}  $\widetilde{SO}(4) \times SO(n_t)$,
as can be seen from the tables of \cite{Sez}, where the
$\widetilde{SO}(4)$ is the R-symmetry of the 6D supergravity
(not to be confused
with the $SO(4)$ R-symmetry of the CFT which is related to the
isometries of the $S^3$) and $n_t$ is the number of tensor multiplets.
On the other hand, the chiral spectrum at the orbifold point is
characterized by the set of integers $n_l, m_l$ and the
type of operator associated with these, as in (\ref{oper-1}).
Furthermore, there is an additional $SO(4)_I$ acting on the chiral
spectrum, related to global rotations of $T^4$ (see, for example,
\cite{Jevicki:1998bm} or the review \cite{David:2002wn}). It is not
immediately clear how the labels $n_l, m_l$ translate in the supergravity
description and what is the relation of $SO(4)_I$ with the supergravity
$\widetilde{SO}(4) \times SO(5)$ ($n_t=5$ for $T^4$).

To get a more precise mapping
let us consider the special case of
chiral primaries with $h=\bar{h}$. We see from
(\ref{oper-2}) that there are 6 such operators for any $h<N/2$, except
when $h=1/2$ in which case there are only 5 operators
(${\cal O}^{(0,0)}_{1}$ is the
identity operator).  In all cases  4 of these operators
form a vector of $SO(4)_I$. On
the supergravity side, the fields $S_{k}^{(r )I}$ and $\S^I_k$
have the correct dimensions and charges to correspond to these
operators. Note that $k>1$ for $\S^I_k$, so we indeed have only 5
fields corresponding to operators of dimension $(1/2,1/2)$.
These fields are singlets under $\widetilde{SO}(4)$ and
$S_{k}^{(r )I}$ transforms in the vector of $SO(5)$. It thus appears
natural to identify $SO(4)_I$ with an $SO(4)$ subgroup of $SO(5)$
and to make the correspondence
\be
{S_{n}^{p(q+6)}} \quad \leftrightarrow \quad {\cal O}^{(1,1) q}_{n}, \quad
q=1,\ldots, 4,  \hsp n \ge 1
\ee
where here and below
the superscript $p$ denotes that the relevant scalar fields are
those for which $j^3=j$ and $\bar{j}^3= \bar{j}$.
The question is then whether ${\cal O}^{(0,0)}_{n+1}$
or ${\cal O}^{(2,2)}_{n-1}$ corresponds to ${S_{n}^{p (6)}}$.
The most natural correspondence seems to be
\bea
{S_{n}^{p (6)}} \quad &\leftrightarrow& \quad {\cal O}_{n+1}^{(0,0)},
\hsp n \ge 1; \la{corr} \\
{\S_{n}^{p}} \quad &\leftrightarrow& \quad {\cal O}_{n-1}^{(2,2)} \hsp
n \ge 2. \nn
\eea
This identification is natural given that there is no $\S_{1}$ in
supergravity but is clearly not unique because $S^p_{n}$ and $\S^p_{n}$
have the same charges so it could be that different combinations
of them correspond to the operators at the orbifold point.

A similar discussion holds for chiral primaries with
$h-\bar{h}=\pm 1/2, \pm 1$. The case of $h-\bar{h}=\pm 1/2$
is not relevant here since we are not considering
solutions associated with odd cohomology in this paper.
The case $h-\bar{h}=\pm 1$ is relevant but most of the points we
want to make can be made using examples that utilize only
chiral primaries with $h = \bar{h}$, so we will not need a
detailed discussion of them. We only mention that the
corresponding supergravity fields are massive vector
fields.

Spectral flow maps these chiral primaries in the NS sector to R ground states, where
\bea
h^R &=& h^{NS} - j_3^{NS} + \frac{c}{24}; \nonumber \\
j_3^R &=& j_3^{NS} - \frac{c}{12}, \label{spe_fl}
\eea
where $c$ is the central charge. Each of the operators in (\ref{oper-1})
is mapped  by spectral flow to an operator of definite R-charge
\be \label{Rop}
\prod_{l=1} ({\cal {O}}^{(p_l+1,q_l+1)}_{n_l})^{m_l} \quad
  \rightarrow \quad {\cal O}^{R(2 j_3^R,2 \bar{j}^R_3)}, \qquad
 j_3^R =\half \sum_l p_l m_l, \
\bar{j}^R_3 =\half \sum_l q_l m_l.
\ee
In particular, for fixed twist $n$ the operators in (\ref{oper-2})
have the following charges after the flow
\bea
{\cal O}^{(0,0)}_{n} & \rightarrow & {\cal O}^{R(-,-)}_{n};
%\hsp j_3^R = \bar{j}^R_3 = - \half;
\label{spe_fl2}\\
{\cal O}^{(2,0)}_{n} & \rightarrow & {\cal O}^{R(+,-)}_{n};
%\hsp j^R_3 = - \bar{j}^R_3 = \half;
\nn \\
{\cal O}^{(0,2)}_n & \rightarrow & {\cal O}^{R(-,+)}_{n};
%\hsp - j^R_3 = \bar{j}^R_3 = \half ;
\nn \\
{\cal O}^{ (2,2)}_n & \rightarrow & {\cal O}^{R(+,+)}_{n};
% \hsp j^R_3 = \bar{j}^R_3 = \half ;
\nn \\
{\cal O}^{(1,1) q}_{n} & \rightarrow & {\cal O}^{R(0,0)q}_{n},
%\hsp j^R_3 = \bar{j}^R_3 =0,
\nn
\eea
where it is understood that each of these operators is tensored
by the appropriate power of the identity operator such
that (\ref{oper-1}) holds. For example, ${\cal O}^{(0,0)}_{n}$
should be tensored by $({\cal O}^{(0,0)}_{1})^{N-n}$, and the
R-symmetry charge of the flown operator
${\cal O}^{R(-,-)}_{n}$ follows from (\ref{spe_fl}) with $c=6n$.
It follows from (\ref{spe_fl2}) that
the operators ${\cal O}^{R(\pm,\pm)}_{n}$
form a $(\half, \half)$ representation
of $SU(2)_L \times SU(2)_R$ whilst the operators ${\cal O}^{R(0,0)q}_{n}$
are $q$ singlets. From the form of the
operators in the NS sector (\ref{oper-1}) it is clear that $j^R \le
\half N$, since one can have at most
$N$ operators in the product. Symmetrization over the copies of the
CFT means that spectral flow in the
left and right moving sectors is not quite independent. When one
has $m$ copies of the same operator
one needs to symmetrize over copies and thus one obtains only states
with $j^R = \bar{j^R} = \half m$ (although the values of
$j_3^R$ and $\bar{j}_3^R$ range independently from $-j^R$ to $j^R$).

We will label by the R-charges the states obtained
by the usual operator-state correspondence,
\be
|j_3^R, \bar{j}_3^R\rangle = {\cal O}^{R(2 j_3^R,2 \bar{j}^R_3)}(0) |0\rangle.
\ee

\subsection{R ground states and vevs}

The R ground states can also be characterized by the expectation
value of gauge invariant operators in them. Since the fuzzball
solutions are conjectured to be dual to R ground states and the
vevs of gauge invariant operators is
the information we extracted from the fuzzball solutions
we would like to see what one can say about them using the dual CFT.
There are two sets of constraints on these vevs: kinematical and
dynamical.

\subsubsection{Kinematical constraints}

The kinematical constraints follow from symmetry
considerations and they have been recently discussed in
\cite{Skenderis:2006ah}. As discussed above
the R ground states in the (usual) basis are eigenstates
of the R-symmetry charge. This implies that only
neutral operators can have a non-vanishing vev,
\be \label{zerov}
\langle -j_3^R,-\bar{j}_3^R| O^{(k_1,k_2)} | j_3^R,\bar{j}_3^R \rangle =0,
\qquad \{k_{1} \neq 0\ or\
k_{2} \neq 0\}
\ee
where $k_1$ and $k_2$ are the R-charges of the operator
and we use the
fact that the bra state has the opposite R charge to the ket state.

\subsubsection{Dynamical constraints and 3-point functions}

The vevs of neutral gauge invariant operators are determined dynamically.
One way to determine them is using 3-point functions at the conformal
point. Let $|\Psi\rangle = O_\Psi(0)|0\rangle$. Then the vev of an operator
$O_k$ of dimension $k$ in the this state is given by
\be
\langle \Psi| O_k(\l^{-1}) |\Psi \rangle
= \langle 0| (O_\Psi(\infty))^\dagger O_k(\l^{-1}) O_\Psi(0)|0\rangle,
\ee
where $\l$ is a mass scale.
For scalar operators the 3-point function is uniquely determined by
conformal invariance and the above computation yields
\be \la{fusion}
\langle \Psi| O_k(\l^{-1}) |\Psi \rangle  = \l^k C_{\Psi k \Psi}
\ee
where $C_{\Psi k \Psi}$ is the fusion coefficient. Similarly, the
expectation value of a symmetry current measures the charge
of the state
\be
\langle \Psi| j(\l^{-1}) |\Psi \rangle
= \langle 0| (O_\Psi(\infty))^\dagger j(\l^{-1}) O_\Psi(0)|0\rangle
= q \l  \langle \Psi|\Psi \rangle
\ee
where $q$ is the charge of the operator $ O_\Psi$ under $j$.

Let us now apply these principles to the cases of interest here.
We will thus need to know the 3-point functions at the conformal
point, which can be
computed in the NS sector and then flowed to the R sector.
A computation of 3-point functions at the orbifold point
has been given in \cite{Jevicki:1998bm,Lunin:2001pw}.
We however need to know the result in the regime
where supergravity is valid. For the theory at hand there is
no known non-renormalization theorem that would protect the
3-point functions. Moreover, as discussed in appendix \ref{apf},
the 3-point functions that can also be computed holographically
(i.e. those involving only operators dual to supergravity fields)
are different from the 3-point functions computed at the orbifold
point.

So the only dynamical tests that one can currently
do must involve states created by operators corresponding
to single particle states. In our case the fuzzball
solutions are meant to correspond to the R ground states
connected with universal cohomology, so only states
created by the operators ${\cal{O}}_n^{R(\pm,\pm)}$
are relevant. For these
cases the corresponding 3-point point functions can be
computed by standard holographic methods using the
results in \cite{Mihailescu:1999cj,Arutyunov:2000by}.

Let $\Phi =(S,A^+,A^-,\S)$ be the fields dual to the operators
${\cal{O}}_n^{R(\pm,\pm)}$.  The three point
functions involving scalar chiral primaries have the following structure
\be %\la{3ptfunc1}
\left < {\cal O}_{\Phi}^{\dagger} {\cal{O}}_{\S} {\cal{O}}_{\Phi} \right >
\neq 0, \hsp
\left < {\cal{O}}_{\Phi}^{\dagger} {\cal {O}}_S  {\cal{O}}_{\Phi} \right > = 0.
\ee
where ${\cal{O}}_{\Phi}^{\dagger}$ denotes the conjugate operator
with $j^3 = - j$, $\bar{j}^3 = - \bar{j}$. Our results for the vevs
include the lowest dimension operators in these towers.

{}From the results of \cite{Arutyunov:2000by} there are however
other non-zero three point functions in supergravity, such as
\be
\left < {\cal O}_{\Phi}^{\dagger} {\cal{O}}_{\tau}
{\cal{O}}_{\Phi} \right > \neq 0, \hsp
\left < {\cal O}_{\Phi}^{\dagger} {\cal{O}}_{\rho^{\pm}}
{\cal{O}}_{\Phi} \right > \neq 0, \hsp
\left < {\cal O}_{\Phi}^{\dagger} {\cal{O}}_{A^{\pm}}
{\cal{O}}_{\Phi} \right > \neq 0, \hsp \cdots
\ee
where the ellipses denote other operators, dual to other vectors and
KK gravitons. These operators all have sufficiently high dimensions
that we did not compute their vevs. Moreover, the vevs of these
operators are not captured at all by the linearized approximation.

\section{Correspondence between fuzzballs and chiral primaries} \la{corresp}

\subsection{Correspondence with circular curves}

Having reviewed the description of the degenerate R ground
states in the CFT we now turn to the connection
with the fuzzball solutions. The basic proposal is that there is a
correspondence between the R ground states
and the curves $F^i(v)$ defining the supergravity solutions. 
Let us consider first states of the specific form
\be \la{px1}
({\cal O}^{R(\pm,\pm)}_{n})^{\frac{N}{n}} | 0 \rangle, \hsp \hsp j_3^R = \pm \frac{N}{2n};
\hsp \bar{j}_3^R = \pm \frac{N}{2n}.
\ee
Then such ground states are proposed to be in one to one
correspondence with circular curves \cite{Lunin:2001jy}:
\be
({\cal O}^{R (+,+)}_{n})^{\frac{N}{n}} | 0 \rangle \hsp \leftrightarrow \hsp F^1 =
\frac{\mu}{n} \cos (\frac{2 \pi n v}{L});
\hsp F^2 = \frac{\mu}{n} \sin (\frac{2 \pi n v}{L}),
\ee
with $F^3 = F^4 = 0$ and
where the parameter $\mu$ is fixed via (\ref{one-brane}) to be
$\sqrt{Q_1 Q_5}/R$, see (\ref{ch-ab}). Similarly
$({\cal O}^{R (-,-)}_{n})^{N/n}$ corresponds to a circle of the same
radius in the 1-2 plane with
the opposite rotation (that is, $F^2 \rightarrow - F^2$) and the operators
$({\cal O}^{R (+,-)}_{n})^{N/n},({\cal O}^{R (-,+)}_{n})^{N/n}$
correspond to circles in the 3-4 plane.

Note the states (\ref{px1}) are generically not dual to supergravity
fields. Only the specific states obtained by flowing the NS
operators $( ({\cao}^{(0,0)}_1)^N, {\cao}^{(p,q)}_N )$ correspond to
supergravity fields. All product operator do not correspond to
supergravity fields, with the exception of $ (\cao^{(0,0)}_1)^N$,
since this is simply the identity operator in the NS sector. Moreover,
whilst the operators ${\cao}^{(p,q)}_N$ are dual to supergravity
fields their special properties (following from having maximal
dimension) are not visible in supergravity computations which
effectively takes $N \rightarrow \infty$.

There are various pieces of evidence for this correspondence
between states and circular curves.
Firstly the rotation charges match, using the discussions in section
\ref{sp-ex}, in particular (\ref{ch-sp}).
Secondly, as first discussed in \cite{Lunin:2001jy}, one can consider
absorption processes in the corresponding geometries, and compare the
scattering behavior with CFT expectations; they agree.
(Note that for a general fuzzball geometry the wave equation for
minimal scalars is not separable,
so the absorption cross-section cannot be computed, and this
comparison cannot be made.)

Our results for the scalar 1-point functions in (\ref{1pt-sp}) (along with (\ref{h-1}))
give more data which can be used to test the proposed
correspondence.
As discussed previously kinematical constraints arise simply
from charge conservation: if the
R ground state is an eigenstate of both $j_3^R$ and $\bar{j}_3^R$ then only scalar
operators with $j^3 = \bar{j}^3 = 0$ can acquire
a vev. These correspond to the $Y^k_0$ harmonics discussed in section
\ref{sp-ex}. Thus the fact that only such
operators appear in (\ref{1pt-sp}) follows solely from kinematics.

\bigskip

Determining which of the (kinematically allowed) operators
actually acquire a vev involves
dynamics also and is rather more subtle. Consider
first the special case where the operator (\ref{oper-1}) determining
the ground state is the product $({\cal{O}}^{(0,0)}_{1})^N$, that is,
the NS vacuum. Then clearly all three point functions
vanish, and thus all 1-point functions (apart from $j$) in the corresponding R
vacuum must vanish.
%This is the result used in section \ref{sp-ex} to fix
%the ambiguity in the vev of ${\cal{O}}_{\S^2_0}$.

Moreover the vanishing of all 1-point functions implies that the
non-linear terms in the vevs of ${\cal{O}}_{\S^k_I}$ in (\ref{h-1})
must contribute. The linear terms in (\ref{h-1}) do give the expected
vanishing vev for ${\cal{O}}_{S^k_I}$  since
the D1-brane and D5-brane densities are constant along the curve.
However, for the circular profile
the linear terms in the ${\cal{O}}_{\S^k_I}$ vevs following from (\ref{h-1}) give
\be \la{circ-vevs}
\left < {\cal{O}}_{\S_0^k} \right > = (-)^{k+1/2} N (\frac{\sqrt{Q_1 Q_5}}{R})^k
\frac{(k-1)^{3/2}}{\pi \sqrt{k(k+1)}} + \cdots
\ee
and therefore the non-linear terms denoted by ellipses
must contribute, to give the expected zero vev.

Next consider the cases where the operator (\ref{oper-1}) determining
the ground state is $({\cal{O}}^{(2,0)}_{1})^N$,
$({\cal{O}}^{(0,2)}_{1})^N$ or $({\cal{O}}^{(2,2)}_{1})^N$. The
supergravity solutions corresponding to these vacua are clearly closely
related to that just discussed: the defining curve is still a circle with
radius $a = 1/ \td{R}$, but the rotation is in the opposite direction
or the circle lies in the 3-4 plane. Therefore the one point functions
should also vanish in these three cases. This is consistent with the
fact that these NS operators are related to the NS vacuum under spectral flow
by an integral parameter (i.e. NS to NS). That is, under a spectral
flow
\be
h' = h - 2 \q j + \frac{c \q^2}{6}; \hsp
j'_3 = j_3 - \frac{c \q}{6}
\ee
with $\q = 1$ the chiral primary with maximal $j^3 = N$ is mapped to
the vacuum.

Now let us move to the more general states of the form (\ref{px1}),
which are conjectured to correspond to circular curves.
Still there are no scalar chiral primary vevs according to
(\ref{circ-vevs}). Kinematics again dictates that only
$j^3 = \bar{j}^3 = 0$ operators acquire a vev, but the fact that
kinematically allowed vevs are zero follows from
dynamical information about three point functions. In particular, one
needs to know the three point functions at the conformal point
for operators ${\cal{O}}_{\Phi}$ which are products
in the CFT, and which therefore do not correspond to single particle
supergravity fluctuations. These are not known, so the results for the
vevs provide a prediction for these correlation functions
at strong coupling, provided the conjectured correspondence is correct.

\subsection{Non-circular curves}

Next we consider the curves corresponding to the most general states
of the form (\ref{oper-1}); it has been conjectured that these
should correspond to connected curves in $R^4$. For example, a state of the form
\be
({\cal{O}}_{n}^{R ++})^{\g N /n} ({\cal{O}}_{n}^{R - -})^{\d N /n}
\hsp \g + \d = 1 \hsp j^R_3 = \bar{j}_R^3 = \half N (\g - \d)/n,
\ee
was conjectured in \cite{Lunin:2001jy} to correspond to an elliptical curve
\be \la{ellipse}
F^1(v) = \mu \frac{a}{n} \cos (\frac{2 \pi n v}{L}); \hsp
F^2(v) = \mu \frac{b}{n} \sin (\frac{2 \pi n v}{L}),
\ee
with $F^3 = F^4 = 0$ and $\mu = \sqrt{Q_1 Q_5}/R$. Provided that
\be
a = \frac{1}{\sqrt{2}} ( \sqrt{1 + (\g - \d)} + \sqrt{1 - (\g - \d)});
\hsp
b = \frac{1}{\sqrt{2}} ( \sqrt{1 + (\g - \d)} - \sqrt{1 - (\g - \d)}),
%A_{\pm} = \frac{1}{\sqrt{2}} \sqrt{1 \pm (\g - \d)},
\ee
the supergravity solution would have the correct angular momenta to
match with the field theory state.

Without any further data to match between supergravity and field
theory one could not check the proposed correspondence further.
The one point functions of chiral primaries computed here, however,
immediately contradict the correspondence between operators
of the form (\ref{oper-1}) and connected curves in $R^4$. The issue is
the following. States of the form (\ref{oper-1}) are eigenstates of
angular momentum operators $j_R^3$ and $\bar{j}_R^3$. This means
that scalar operators can acquire a vev only if $j_R^3 = \bar{j}_R^3 =
0$, following (\ref{zerov}). Note that this is again purely kinematical,
with dynamical information determining precisely which of these
operators actually acquire a vev.

However, the supergravity solution generated by a connected curve
will, according to the formulae, give rise to non-zero
vevs for operators with $(j_R^3, \bar{j}_R^3) \neq 0$ whenever the
curve is not circular. Put differently, a non-circular curve
explicitly breaks the $SO(2) \times SO(2)$ symmetries, with the
symmetry breaking characterized by the vevs for operators
with non-zero $(j_R^3,\bar{j}_R^3)$.

One might wonder whether a non-circular curve could nonetheless give
rise to vevs only for $j^3_R = \bar{j}_R^3 = 0$ operators. That is,
although the curve is non-circular in flat coordinates on $R^4$, it
might be circular in another coordinate system, and the vevs might
be related to multipole moments in that coordinate system. This
however contradicts the explicit formulae for the vevs, exemplified by
the case of an ellipsoidal curve, whose vevs are given in
(\ref{1pt-sp-ell}). More generally, the vevs will clearly involve
the multipole moments of the charge distribution on the $R^4$.

\subsection{Testing the new proposal}

Now consider the proposal made in \cite{Skenderis:2006ah} and here, that the supergravity solution
defined by a given curve is dual to a linear superposition of states
with coefficients following from those in the coherent state in the
dual FP system. In particular, according to (\ref{ellp-01}) and
(\ref{ellp-02}) the ellipse (\ref{ellipse}) would be dual to the
linear superposition 
\be \la{prop-linear} | ellipse ) =
\sum_{k=0}^{N/n} \frac{1}{2^{\frac{N}{n}}} \sqrt{\frac{
(\frac{N}{n})!}{(\frac{N}{n} - k)! k!}} (a + b)^{\frac{N}{n} - k} 
(a-b)^k ({\cal{O}}_{n}^{R ++})^{(\frac{N}{n} - k)} 
({\cal{O}}_{n}^{R --})^{k}; 
\ee 
note that $(a^2 + b^2) = 2$ and that $(a,b)$ are both
real.

The issue is whether this proposal is consistent with the vevs
extracted from the corresponding geometry in section (\ref{ellipsoid}). Again
this question is divided into kinematical and dynamical
parts. The fact that operators with equal and opposite $J^{12}$ charge
acquire equal values in section (\ref{ellipsoid}) follows from the orientation
of the ellipse and is a kinematical constraint which must also be
implicit in the dual description. (That operators with non-zero
$J^{34}$ charge do not acquire a vev is also a kinematical constraint,
of course, but this is automatically satisfied for any proposed dual
involving only operators of zero $J^{34}$ charge.) The actual
non-zero values for the vevs in section (\ref{ellipsoid}) require
dynamical information.

So does the proposed linear superposition satisfy the kinematical
constraints? We can prove that it does as follows. Let us write
(\ref{prop-linear}) as
\be
| ellipse ) = \sum_{k=0}^{N/n} a_k | (\frac{N}{n} - k); k \rangle,
\ee
where $|(\frac{N}{n} - k); k \>$ 
is shorthand for the state created by
$({\cal{O}}_{n}^{R ++})^{(\frac{N}{n} - k)}
({\cal{O}}_{n}^{R --})^{k}$  
and $a_k$ are real
coefficients (that can be read-off from (\ref{prop-linear})). 
Now consider a general
$J^{12}$ charged operator ${\cal O}_{m,m}$. Its vev is given by
\be \la{py1}
( ellipse | {\cal O}_{m,m} | ellipse ) = \sum_{k=0}^{N/n - m}
a_{k}^{\ast} a_{m+k} \langle (\frac{N}{n} - k) ;k | {\cal
  O}_{m,m} 
|(\frac{N}{n} - k - m); k + m  \rangle,
\ee
whilst the corresponding operator with opposite charge  ${\cal
  O}_{-m,-m}$ acquires a vev
\be \la{py2}
( ellipse | {\cal O}_{-m,-m} | ellipse ) = \sum_{k=0}^{N/n - m}
a_{m+k}^{\ast} a_k  
\left ( \langle (\frac{N}{n} - k) ;k | {\cal
  O}_{m,m} |(\frac{N}{n} - k - m ); k + m  \rangle \right )^{\dagger},
\ee
Given that the coefficients $a_m$ are real, the vevs (\ref{py1}) and
(\ref{py2}) will be the same provided that the overlaps are real; the
fusion coefficients for the corresponding extremal three point
functions do indeed have this property.

To test the values of the non-zero vevs in (\ref{1pt-sp-ell}) one
needs dynamical information. One can check that the R charges
are in agreement with those of the superposition (\ref{prop-linear})
as follows. The state $|(\frac{N}{n} - k); k \>$  is an eigenstate of both
$j^3$ and $\bar{j}^3$ with (equal) eigenvalues $(N/2n - k)$. Then
\bea
( ellipse | j^3 | ellipse ) &=& 
\sum_{k=0}^{N/n} \frac{1}{2^{2 \frac{N}{n}}} \frac{
(\frac{N}{n})!}{(\frac{N}{n} - k)! k!} (a + b)^{2 \frac{N}{n} - k}
(a-b)^{2 k} (\frac{N}{2 n}- k) \\ 
&=&- \frac{ (a^2 - b^2)^{\frac{N}{n}} } {2^{\frac{2N}{n}+1}} 
z \frac{\pa}{\pa z} ( z
+ \frac{1}{z})^{\frac{N}{n}} = \frac{N}{2 n} ab; 
\hsp z = \frac{(a-b)} {(a+b)}; \nn 
\eea
with the same result for $\bar{j}^3$. This is in exact agreement with 
the result of (\ref{1pt-sp-ell}).

The remaining non-zero vevs of (\ref{1pt-sp-ell}) are the vevs
of the charged operators 
${\cal O}_{1,1}\equiv\{{\cal O}_{S_{1,1}}, {\cal O}_{\S_{1,1}}\}$,
and the neutral operators, ${\cal O}_{0,0}\equiv 
\{{\cal O}_{S_{0,0}}, {\cal O}_{\S_{0,0}}\}$,
where $(m,n)$ denote the $(SU(2)_L, SU(2)_R)$ charges.
To test whether the proposal is consistent with these vevs
is far more difficult: we would need to know the three
point functions between all operators occurring in (\ref{prop-linear}) 
and the dimension two operators. Given that the former are not dual to
supergravity fields, we do not have any information about the relevant 
three point functions and thus cannot check the vevs. That said, a 
well motivated guess for the structure of the three point
functions leads to vevs which agree remarkably well with those in
(\ref{1pt-sp-ell}). 

Note that in (\ref{1pt-sp-ell}) the vevs of the operators with the 
same charges are the same up to overall numerical coefficients. 
We aim here to derive the universal behavior. For simplicity we set $n=1$.
The corresponding state $| N-k; k \rangle$ in the FP system 
is a multiparticle state, built out of free harmonic oscillators, as
in (\ref{harm-osc}), containing $(N-k)$ quanta of
negative angular momentum and $k$ quanta of positive angular
momentum. We will assume that the same picture holds
in the D1-D5 system, at least in the
large $N$ limit, where the negative (positive) angular momenta
quanta are now positive (negative) R-charge quanta.

We now treat $ {\cal O}_{1,1}$ and ${\cal O}_{0,0}$ in similar way.
$ {\cal O}_{1,1}$ creates a quantum of positive R-charge 
and destroys a quantum of negative R-charge, so
\be
{\cal O}_{1,1} \sim  (a^{-12})^{\dagger} a^{+12},
\ee
and ${\cal O}_{0,0}$ is the product of number operators for
positive and negative R-charge quanta,
\be
{\cal O}_{0,0} \sim \frac{1}{N} \left((a^{+12})^{\dagger} a^{+12}\right) 
\left((a^{-12})^{\dagger} a^{-12}\right),
\ee
where the normalization factor is introduced for later convenience.

Using standard harmonic oscillator relations then yields
\be \la{vev-approx}
\langle  N -k; k | {\cal O}_{1,1} | N - k - 1 ; k + 1  \rangle \sim
\sqrt{(N -k)(k+1)} \mu^2,
\ee
with the scale $\mu^2$ appropriate to a dimension two operator inserted, as in
(\ref{fusion}). Then the total vev for the ellipse is
\bea
( ellipse | {\cal O}_{1,1} | ellipse ) & \sim & \sum_{k=0}^{N-1}
\frac{ \mu^2 }{2^{2N}} \frac{N!}{(N-1 - k)! k!} (a+b)^{2N - 2 k -1} (a-b)^{2
  k + 1}; \nn \\
& = & \frac{N \mu^2 }{2^{2N}} ( 2 (a^2 + b^2))^{N-1} (a^2 - b^2)
= \qu N \mu^2 (a^2 - b^2),
\eea
which indeed agrees in form with the vevs of charged operators in 
(\ref{1pt-sp-ell}). The fact that such a simple approximation for the
three point functions works so well merits further investigation. 

For the neutral operators we obtain
\be \la{why}
\langle N-k;k | {\cal O}_{0,0} | N-k,k \rangle \sim 
\frac{1}{N} \mu^2  (N-k) k,
\ee
and the corresponding total vev for this neutral operator is
\bea
( ellipse | {\cal O}_{0,0} | ellipse ) & \sim & \sum_{k=1}^{N-1} 
\frac{\mu^2}{2^{2N}} \frac{(N-1)!}{(k-1)! (N - (1+k))!} 
(a+b)^{2(N-k)} (a-b)^{2k}; \\
&=& \frac{1}{2^{2 N}} (N-1) \mu^2 (a^2 - b^2)^2 (2 (a^2 + b^2))^{N-2} \sim
\frac{1}{4} N \mu^2 (1 - a^2 b^2), \nn
\eea
in agreement with the vevs for uncharged operators 
given in (\ref{1pt-sp-ell}). Note that (\ref{why}) also gives zero
for $k=0$ and $k=N$, in agreement with the vanishing vevs of the neutral
operators for the circular case. 

Now consider the more general ellipse of (\ref{fell2}). The proposed
dual in this case would be
\bea
| a, b ,c ,d ) &=& \sum_{k=0}^{N/n} \frac{1}{2^{\frac{N}{n}}} \sqrt{\frac{
(\frac{N}{n})!}{(\frac{N}{n} - k)! k!}} (A_+)^{\frac{N}{n} - k}
(A_-)^k 
({\cal{O}}_{n}^{R ++})^{(\frac{N}{n} - k)} ({\cal{O}}_{n}^{R -
  -})^{k}, \nn \\
A_{\pm} &=& (a \pm b) + i (c \mp d), \la{ellp-2-prop}
\eea
with $(a^2 + b^2 + c^2 + d^2) = 2$. Following the same steps as above,
one finds exactly the R charges as in
(\ref{1pt-sp-ell2}), supporting the proposal. As discussed below 
(\ref{1pt-sp-ell2}), charged operators ${\cal O}_{1,1}$ and 
${\cal O}_{-1,-1}$ no longer have equal vevs. Repeating the steps
which led to (\ref{py1}) and (\ref{py2}) one finds that
\be
(\frac{A_+}{A_-})^m \langle {\cal O}_{m,m}  \rangle =
(\frac{A^{\ast}_+}{A^{\ast}_-})^m \langle {\cal O}_{-m,-m} \rangle.
\ee
Taking the case $m=1$ 
this is indeed the relationship between the vevs $ \langle {\cal
  O}_{\S^2_{\pm 1, \pm1}} \rangle $ and $ \langle {\cal O}_{S^2_{\pm
    1, \pm1}} \rangle $ in (\ref{1pt-sp-ell2}), thus demonstrating
that the proposal passes kinematical checks. Now let us compute the 
vevs of the dimension two charged operators using the same
approximation (\ref{vev-approx}) as before; this gives
\be
\langle {\cal O}_{\pm 1, \pm 1} \rangle \sim N \mu^2 ((a \pm i d)^2 +
(c \pm i b)^2), 
\ee
in agreement with (\ref{1pt-sp-ell2}). There is similar agreement
for the behavior of the vevs of neutral operators ${\cal O}_{0,0}$.
Of course, given the agreement
for the ellipse above, there must be agreement for the rotated ellipse
if the proposed dual captures correctly the orientation of the curve in the
1-2 plane. Nonetheless, this example clearly demonstrates how the parameters
of the curve are captured by the (complex) coefficients in the linear
superposition.  

So to summarize: we have tested the proposed field theory dual in the
case of elliptical curves. We find perfect agreement for all
kinematically determined quantities, thus demonstrating the
consistency of the proposal. We also find exact matching for the
R charges and qualitative agreement for the vevs of the scalar 
operators. To test the correspondence further would require
knowledge of three point functions involving multiparticle states at
the conformal point.

\section{Symmetric supergravity solutions} \la{symmetric-s}

We next move to the question of whether one can find geometries which
are dual to a single chiral primary, rather than a superposition of
chiral primaries. As has already been discussed,
a geometry which is dual to a chiral primary must preserve the
$SO(2) \times SO(2)$ symmetry. This immediately implies that the asymptotics
must be of the following form:
\bea
f_5 &=& \frac{Q_5}{r^2} \sum_{k=2l} \frac{f^5_{k0}}{r^k} Y^{0}_{k};
\la{av-1} \\
f_1 &=& \frac{Q_1}{r^2} \sum_{k=2l} \frac{f^1_{k0}}{r^k} Y^{0}_{k}, \nn
\eea
where the scalar spherical harmonics $Y^0_{2l}$ which are
singlets under $SO(2) \times SO(2)$ are defined in (\ref{singlet}).
The forms $(A,B)$ must similarly admit an asymptotic expansion of the
form:
\bea
A_a &=& \sum_{k} \frac{Q_5}{r^{k+1}} (A_{k0+} Y^{0+}_{ka} + A_{k0-}
Y^{0-}_{ka}); \la{av-2} \\
B_a &=&
\sum_{k} \frac{Q_5}{r^{k+1}} (- A_{k0+} Y^{0+}_{ka} + A_{k0-}
Y^{0-}_{ka}), \nn
\eea
where the vector spherical harmonics $Y^{0 \pm}_{ka}$ of degree $k$
($k$ odd) whose Lie derivatives along the $SO(2)^2$ directions are zero
are defined in (\ref{vec-sing}). Note that these forms have only
components along the $(\phi,\psi)$ directions.
We will now give several examples of solutions which have asymptotics
of this form, and discuss their interpretations.

\subsection{Averaged geometries} \la{aver}

Here we discuss a way to construct supergravity solutions based on a general
closed curve $F^i$ which are symmetric under $SO(2) \times SO(2)$
and thus have vanishing vevs for all charged operators.
Let us first discuss the construction for
arbitrary planar curves in the 1-2 plane.
Starting from a general curve $(F^1, F^2, 0,0)$ we construct
a rotated curve,
\be \label{rot}
\tilde{F}^1 = \cos \a F^1 + \sin \a F^2, \qquad
\tilde{F}^2 = -\sin \a F^1 + \cos \a F^2,
\ee
and then superimpose the solutions. This leads to a new harmonic
function,
\be \label{f5}
f_{5} =\int_0^{2 \pi}
\frac{d \a}{2 \pi} \frac{Q_5}{L} \int^{L}_{0}
\frac{dv}{| x - \tilde{F}|^2}
=\frac{Q_5}{L}
\int^{L}_{0}
\frac{dv}{\sqrt{(r^2 + |F|^2)^2 - 4 r^2 |F|^2 \sin^2 \q}}
\ee
where we use coordinates on $R^4$ such that $(x^1)^2 + (x^2)^2 = r^2 \sin^2 \q,
(x^3)^2 + (x^4)^2 = r^2 \cos^2 \q$. The harmonic function for $f_1$ is the
same as $f_5$ in (\ref{f5})
 but with the numerator on the rhs multiplied by $|\dot{F}|^2$. The
non-vanishing part of the gauge field is given by
\be
A_\phi = \frac{Q_5}{L}
\int_{0}^{L}
\frac{\dot{F}^{[1} F^{2]} dv}{|F|^2}
(1-\frac{r^2 +|F|^2}{\sqrt{(r^2 + |F|^2)^2 - 4 r^2 |F|^2 \sin^2 \q}}), \la{a-phi}
\ee
where $\f$ is a polar coordinate in the 1-2 plane and square brackets
indicate antisymmetrization with unit strength. The only non-vanishing component of the
dual form $B$ is
\be
B_{\psi} = \frac{Q_5}{L} \int_{0}^{L} \frac{\dot{F}^{[1} F^{2]} dv}{|F|^2}
(\frac{r^2 - |F|^2}{\sqrt{(r^2 + |F|^2)^2 - 4 r^2 |F|^2 \sin^2 \q}} - 1). \la{b-psi}
\ee
where $\psi$ is a polar coordinate in the 3-4 plane.
For a general curve $(F^1, F^2, F^3, F^4)$ we can proceed
analogously by considering solutions rotated by angle $\a$ in the 1-2 plane
and by angle $\b$  in 3-4 plane and then averaging over
$\a$ and $\b$. For example, the function $f_5$ would be given by
\bea
f_{5} &=& \int_0^{2 \pi}
\frac{d \b}{2 \pi} \frac{Q_5}{L} \int^{L}_{0}
\frac{dv}{\sqrt{(r^2 + |F|^2 - 2 r
\cos \q g(\b))^2
- 4 r^2 ((F^1)^2 + (F^2)^2) \sin^2 \q}}; \nn \\
g(\b) &=& (F^3 \cos (\psi + \b) + F^4 \sin (\psi + \b)).
\eea
This integral can be expressed in terms of elliptic integrals,
although we have not obtained the exact form. 
The asymptotics are however given by:
\bea
f_5 &=& \frac{Q_5}{L r^2} \int^{L}_{0}
\sum_{l \ge 0} \frac{dv}{r^{2l}} P_{l}(\cos (2\q)) P_l (Z(F (v))); \nn
\\
f_1 &=& \frac{Q_5}{L r^2} \int^{L}_{0}
\sum_{l \ge 0} \frac{dv}{r^{2l}} \left | \pa_v F \right |^2
P_{l}(\cos (2\q)) P_l (Z(F (v))); \\
A &=&  \frac{Q_5}{L} \int^L_0 \sum_{k} \frac{dv}{\sqrt{2} (k+1) r^{k+1}}
\left ( p_k(F) (\dot{F}^1 F^2 - \dot{F}^2 F^1) (Y^{0+}_{ka} -
Y^{0-}_{ka}) \right . \nn \\
&& \hspace{40mm} \left .
+ q_k(F) (\dot{F^3} F^3 - \dot{F}^4 F^3) (Y^{0+}_{ka} + Y^{0-}_{ka})
\right ); \nn \\
Z(F) &=& (F^3)^2 + (F^4)^2 - (F^1)^2 - (F^2)^2, \nn
\eea
where $P_l(x)$ are Legendre polynomials of degree $l$ and $ p_k(F)$
and $q_k(F)$ are defined in (\ref{charg-har})-(\ref{lift}).
These asymptotics are manifestly of the form given in (\ref{av-1}) and
(\ref{av-2}). Setting $F^3 = F^4 = 0$ gives the asymptotic expansion
of the expressions given in (\ref{f5}) and (\ref{a-phi}).

\subsubsection{Example 1: the averaged ellipse}

Consider the case of an ellipse, so that the defining curve is
\be \la{ellp3}
F^1 = \mu a \cos \frac{2 \pi v}{L}, \hsp
F^2 = \mu b \sin \frac{2 \pi v}{L},
\ee
with $\mu = \sqrt{Q_1 Q_5}/R$ and $(a^2 + b^2) = 2$.
(For simplicity we choose the frequency $n$ to be one.) In this
example the integral over the curve in (\ref{f5}) can be carried out
explicitly to give
\bea
f_5 &=& \frac{2 Q_5}{\pi z} K(w); \\
z^4 &=& (C^2 + D^2); \hsp
w = \frac{\sqrt{(z^2 - C)}}{\sqrt{2} z}; \nn \\
C &=& (r^4 + 2 \mu^2 r^2 \cos 2 \q + \mu^4 a^2 b^2); \nn \\
D &=& \mu^2 r^2 \sin 2 \q (a^2 - b^2), \nn
\eea
where $K(w)$ is the complete elliptic integral of the first kind.
Then $f_5$ has poles only where $z$ has zeroes, namely at $\q = \pi/2$
and $r = \mu a$ or $r= \mu b$. This suggests that any singularities of the
metric are confined to these locations, namely circles of radius $a$
and $b$ in the 1-2 plane, and indeed one finds that the other defining
functions $(f_1,A,B)$ only have poles at these locations. Thus the
geometry is less singular than one might have anticipated. The
integrands have singularities in the annular region defined by  
$\q = \pi/2$ and $\mu a \le r \le \mu b$  (assuming $a \le b$) but the
integrated functions only have singularities on 
the circles bounding this annulus. Moreover these singularities seem
to be such that the only singularities of the resulting metric
are conical.

\subsubsection{Example 2: Aichelburg-Sexl metric}

The Aichelburg-Sexl metric was also obtained by the procedure of
averaging over curve orientations in \cite{Lunin:2002bj}. The defining
curve has a section which is constant:
\bea
F^1 &=& a \cos (\frac{2 \pi v}{\xi L}); \hsp
F^2 = a \sin (\frac{2 \pi v}{\xi L}), \hsp 0 \le v \le \xi L; \\
F^1 &=& a, \hsp \xi L \le v \le L, \nn
\eea
with all other $F^i(v) = 0$ and $\xi < 1$. Such a profile gives rise to the
following harmonic functions:
\bea
f_5 &=& \left (1 + \frac{Q_5 \xi}{\hat{r}^2 + a^2 \cos^2 \hat{\q}}
+ \frac{Q_5 ( 1- \xi)}{(x_1-a)^2 + x_2^2 + x_3^2 + x_4^2}
\right ); \\
f_1 &=& \left (1 + \frac{Q_1}{\hat{r}^2 + a^2 \cos^2 \hat{\q} } \right
); \nn \\
A_{\phi} &=& a \sqrt{\xi} \frac{ \sqrt{Q_1 Q_5}}{\hat{r}^2 + a^2 \cos^2
  \hat{\q}}, \nn
\eea
where $Q_1 = Q_5 a^2 (2 \pi/L)^2/\xi$ and as in (\ref{non-stand})  we introduce
non-standard polar coordinates on $R^4$ to simplify the harmonic
functions. Now we take the $SO(2)$ orbit of the defining curve, thus
averaging over the location of the constant section in the 1-2
plane. This leads to the $SO(2)$ symmetric harmonic functions
\bea
f_5 &=& \left (1 + \frac{Q_5}{\hat{r}^2 + \xi \mu^2 \cos^2 \hat{\q}} \right ); \\
f_1 &=& \left (1 + \frac{Q_1}{\hat{r}^2 + \xi \mu^2 \cos^2 \hat{\q}
} \right
); \nn \\
A_{\phi} &=& \frac{ \xi \mu \sqrt{Q_1 Q_5}}{\hat{r}^2 + \xi \mu^2
\cos^2 \hat{\q}}, \nn
\eea
which are those of the Aichelburg-Sexl metric
\bea ds^2 &=& f_{1}^{- 1/2} f_{5}^{- 1/2} \left ( - (dt -
\frac{\xi \mu
  \sqrt{Q_1 Q_5}}{\hat{r}^2 + \xi \mu^2 \cos^2 \hat{\theta}} \sin^2 \hat{\q}
d \phi)^2 + (dy -
\frac{\xi \mu  \sqrt{Q_1 Q_5}}{\hat{r}^2 + \xi \mu^2
\cos^2 \hat{\theta}} \cos^2 \hat{\q} d \psi)^2
\right ) \nn \\
&& + f_{1}^{1/2} f_{5}^{1/2} \left ( (\hat{r}^2 + \xi \mu^2 \cos^2 \hat{\q}) (
\frac{d\hat{r}^2}{\hat{r}^2 + \xi \mu^2} + d \hat{\q}^2) + \hat{r}^2 \cos^2
\hat{\q} d \psi^2 + (\hat{r}^2 + \xi \mu^2) \sin^2 \hat{\q} d \phi^2 \right ).
\nn
\eea
Here $\mu = \sqrt{Q_1 Q_5}/R$.
This solution is clearly very similar to those based on circular
curves, discussed in section \ref{sp-ex}.
The non-zero vevs extracted from the decoupled part of the geometry follow
from (\ref{vv2}) and are given by
\bea
\left < J^{+3} \right > &=&  \frac{N}{4 \pi} \mu \xi (d {y} -
d {t}); \hsp
\left < J^{-3} \right > =  \frac{N}{4 \pi} \mu \xi (d {y} +
d {t}); \\
\left < {\cal O}_{\S^2_0} \right > &=& \frac{N \sqrt{2} \xi \mu^2 }{2 \pi
    \sqrt{3}} (1 - \xi). \nn
\eea
These clearly reduce to those for the case of the circular curves when
$\xi = 1$. Note that the Aichelburg-Sexl metrics do not have conical
singularities, and are therefore actually less singular than the
unaveraged geometries. However, whilst the
Aichelburg-Sexl metrics do have the correct asymptotics to correspond to chiral
primaries, they are based on averaging curves with straight
sections. The interpretation of these straight sections from the dual
perspective is rather unclear, given the proposed correspondence
between frequencies on the curve and twists of the dual operators.

\subsection{Disconnected curves} \la{discon}

Another way to obtain solutions which preserve the $SO(2) \times SO(2)$
symmetry is to consider curves made up of disconnected circles.
There exist supergravity
solutions defined by the following functions
\bea \la{disc}
f_5 &=& \sum_{l=1}^{I} \frac{Q_5 N_l}{N L} \int_{0}^{L}
\frac{dv_l}{\left | x -  F_l \right |^2}; \\
f_1 &=& \sum_{l=1}^{I} \frac{Q_5 N_l}{N L} \int_{0}^{L_l} \frac{dv_l
  (\pa_{v_l} F_l)^2}
{\left | x - F_l \right |^2}; \nn \\
A_i &=& \sum_{l=1}^I \frac{Q_5 N_l}{N L} \int_{0}^{L_l} \frac{dv_l
  \pa_{v_l} F_l^i}
{\left | x - F_l \right |^2}, \nn
\eea
where the $l$th curve is parametrized by $v_l$ with $\sum_{l} N_l = N$
and is circular within either the 1-2 or 3-4 plane. That is,
the curve defining the $l$th circle is given by
\be \la{curve1}
F^{1}_l = \frac{\sqrt{Q_1 Q_5}}{R n_l} \cos \left ( \frac{ 2 \pi n_l
  v_l}{L} \right ); \hsp
F^2_l = \pm \frac{\sqrt{Q_1 Q_5}}{R n_l} \sin \left ( \frac{2 \pi n_l
  v_l}{L} \right ),
\ee
assuming the circle lies in the 1-2 plane; the sign determines the
direction of rotation. A curve lying in the 3-4 plane will take an
analogous form. Such a linear superposition of sources solves the
field equations and is supersymmetric.
By construction the total D5-brane and D1-brane charges are $Q_5$ and
$Q_1$ respectively, with the $l$th curve sourcing a fraction $N_l/N$
of (both) the total charges. The related radii and frequencies in
(\ref{curve1}) ensure that the D1-brane charge of each curve is a
fraction $N_l/N$ of the total. This prescription also reduces to that
given for the curves corresponding to the operators (\ref{px1});
in that case one lets $I = N/n$ and $N_l = n$ in the supergravity
solution above and takes the circles to be coincident.
Furthermore the total R-charges will be given by
\be
j_3 = \half \sum_{l=1}^{I} \ep_l m_l; \hsp
\bar{j}_3 = \half \sum_{l=1}^I \bar{\ep}_l m_l,
\ee
where $m_l = N_l/n_l$. Here $(\ep_l,\bar{\ep}_l)$ = $(\pm 1, \pm1)$
depending on the orientation and rotation of the curve.

Since the sources are located on disconnected circles, the singularity
structure of these geometries is similar to that discussed in section
\ref{sp-ex}. Namely, there are conical singularities whenever $n_l
\neq 1$. Thus, these solutions are no more singular than the
geometries based on a single circle, although they are more singular
than a geometry based on a general non-intersecting curve.

\subsection{Discussion}

These are not the only symmetric geometries. For example, one could
consider more general superpositions of curves, superposing not just
different orientation curves but also different shape curves. However, 
the procedure we outlined above does illustrate how
symmetric geometries can be obtained from those defined in terms of
a single curve. The symmetrization we used is the simplest, in that the
measure for each curve is the same. The field theory dual suggests 
that symmetrizing over shapes of curves should involve a non-trivial
measure. That is, if one has an ellipse with parameters $(a,b)$ so
that the proposed dual is
\be
| ellipse )_{a,b} = \sum_{k=0}^{N/n} (a_{k})_{a,b} | \frac{N}{n} - k;
  k \rangle,
\ee
then one can formally invert the relation to give
\be
| \frac{N}{n} - k;
  k \rangle = \sum_{a,b} (a_{k})^{-1}_{a,b} | ellipse )_{a,b}.
\ee
This suggests that to obtain a geometric dual for a given chiral primary
one could consider a linear superposition of
curves with different parameters $(a,b)$ using a measure which is
related to $(a_{k})^{-1}_{a,b}$. Precisely what the measure should be
is not however immediately apparent, because, as we will discuss
below, such a symmetrization via linear superposition 
may in fact be rather too naive, because
of the non-linear relationship between harmonic functions and vevs.
To test whether a given symmetric geometry does indeed
have the correct properties to correspond to a given chiral primary,
one will need to use the actual values of the 
kinematically allowed vevs, as we will now discuss.

\section{Dynamical tests for symmetric geometries} \la{dyn-2}

The geometries in sections \ref{aver} and \ref{discon} have the
correct asymptotics to correspond to chiral primaries. Since the
geometries in section \ref{discon} are based on separated sources, one would
not however anticipate that these correspond to Higgs branch vacua;
the more natural proposal would be that they relate to Coulomb branch
vacua. By extracting all vevs and $n$-point functions from each geometry
one could in principle identify the field theory dual uniquely.

Furthermore, given any proposed correspondence between geometries and
field theory vacua, we can use dynamical information for the
kinematically allowed vevs to test it. In particular, let us consider
the averaged geometries, focusing on the example of the averaged ellipse. 
In this case, we consider a defining curve (\ref{ellp3}) with corresponding 
rotated curve $\td{F}^1, \td{F}^2$ defined in (\ref{rot}). The geometry based on
the latter is proposed to correspond to the linear superposition 
(\ref{ellp-2-prop}) with 
\be
A_{+} = (a + b) e^{i \a}; \hsp  A_{-} = (a-b) e^{- i \a}.
\ee
This means that the superposition dual to the rotated ellipse is
\be
| ellipse )_{\a} = \sum_{k=0}^{N/n} e^{i \a ( \frac{N}{n} - 2 k)} 
\frac{1}{2^{\frac{N}{n}}} \sqrt{\frac{
(\frac{N}{n})!}{(\frac{N}{n} - k)! k!}} (a + b)^{\frac{N}{n} - k} (a
-b)^k ({\cal{O}}_{n}^{R ++})^{(\frac{N}{n} - k)} ({\cal{O}}_{n}^{R -
-})^{k}. 
\ee
Averaging over the angle $\a$ clearly picks out the $k = N/2n$ term in
the superposition, which is a state of zero angular momentum. However,
the geometry obtained by averaging over rotated ellipses does not have
zero angular momentum, but rather the same angular momentum as the
original geometry. This suggests that this geometric averaging
might actually average over vevs, rather than over states, and thus not
pick out a geometry dual to a single chiral primary. Given that the
averaging linearly superposes harmonic functions, however, and the
vevs are non-linearly related to the harmonic functions, the geometric
averaging probably does not lead to just an overall averaging over the
vevs. One will have to use the actual vevs for the neutral operators
to see what the geometry describes.      

So now let us discuss how one would use information about three point
functions at the conformal point to test whether a given 
geometry corresponds to a
chiral primary. Let us work with an example: 
consider the R vacuum corresponding to the
operator $({\cal O}_{S^p_n})_R$  obtained by spectrally
flowing the operator ${\cal O}_{S^p_n}$
dual to the supergravity field $S^{p(6)}_n$. (Recall that the
superscript $p$ denotes that it is primary, $j^3 = j$ and $\bar{j}^3 =
\bar{j}$.) Next suppose that there is a candidate dual geometry, which
has the correct symmetries and R-charges, the latter being $(\half
(n-N), \half (n-N))$. This means that the holographic vevs for the R
symmetry currents must be
\be \la{ex-vev}
\langle J^{\pm 3} \rangle = \frac{\mu}{4 \pi} (n-N) (dy \mp dt),
\ee
where $y$ has periodicity $2 \pi \td{R} = 2 \pi R/\sqrt{Q_1 Q_5}$
and $\mu = \sqrt{Q_1 Q_5}/{R}$.

Now let us consider how we can relate this vev to the normalized three point
function at the conformal point. That is,
\be
\langle J^{\pm 3} \rangle_{\Psi_{S_n}} =
\< (\cao_{S^{n}})_R^{\dagger}  J^{\pm 3} (w_0) (\cao_{S^n})_R \>  \equiv
\frac{\< (\cao_{S^{n}})_R^{\dagger} (\infty) J^{\pm3} (w_0) (\cao_{S^k})_R (0) \>
}{ \< (\cao_{S^{n}})_R^{\dagger} (\infty) (\cao_{S^n})_R (0) \>},
\ee
where $\Psi_{S_n}$ denotes that the theory is in the vacuum created
by $({\cal O}_{S^p_n})_R$. The scale $w_0$ at which the current is
inserted is found by comparing the vevs (\ref{ex-vev}) with the
normalized three point functions, computed in (\ref{jj5}). The latter
give
\be
\langle J^{+ 3} \rangle_{\Psi_{S_n}} = \frac{(n-N)}{4 \pi w_0}; \hsp
\langle J^{- 3} \rangle_{\Psi_{S_n}} = \frac{(n-N)}{4 \pi \bar{w}_0},
\ee
which comparing with (\ref{ex-vev}) implies that the inserted scale
must be $w_0 = \bar{w}_0 = \mu^{-1}$.

We can now use the three point functions between ${\cal O}_{S^p_n}$
and neutral dimension two operators to predict the vevs for the
latter. This gives
\bea
\langle {\cao}_{S^2_0} \rangle_{\Psi_{S_n}} &=& 0; \la{cancel} \\
\langle  {\cao}_{\S^2_0} \rangle_{\Psi_{S_n}} &=&
\< (\cao_{S^{n}})_R^{\dagger} {\cao}_{\S^2_0}
(\mu^{-1}) (\cao_{S^n})_R  \> = \frac{\sqrt{3} n^3 \mu^2}{\sqrt{2} \pi (n-1)^2}. \nn
\eea
where the normalized three point function is 
defined in (\ref{norm-tpf}) and the
inserted scale is as before $w_0 = \bar{w}_0 = \mu^{-1}$. Note that 
$\mu^2 \sim N$, so the vev has the correct large $N$ behavior (for our
choice of normalization). From the
expressions given in (\ref{vv2}) for the vevs of these operators in
terms of the asymptotics we can determine the degree two coefficients
in (\ref{av-1}). The vanishing of $\langle {\cao}_{S^2_0}
\rangle_{\Psi_{S_n}}$ implies that $f^{1}_{20} = f^{5}_{20}$ whilst the
expression for the vev $\langle {\cao}_{\S^2_0} \rangle_{\Psi_{S_n}}$
in (\ref{vv2}) implies that
\bea \label{hol_f}
f^{1}_{20} &=& - \frac{\mu^2}{\sqrt{3} N^2} \left ( (n-N)^2 + 
\frac{3 n^3 N}{(n-1)^2} \right ) \\
&=& f^{1}_{20}( circ) \left (1 + \frac{n}{N} 
+ \cdots \right ), 
\nn
\eea
where $f^{1}_{20}(circ) = - \mu^2/\sqrt{3}$ is the value of $f^{1}_{20}$
for the circular solution.  The $(n-N)^2$ contribution on the rhs 
is due to the non-linear contribution 
$8 a^{\a -} a^{\b +} f_{I\a\b}$ and in the second equality
we use $1 \ll n \ll N$. The upper limit on $n$
follows from the fact that the supergravity
three point functions are known only to leading order in $N$ and do
not apply for operators with dimensions comparable to $N$. The lower
limit is unnecessary and is imposed only to simplify the formula.

By extending the computation of the vevs to higher dimension operators
and comparing with those predicted from three point functions at the
conformal point, one could in principle extract the higher degree
coefficients in (\ref{av-1}) and resum the asymptotic
series to obtain the full geometry.

There is an important caveat, however. In all computations so far we
have worked in the $N \rightarrow \infty$ limit, retaining only the
leading terms. This applies both to the computation of the vevs and to
the computation of three point functions. For the computation 
of the 3-point function to be valid we need $N \gg n$, but then the
``holographically engineered'' $f_{20}^1$ in (\ref{hol_f}) 
differs from the answer for the circle only by terms subleading 
in $n/N$. In other words, the holographically engineered geometry 
would be that of the circular solution up to $1/N$ corrections.

Next consider R vacua corresponding to operators obtaining by
spectral flow on operators which are either of high dimension
(comparable to $N$) or multiparticle. The latter include 
operators of the form 
$({\cal O}_n^{R++})^{N/n -k} ({\cal O}_{n}^{R --})^k$ for which the
duals may be related to averaged ellipses. Since there is 
no information about three point functions of these operators at
strong coupling, we have no precise predictions for the vevs of neutral
operators and thus cannot currently test whether a given geometry 
is indeed dual to such a state. Given any future progress on computing
the relevant fusion coefficients via string theory, one could however
test the correspondence further. 

To summarize: a geometry with $SO(2) \times SO(2)$ symmetry can be
characterized by its angular momentum and vevs of neutral operators. 
The latter can in principle be used to determine the corresponding
dual, but to implement this program will in general require going
beyond the leading supergravity approximation.

\section{Including the asymptotically flat region} \la{flat-reg}

In this section we will discuss how the asymptotically flat region of
the geometry may be interpreted using the AdS/CFT dictionary. Our
discussion will parallel an analogous discussion for D3-branes given
in section 6 of \cite{Skenderis:2006di}.

The six-dimensional metric of (\ref{six-met}) along with the scalar
and tensor field of (\ref{ez1}) are characterized by two harmonic
functions $(f_1,f_5)$ and a harmonic form $A_i$. The field equations
are satisfied for any choice of harmonic functions. The specific
choices in (\ref{mathu1}) correspond to (part of) the (supersymmetric) Higgs
branch of the D1-D5 system. Multi-centered harmonic functions for
$(f_1,f_5)$ with $A_i =0$ are also well-known supergravity solutions,
corresponding to part of the Coulomb branch.

In (\ref{p0}) we gave the most general form for the asymptotic
expansions of $(f_1,f_5,A_i)$ under the condition that the solution is
asymptotically $AdS_3 \times S^3$. The asymptotically flat
region may be included by adding constant terms to the $(f_1,f_5)$
harmonic expansions, namely
\be
f_1 = \ep_1 + \frac{Q_1}{r^2} \sum_{k,I} \frac{f^1_{kI}
  Y^I_k(\q_3)}{r^k}; \hsp
f_5 = \ep_5 +  \frac{Q_5}{r^2} \sum_{k,I} \frac{f^5_{kI}
  Y^I_k(\q_3)}{r^k},
\ee
whilst keeping the large radius expansion for $A_i$ as in (\ref{p0}).
To include all of the asymptotically flat region, the parameters
$\ep_1$ and $\ep_5$ clearly need to be finite. However, let us take
the parameters to be infinitesimal so that the solution remains
asymptotically $AdS_3 \times S^3$. Since we have discussed already the
terms in the harmonic expansion behaving as $r^{- k}$ with $k \ge  3$,
we consider only the new terms as a perturbation to the $AdS$
background. That is, we let
\be
f_1 = \ep_1 + \frac{Q_1}{r^2}; \hsp
f_5 = \ep_5 + \frac{Q_5}{r^2},
\ee
with $A_i = 0$ and then identify the terms induced in the harmonic expansion of the
fluctuations (\ref{flc1}). The field fluctuations are
\bea
- h_{tt} &=& h_{yy} = - \half r^4 (\hat{\ep}^1 + \hat{\ep}^5); \hsp
h_{rr} = \half (\hat{\ep}^1 + \hat{\ep}^5); \\
h_{ab} &=& \half r^2 (\hat{\ep}^1 + \hat{\ep}^5); \hsp
\phi^{(56)} = \half r^2 (\hat{\ep}^1 - \hat{\ep}^5); \nn \\
g^{5}_{tyr} &=& - r^3 (\hat{\ep}^1 + \hat{\ep}^5); \hsp
g^{6}_{tyr} = - r^3 (\hat{\ep}^1 - \hat{\ep}^5), \nn
\eea
where we define $\hat{\ep}^1 = \ep^1/Q_1$ and $\hat{\ep}^5 =
\ep^5/Q_5$. Thus the only non-vanishing dynamical fields are those from
(\ref{diageqm})
\be
\t_0 \equiv \frac {\pi^0}{12} = \frac{1}{8}
r^2 (\hat{\ep}^1 + \hat{\ep}^5); \hsp
t_0 \equiv \qu \phi_{0}^{(56)}  = \frac{1}{8}
r^2 (\hat{\ep}^1 - \hat{\ep}^5).
\ee
(The other non-vanishing components are induced by constraint
equations and do not correspond to dynamical fields.) Since both
$\t_0$ and $t_0$ couple respectively to the dimension
four operators ${\cal{O}}_{\t_0}$ and
${\cal O}_{t_0}$, the radial dependence of these fields corresponds
to source behavior. Thus the CFT lagrangian is deformed by the terms
\be \la{deform}
\int d^2z \left ( (\hat{\ep}^1 + \hat{\ep}^5) {\cal{O}}_{\t_0} +
(\hat{\ep}^1 - \hat{\ep}^5) {\cal O}_{t_0} \right ).
\ee
Note that the operators $({\cal{O}}_{\t_0},{\cal O}_{t_0})$ are the top components of
the short multiplets generated from the chiral primaries
$({\cal{O}}_{\S_2},{\cal O}_{S_2})$ respectively through the action of
the supercharges. That is, they are given by
\be
G^{1 \dagger}_{-1/2} G^{2}_{-1/2} \td{G}^{1 \dagger}_{-1/2} \td{G}^{2}_{-1/2} \left |
CPO \right >,
\ee
where $(G^{a}_{\pm 1/2}, \td{G}^{a}_{\pm 1/2})$ with $a =1,2$ are left and right
supercharges. Here $(G^{1 \dagger}_{-1/2}, G^{2}_{-1/2})$ and corresponding
right moving charges act as raising operators on the $\D =2 $ chiral
primaries. The latter have $h = j = j^3 = \bar{h} = \bar{j} =
\bar{j}^3  =1 $. Computing two point functions in the presence of the
deformation (\ref{deform}) may capture scattering into the
asymptotically flat part of the D1-D5 geometry.

\section*{Acknowledgments}

The authors are supported by NWO, KS via the Vernieuwingsimplus grant
``Quantum gravity and particle physics'' and
IK, MMT via the Vidi grant ``Holography,
duality and time dependence in string theory''. This work was also
supported in part by the EU contract MRTN-CT-2004-512194. KS and MMT would like
to thank both the 2006 Simons Workshop and the theoretical physics
group at the University of Crete, where some of this work was completed.

\appendix

\section{Properties of spherical harmonics} \la{apa}

Scalar, vector and tensor spherical harmonics satisfy the following
equations
\bea
\Box Y^{I} &=& - \Lambda_{k} Y^{I}, \\
\Box Y_{a}^{I_v} &=& (1 - \Lambda_{k}) Y_{a}^{I_v}, \hsp D^{a}
Y_{a}^{I_v} = 0, \nn \\
\Box Y_{(ab)}^{I_t} &=& (2 - \Lambda_{k}) Y_{(ab)}^{I_t}, \hsp D^{a}
Y_{k(ab)}^{I_t} = 0, \nn
\eea
where $\Lambda_k = k (k+2)$ and the tensor harmonic is traceless. It
will often be useful to explicitly indicate the degree $k$ of the harmonic;
we will do this by an additional subscript $k$, e.g. degree $k$ spherical
harmonics will also be denoted by $Y_k^I$, etc.
$\Box$ denotes the d'Alambertian along the three sphere. The
vector spherical harmonics are the direct sum of two irreducible
representations of $SU(2)_L \times SU(2)_R$ which are characterized by
\be
\ep_{abc} D^b Y^{c I_v \pm} = \pm (k+1) Y_{a}^{I_v \pm}
\equiv \l_k Y_{a}^{I_v \pm}. \la{vec-dual}
\ee
The degeneracy of the degree $k$ representation is
\be
d_{k,\ep} = (k+1)^2 - \ep,
\ee
where $\ep = 0,1,2$ respectively for scalar, vector and tensor harmonics.
For degree one vector harmonics $I_v$ is an adjoint
index of $SU(2)$ and will be denoted by $\a$.

We use normalized spherical harmonics such that
\be
\int Y^{I_1} Y^{J_1} = \Omega_3 \d^{I_1 J_1}; \hsp
\int Y^{a I_v} Y_a^{J_v} = \Omega_3 \d^{I_v J_v}; \hsp
\int Y^{(ab) I_t} Y_{(ab)}^{J_t} = \Omega_3 \d^{I_t J_t},
\ee
where $\Omega_3 = 2 \pi^2$ is the volume of a unit 3-sphere.
Then
\be
\int D_a Y^{I_1} D^a Y^{J_1} = \Omega_3 \L^{I_1} \d^{I_1 J_1}; \hsp
\int D^{(a} D^{b)} Y^{I_1} D_a D_b Y^{I_2} = \Omega_3 \frac{2}{3} \L^{I_1}
(\L^{I_1} -3) \d^{I_1 J_1}.
\ee
The following identities are useful
\bea
\frac{1}{\Omega_3}
\int Y^{I} D^{a}Y^J D_a Y^{K} &\equiv & b_{IJK} = \half (\L^{J} + \L^{K} - \L^{I})
a_{IJK}; \la{ap-ov-1} \\
\frac{1}{\Omega_3} \int D^{(a} D^{b)} Y^{I} D_{a} D_{b} Y^{J} Y^{K}
&\equiv& c_{IJK} = ( \qu
\Lambda_{IJK}(\Lambda_{IJK}- 4) - \frac{1}{3} \Lambda^{I} \Lambda^{J}) a_{IJK}; \nn \\
\frac{1}{\Omega_3}\int D_{(a} Y^{I} D_{b)} Y^{J} D^{a} D^{b} Y^{K} &\equiv& d_{IJK} = ( \qu
\L_{IKJ} \L_{JKI} + \frac{1}{6} \Lambda^{K} \L_{IJK}) a_{IJK}, \nn
\eea
where $\L_{IJK} = (\L^{I} + \L^{J} - \L^{K})$.
We define the following triple integrals as
\bea
\int Y^{I} Y^{J} Y^K &=& {\Omega_3} a_{IJK}; \la{ap-ov0} \\
\int (Y^{\a \pm}_{1})^a Y^{j}_{1} D_a Y^{i}_1 &=& \Omega_3 e^{\pm}_{\a ij};
\la{ap-ov3} \\
\int Y^{I} (Y_{1}^{\a -})_a (Y_{1}^{\b+})^a &=& {\Omega_3} f_{I \a \b};
\la{op4} \\
\int (Y^{I_v \pm }_{k_v})^a Y^{I}_{k} D_a
Y^{i}_1 &=& \Omega_3 E^{\pm}_{I_v
  I i}; \la{gt1} \\
\int (Y^{I_v \pm }_{k_v})^a Y^{I}_{k} D_a
Y^{J}_{l} &=& \Omega_3 E^{\pm}_{I_v
  I J}; \la{gt2}
\eea
We also use specific identities for harmonics of low degree. The
degree one vector harmonics $Y^{\a}_{1 \pm}$ transform in the $(1,0)$
and $(0,1)$ representation of $(SU(2)_L, SU(2)_R)$ whilst the degree
$k$ scalar harmonics transform in the $(\half k, \half k)$
representation. This immediately implies that the following triple
overlaps are zero:
\be
\int Y_{2}^I (Y^{\a +}_{1})_{a} (Y^{\b +}_{1})^{a}) = \int Y_{2}^I
(Y^{\a -}_{1})_a (Y^{\b -}_{1})^a) = \int Y_{0} (Y^{\a +}_{1})_a
(Y^{\b -}_{1})^a) = 0. \la{op5}
\ee
Using the following explicit coordinate system on the sphere
\be
ds^2_3 = d\q^2 + \sin^2 \q d \phi^2 + \cos^2 \q d \psi^2,
\ee
with volume form $\eta_3 = \sin \q \cos \q d\q \wedge d\phi \wedge d \psi$
the following are normalized Killing forms
\be \la{kform}
Y^{3+}_{1} = (\sin^2 \q d \phi + \cos^2 \q d \psi); \hsp
Y^{3-}_{1} = - (\sin^2 \q d \phi - \cos^2 \q d \psi),
\ee
which generate the Cartan of the $SO(4)$ symmetry group. The remaining
Killing forms are
\bea \la{kform2}
Y^{1+}_{1} &=& (\cos(\psi + \phi) d \q + \sin (\psi + \phi) \sin \q
\cos \q d(\psi - \phi)); \nn \\
Y^{2+}_{1} &=& (- \sin(\psi + \phi) d \q + \cos (\psi + \phi) \sin \q
\cos \q d(\psi - \phi)); \nn \\
Y^{1-}_{1} &=& ( \cos (\psi - \phi) d \q + \sin (\psi - \phi) \sin \q
\cos \q d(\phi + \psi)); \nn \\
Y^{2-}_{1} &=& (- \sin(\psi - \phi) d \q + \cos (\psi - \phi) \sin \q
\cos \q d(\phi + \psi)). \nn
\eea
The $SU(2) \times SU(2)$ algebra realized by the Killing vectors
is normalized such that
\be
[Y^{\a +}_{1}, Y^{\b +}_{1}] = 2 \ep_{\a\b\g} Y^{\g +}_{1}; \hsp
[Y^{\a -}_{1}, Y^{\b -}_{1}] = 2 \ep_{\a\b\g} Y^{\g-}_{1}; \hsp
[Y^{\a +}_{1}, Y^{\b -}_{1}] = 0.
\ee
Furthermore
\be
Y^{\a \pm}_{1} \wedge Y^{\b \pm}_{1} \wedge Y^{\g \pm}_{1} = \mp \ep_{\a
  \b \g} \eta_3,
\ee
which implies that
\be
\int \ep^{abc} Y^{\a \pm}_{a} Y^{\b \pm}_{b} Y^{\g \pm}_{c}
%= {\Omega_3} j^{\pm}_{\a \b \g}
= \mp  {\Omega_3} \ep_{\a  \b \g} \la{vector}
\ee
In the same coordinate system $Y_2^0 = \sqrt{3} \cos 2 \q$ is the
normalized degree 2 spherical harmonic which is a singlet under the $SO(2)^2$
Cartan, with the following triple overlap
\be \la{o2}
\int Y_2^0 (Y^{3 +}_{1})^a (Y^{3-}_{1})_a = \frac{1}{\sqrt{3}} \Omega_3.
\ee
Thus, $f_{0 3 3}=1/\sqrt{3}$ in this specific case.
More generally the normalized spherical harmonics which are
singlets under the Cartan can be expressed as
\be \la{singlet}
Y_{2l}^{0} = \sqrt{2l + 1} P_{l} (\cos 2 \q),
\ee
where $P_{l}(x)$ is a Legendre polynomial of degree $l$, normalized so
that $P_{l}(1) = 1$ and $P_{l}(-1) = (-1)^l$.

In this coordinate system normalized degree one spherical
harmonics are
\bea
Y^1_1 &=& 2 \sin \q \cos \phi; \hsp
Y^2_1 = 2 \sin \q \sin \phi; \\
Y^3_1 &=& 2 \cos \q \cos \psi; \hsp
Y^4_1 = 2 \cos \q \sin \psi. \nn
\eea
Defining $Y^{ij} \equiv \half (Y^j_1 d Y^1_1 - Y^1_1 d Y^j_1)$,
\bea
Y^{12} &=& (Y^{3-}_{1} - Y^{3+}_{1}); \hsp
Y^{34} = - (Y^{3+}_{1} + Y^{3-}_{1}); \hsp
Y^{13} =  (Y^{1+}_{1} + Y^{1-}_{1}); \\
Y^{34} &=& (Y^{1-}_{1} - Y^{1+}_{1}); \hsp
Y^{14} = - (Y^{2+}_{1} + Y^{2-}_{1}); \hsp
Y^{23} = ( Y^{2+}_{1} - Y^{2-}_{1}), \nn
\eea
and therefore the explicit values for the overlaps $e^{\pm \a}_{i j}$
defined in (\ref{ap-ov3}) are
\bea \label{ecof}
e^{+3}_{12} &=& -1; \hsp e^{-3}_{12} = 1; \hsp e^{+3}_{34} = -1; \hsp
e^{-3}_{34} = -1; \hsp e^{+1}_{13} = 1; \hsp e^{-1}_{13} = 1; \\
e^{+1}_{24} &=& -1; \hsp e^{-1}_{24} = 1; \hsp e^{+2}_{14} = -1; \hsp
e^{-2}_{14} = -1; \hsp e^{+2}_{23} = 1; \hsp e^{-2}_{23} = -1. \nn
\eea
Note that $e^{\pm \a}_{ij} = - e^{\pm \a}_{ji}$.

We will also make use of normalized degree $k$ scalar harmonics with
maximal $(m, \bar{m})$ $(SU(2)_L,SU(2)_R)$ charges:
\bea
Y_{k}^{\pm \half k, \pm \half k} &=&
\sqrt{k+1} \sin^{k} \q e^{\pm i k \phi}; \la{ch-sp-ha} \\
Y_{k}^{\pm \half k, \mp \half k} &=& \sqrt{k+1} \cos^{k} \q e^{\pm i k \psi}.
\nonumber
\eea
The triple overlap between two such harmonics of opposite charges with
the neutral harmonic of degree two given in (\ref{singlet}) is given
by
\be
\frac{1}{2 \pi^2} \int Y_{k}^{\half k, \half k} Y_{k}^{- \half k,
  -\half k} Y^{0}_{2} = - \frac{\sqrt{3} k}{k+2}. \la{sp-overlap}
\ee
We will also need the explicit values of the overlaps between two such
harmonics of opposite charges and the commuting Killing vectors:
\bea
E^{\pm}_{3 (--) (++)} \equiv \frac{1}{2\pi^2} \int  D^{a}
Y_{k}^{\half k, \half k} Y_{k}^{- \half k, -\half k} Y^{3 \pm}_a = \pm
i k; \la{ov-z1} \\
E^{\pm}_{3 (+-) (-+)} \equiv \frac{1}{2\pi^2} \int  D^{a}
Y_{k}^{-\half k, \half k} Y_{k}^{\half k, -\half k} Y^{3 \pm}_a = i k.
\eea
Vector spherical harmonics $Y^{0
  \pm}_{ka}$ whose Lie derivatives along the $SO(2)$ directions are
zero can be expressed as
\bea
Y^{0+}_{k} &=& \frac{1}{\sqrt{2}} (\sin^2 \q p_l(\q) d \phi + \cos^2 \q
q_l(\q) d \psi); \la{vec-sing} \\
Y^{0-}_{k} &=& \frac{1}{\sqrt{2}} (- \sin^2 \q p_l(\q) d \phi + \cos^2 \q
q_l(\q) d \psi),
\eea
where $k = 2l +1$ and $l$ is an integer.
The functions $p_l(\q)$ and $q_l(\q)$ of degree $2l$ are related to
degree $k = 2l+1$
scalar harmonics with $SO(2) \times SO(2)$ charges $(\pm \half, \pm \half)$. That is,
\be \la{charg-har}
Y_{k}^{\pm \half, \pm \half} (\q) = e^{\pm i \phi} \sin \q p_l(\q); \hsp
Y_{k}^{\pm \half, \mp \half} (\q) = e^{\pm i \psi} \cos \q q_l(\q),
\ee
are normalized degree $k$ spherical harmonics. Explicit series
representation of these functions are
\bea
p_l(\q) &=& \sqrt{k+1} \left( \sum_{m=0}^{l} (-)^{m}
\left (
\begin{array}{c}
l \\
m
\end{array}
\right )
\left (
\begin{array}{c}
l + m + 1 \\
l + 1
\end{array}
\right )
(\cos\q)^{2m} \right ); \\
q_k(\q) &=& \sqrt{k+1} \left( \sum_{m=0}^{l} (-)^{m}
\left (
\begin{array}{c}
l \\
m
\end{array}
\right )
\left (
\begin{array}{c}
l + m + 1 \\
l + 1
\end{array}
\right )
(\sin \q)^{2m} \right ). \nn
\eea
Finally, let us make explicit the relation between spherical harmonics
and traceless symmetric tensors on $R^4$. There is a one to one map
between scalar spherical harmonics of degree $k$  and rank $k$
symmetric traceless tensors. Given the spherical harmonic, one can
read off the associated tensor by lifting it onto a sphere in
$R^4$. For example, for the charged harmonics (\ref{charg-har}), we
get
\bea \label{lift}
Y_k^{\pm \half, \pm \half}(\q) & \rightarrow & C_k^{\pm \half, \pm \half} =
(x^1 \pm i x^2) p_l(x); \\
p_l(x) &=& \sqrt{k+1} \left( \sum_{m=0}^{l} (-)^{m}
\left (
\begin{array}{c}
l \\
m
\end{array}
\right )
\left (
\begin{array}{c}
l + m + 1 \\
l + 1
\end{array}
\right )
( (x^1)^2 + (x^2)^2)^{m} (\sum_i (x^i)^2)^{l-m} \right ). \nn
\eea

\section{Proof of addition theorem for harmonic functions on $R^4$} \la{apa1}

To prove the addition theorem one first writes
\be \la{wer1}
\left | x - y \right |^{-2} = \frac{1}{r^2} \sum_{n=0}^{\infty}
\sum^{n}_{m \ge 0} (-1)^{n+m}
\frac{n!}{m! (n-m)!} \frac{y^{2n-m}}{r^{2n-m}} (2 \hat{x} \cdot \hat{y}
)^m,
\ee
where $x^{i} = r \hat{x}^i$ and $y^i = y \hat{y}^i$ with
$(\hat{x}^i,\hat{y}^i)$ unit vectors. Collecting together terms of the same
radial power and summing the finite series one finds
\be
\left | x - y \right |^{-2} = \sum_{k \ge 0} \frac{y^k}{r^{2+k}}
\frac{\sin( (k+1) \g)}{\sin(\g)},
\ee
where the angle $\g$ is defined as $\cos \g = \hat{x} \cdot \hat{y}$.

Now at each degree $k$ there is precisely one $SO(3)$ invariant spherical
harmonic and the normalized such harmonic is given by
\be \la{j9}
Y^{0}_{k}(\g) = \sin( (k+1) \g) /\sin(\g).
\ee
One can show this using spherical coordinates adapted to the $SO(3)$
symmetry group, namely
\be \la{coor5}
ds_3^2 = d \hat{\q}^2 + \sin^2 \hat{\q} d \W_2^2.
\ee
Then $Y^{0}_{k}(\hat{\q})$ satisfies the degree $k$ $SO(3)$ invariant
spherical harmonic equation
\be \la{difeq}
\left ( \frac{1}{\sin^2 \hat{\q}} \pa_{\hat{\q}} (\sin^2 \hat{\q}
\pa_{\hat{\q}} ) + k (k+2) \right ) Y^{0}_k(\hat{\q}) = 0,
\ee
and is normalized as in the previous section. Therefore
the addition theorem amounts to proving the following identity
\be \la{add2}
Y^{0}_{k} (\g) = \a_k \sum_{I}
Y^{I}_{k}(\q^x_3) Y^{I}_{k} (\q^y_3),
\ee
where $Y^{I}_k (\q_3)$ are (normalized) spherical harmonics of degree $k$ on the
$S^3$ and $\a_{k} = 1/(k+1)$.  First note that in the coordinate system (\ref{coor5}) on
the sphere
\be
\cos \g = \cos \q_x \cos \q_y + \sin \q_x \sin \q_y (\cos \g_{2}),
\ee
where $\g_{2}$ is the angle separating the vectors on the $S^2$. Thus
when $\q_y = 0$ (it lies on the ``axis'') $\cos \g = \cos \q_x$.
Since the $SO(3)$ singlet harmonic is the only harmonic at level $k$
which is non-vanishing on the axis (\ref{add2}) collapses to
\be
Y^{0}_{k} (\g) = \a_k
Y^{0}_{k}(\q_x) Y^{0}_{k} (0),
\ee
which is true if $\a_{k} = 1/(k+1)$ since from (\ref{j9}) $Y^{0}_{k}(0) = (k+1)$.

Now consider rotating the axes so that $\q_y$ is no longer zero. Then
the function $ Y^{0}_{k} (\g)   $ still satisfies the
covariant version of (\ref{difeq}), namely
\be
( \Box_x + k(k+2)) Y^{0}_{k} (\g) = 0,
\ee
where $\Box_x$ is the Laplacian on the $S^3$ with coordinates $\q_3^x$. In
other words, the function can always be expanded in spherical harmonics of
rank $k$ as
\be
 Y^{0}_{k} (\g) = \sum_{I} \a^{I}_{k}( \q^y_3) Y^{I}_{k}
 (\q^x_3),
\ee
where the coefficients are given by
\be \la{r1}
\a^{I}_{k}( \q_3^y) = \int_{S^3} d\Omega_3
Y^{I}_{k} (\q_3^x ) Y^{0}_k (\cos \g).
\ee
However, a generic function can be expanded in terms of spherical
harmonics as
\be
f(\q_3^x) = \sum_{k,I} \beta_{kI} Y^{I}_{k} (\q_3^x),
\ee
where
\be
\beta_{kI} = \int_{S^3} d\Omega_3 f(\q_3^x) Y^I_k(\q_3^x),
\ee
and in particular for the $SO(3)$ singlet coefficients
\be
\beta_{k} = \int_{S^3} d\Omega_3 f(\q_3^x) Y^{0}_{k} (\q_x),
\ee
so that $f(\q_x = 0) = \sum_{k} \beta_k (k+1) $. Then (\ref{r1}) is the
$SO(3)$ singlet coefficient in an expansion of the function $Y^{I}_{k}
(\q_3^x)$ in a series of $Y^{I}_{k}(\g, \cdots)$ (i.e. with respect
to the rotated axis discussed earlier). One can thus read off the
coefficient (\ref{r1}) as
\be
\a^{I}_{k}( \q_3^y) = (k+1)^{-1} Y^{I}_{k} (\q_3(\g,
\cdots))_{\g = 0} = (k+1)^{-1}  Y^{I}_{k} (\q_3^y),
\ee
since in the limit $\g \rightarrow 0$ the angles $(\q,\cdots)$ go over
into $(\q_y,\cdots)$. This completes the proof of (\ref{add2}).

\section{Six dimensional field equations to quadratic order} \la{apb}

In this appendix we summarize the computation of the relevant quadratic
corrections to
the six-dimensional field equations using the results of \cite{Arutyunov:2000by,
  Pank}. Expanding the Einstein equation (\ref{x2}) up to second order
in fluctuations gives
\bea
R^{(1)}_{MN} + R^{(2)}_{MN} &=& H^{A}_{MPQ} {H^A_N}^{PQ} - 2 (h^{KL} -
h^{KP} {h^{L}}_{P}) H^{A}_{MKQ} {H^{A}_{NL}}^{Q} \\
&& + h^{KL} h^{PQ} H^{A}_{MKP} H^{A}_{NLQ} + D_{M} \Phi D_{N} \Phi,
\nn \\
& \equiv & (E^{(1)}_{MN} + E^{(2)}_{MN}) \la{ein1}
\eea
where
\bea
R^{(1)}_{MN} &=& D_{K} h^{K}_{MN} - \half D_{M} D_{N} (h^{L}_L); \\
R^{(2)}_{MN} &=& -D_{K} (h^{K}_{L} h^{L}_{MN}) + \qu D_{M} D_{N}
(h^{KL} h_{KL}) + \half h^{K}_{MN} D_{K} (h^{L}_{L}) - h^{K}_{ML}
h^{L}_{KN}; \nn \\
h^{K}_{MN} & \equiv &  \half (D_{M} h^{K}_{N} + D_{N} h^{K}_{M} -
D^{K} h_{MN}). \nn
\eea
The quantities $(E^{(1)}_{MN}, E^{(2)}_{MN})$ are defined to be
linear and quadratic in fluctuations respectively. The expansion of
the scalar field and the three forms $G^{A}$ (\ref{dq2}) implies the
following expansion for the three forms $H^A$ up to quadratic
order in fluctuations:
\bea
H^{5} &=& g^{o} + g^{5} + \Phi g^{6} + \half g^{o} \Phi^2; \\
H^{6} &=& g^{6} + g^{o} \Phi + g^{5} \Phi, \nn
\eea
where $(g^5,g^6)$ are the (closed) three form fluctuations given in
(\ref{dq2}) and $g^{o}$ is the background three form.

The scalar field equation up to second order is
\bea
(\Box + \Box_a)  \Phi &\equiv& E^{(1)} + E^{(2)};  \la{scal1} \\
&=& D_{K} \Phi (D_{L} h^{KL} - \half D^{K}(h^L_L)
+ h^{KL} D_{K} D_{L} \Phi + \frac{2}{3} H^{5}_{KLM} (H^{6 KLM} - 3
h_{S}^{K} H^{6 SLM}), \nn
\eea
where $E^{(1)}$ is the part linear in fluctuations and $E^{(2)}$ is
quadratic part. Recall that $\Box$ is the d'Alambertian on $AdS_3$
and $\Box_a$ is the d'Alambertian on $S^3$.

The (anti)-self duality equation is
\be
H \mp \ast H \pm S^{(1)} \pm S^{(2)} \equiv T^{(1)} + T^{(2)} = 0, \la{form1}
\ee
where
\bea
S^{(1)}_{KLM} &=& \half h (\ast H)_{KLM} - 3 h^P_{[K} (\ast H)_{LM]
  P}; \\
S^{(2)}_{KLM} &=&  \frac{3}{2} h^{P}_{P} h^{Q}_{[K} (\ast H)_{LM] Q} -
(\frac{1}{8} h^2 + \qu h^{PQ} h_{PQ}) (\ast H)_{KLM} - 3 h^{P}_{[K}
  h^{Q}_{L} (\ast H)_{M] PQ}, \nn
\eea
and $(T^{(1)},T^{(2)})$ are the parts linear and quadratic in fluctuations
respectively.

We are interested in corrections to the
$(s^2,\s^2,H_{\m\n},A^{\pm}_{\m})$ field equations quadratic in the scalar field
$s^1$ and the gauge field $A^{\pm}$. Consider first the $s^2$ field
equations. The linearized field equation is given by a combination of
the scalar field equation (\ref{scal1}) and components of the
anti-self-duality equation (\ref{form1}). That is,
\be
\Box s^2_I \equiv \frac{1}{12} \left ( (\Box + \Box_a) \Phi - E^{(1)}
- \ep^{abc} ( \half D^{\m} D^{a} T^{(1)}_{\m bc} + \frac{2}{3}
T^{(1)}_{abc}) \right )_{Y^I_2} = 0,
\ee
where $A_{Y^I_2}$ denotes the projection of $A$ onto the $Y^I_2$
harmonic. For the quadratic corrections to this equation first
define the following quantities
\be
q_1 = E^{(2)}; \hsp
q_{2\m a} = - \half \ep^{abc} T^{(2)6}_{ \m bc}; \hsp
q_{3} = \frac{1}{6} \ep^{abc} T_{ a b c}^{(2) 6},
\ee
then the correction to the $s^2_I$ equation is given by
\be
\Box s^2_I = \frac{1}{12} ((q_1) + D^{\m} D^{a} (q_{2\m a}) + 4
(q_{3}))_{Y^I_2}.
\ee
Now the explicit computations of \cite{Pank} show that there are no
such correction terms quadratic in $S^1_i$ and $A^{\pm \a}$. Therefore
the linearized equation remains uncorrected to quadratic order.

Next consider the $\s^2_I$ equation. Here the linearized equation is a
specific combination of the components of the Einstein equation
(\ref{ein1}) along the sphere with components of the self-duality
equation. Namely
\be
\Box \s^2_I \equiv \frac{1}{6} \left ( \frac{1}{3} (E_{a}^{(1)a} -
R^{(1)a}_a) + \qu (E^{(1)}_{(ab)} - R^{(1)}_{(ab)}) - \qu \ep^{\m \n
  \r} D_{\m } D^{a} T^{(1)5}_{\n \r a} + \frac{2}{3} \ep^{abc} T^{(1)
  5}_{abc} \right )_{Y^I_2} = 0.
\ee
For the quadratic corrections to this equation define
\bea
Q_1 &=& \frac{1}{3} (E_{a}^{(2)a} - R^{(2)a}_a); \hsp
Q_{2 (ab)} = (E^{(2)}_{(ab)} - R^{(2)}_{(ab)}); \\
Q_{3 a}^{\m} &=& \frac{1}{2} \ep^{\m \n \r} T_{\n \r a}^{(2)5}; \hsp
Q_{4} = \frac{1}{3!} \ep^{abc} T_{abc}^{(2) 5}, \nn
\eea
and again denote as $(Q)_{Y^I_k}$ the projection of $Q$ onto $Y^I_k$. Then
\be \la{ss2}
\Box \s^2 = \frac{1}{6} (Q_1 + \qu D^{a} D^{b} Q_{2 (ab)} - \half
D^{\m} D^{a} Q_{3 a \m} + 4 Q_4)_{Y^I_2}.
\ee
Now the terms quadratic in the scalar fields $s^1$ were
computed in \cite{Pank}
\bea
(Q_1)_{Y^I} &=& - 14 s^1_i s^{1}_j a_{I ij} + \frac{2}{3} (D_{\m} s^1_i
D^{\m} s^{1}_j + 2 s^{1}_i s^{1}_j) b_{I ij};  \la{b12} \\
(D^{a} D^{b} Q_{2 (ab)})_{Y^I} &=&  4 \left ( s^{1}_i s^{1}_j -
D_{\m} s^{1}_i D^{\m} s^{1}_j \right ) d_{ij  I}; \nn \\
(D^{\m} D^{a} Q_{3 a \m})_{Y^I} &=& - 4 \left ( s^{1}_i s^{1}_j -
D_{\m} s^{1}_i D^{\m} s^{1}_j \right ) b_{i I j}; \nn \\
(Q_4)_{Y^I} &=& 4 s^{1}_i s^{1}_j a_{I ij}. \nn
\eea
The relevant spherical harmonic triple overlaps are defined in
appendix \ref{apa}. We should mention here that
there are also contributions to (\ref{ss2}) quadratic in the gauge
field which were not explicitly computed in \cite{Pank}. These are
given by
\bea
(Q_1)_{Y^I} &=& -\frac{1}{8} F_{\m \n}(A^{+\a}) F^{\m\n}(A^{-\b})
f_{I \a \b} + \cdots; \la{b13} \\
(D^{a} D^{b} Q_{2 (ab)})_{Y^I} &=& - \frac{5}{2} F_{\m \n}(A^{+\a})
F^{\m\n}(A^{-\b}) f_{I  \a \b} + \cdots; \nn \\
(D^{\m} D^{a} Q_{3 a \m})_{Y^I} &=& \frac{3}{4} D_{\m} \left ( F^{\m \n}(A^{+\a})
A^{-\b}_{\n} + F^{\m \n} (A^{-\b}) A^{+\a}_{\n} \right )
f_{I \a \b} + \cdots. \nn
\eea
The spherical harmonic triple overlap $f_{I a\b}$ is defined in
(\ref{op4}). Terms quadratic in two
$SU(2)_L$ gauge fields or two $SU(2)$ right gauge fields are
projected out via the identities (\ref{op5}).
The ellipses denote terms quadratic in the gauge field rather than its
field strength, that is,
proportional to $A^{\pm \a}_{\m} A^{\m \pm \b}$. These terms cancel
out when combined in (\ref{ss2}) leaving only a contribution involving
field strengths. The latter however vanish when one imposes the
leading order field equations, and thus
the combination of the corrections (\ref{b12}) and (\ref{b13})
gives the $\s^2$ field equation (\ref{cor-s2}), containing only scalar
field corrections.

Next consider the corrections to the Einstein equation. Recall that
the three dimensional metric to quadratic order in the fields is
\be
H_{\m \n} = h^{0}_{\m \n} + \pi^{0} g^{o}_{\m \n} - h_{\m}^{\pm \a}
h_{\n}^{\pm \a} \equiv \hat{H}_{\m \n} - h_{\m}^{\pm \a} h_{\n}^{\pm \a}.
\ee
Then one can show that
\be
({\cal{L}_E} +2 ) \hat{H}_{\m \n} = (E^{(2)}_{\m \n} - R^{(2)}_{\m \n})_{Y_0} + (3
Q_1 + 4 Q_4)_{Y_0} g^{o}_{ \m \n},
\ee
where the linearized Einstein operator is defined in
(\ref{Ein-eq}). The following terms which are quadratic in the scalar fields
\be
(E^{(2)}_{\m \n} - R^{(2)}_{\m \n})^0 = (- 2 s^{1}_i s^{1}_j g^{o}_{\m\n}
+ 16 D_{\m} s^1_i D_{\n} s^{1}_j - 6 D_{\r} s^{1}_i D^{\r} s^{1}_j g^{o}_{\m
  \n})\d^{ij},
\ee
in combination with those contained in (\ref{b12}) give
\be
({\cal{L}_E} + 2 ) \hat{H}_{\m \n} = 16 (D_{\m} s^1_i D_{\n} s^{1}_i
- g^{o}_{\m \n}s^1_i s^1_i).
\ee
There are also contributions quadratic in the gauge fields to both
$({\cal{L}_E} + 2 ) \hat{H}_{\m
  \n}$ and $({\cal{L}_E} + 2 ) h_{\m}^{\pm \a} h_{\n}^{\pm \a}$. These
contributions involve both the gauge fields and their field
strength, and in particular do not vanish for flat connections. This
is unsurprising, since we know from general arguments that
$\hat{H}_{\m \n}$ on its own does not transform correctly under gauge
transformations.
However the gauge field contributions to $({\cal{L}_E} + 2 ) {H}_{\m \n}$,
where $H_{\m \n}$ is the three dimensional metric (\ref{3dmet})
that transforms correctly under diffeomorphisms,
do vanish for flat connections, as indeed they should, and thus are zero
when one imposes the leading order gauge field equations. The corrected
Einstein equation is therefore that given in (\ref{cor-Ein}).

\section{3-point functions} \la{apd}

In this appendix we discuss the supergravity
computation of certain 3-point functions.

\subsection{Extremal scalar three point functions}

First we will consider the computation of the 3-point function
between two operators of dimension 1 and one operator of dimension $k$.
The operators of dimension 1 may be the same or different and are
dual to the fields $S^1$; there are four such operators
corresponding to the four scalar harmonics of degree 1 which are
labeled by $i,j$.
The operator $\cao_{{\S}^k_I}$
of dimension $k$
is dual to the field $\S^k_I$ (there are $(k+1)^2$ such operators
labeled by $I$).
The $k=2$ case is special in that the correlator is extremal
\cite{D'Hoker:1999ea}. As in the five dimensional case, the computation
of extremal correlators is subtle. The bulk coupling vanishes but the
spacetime integral diverges when $k \to 2$ in such way that
the corresponding 3-point function is finite. We will take this
value to be the correct extremal correlator and this will allow
us to fix the coefficient of the relevant terms non-linear in
momentum in the 1-point function of $\S^2$.

The three dimensional field equations to quadratic order were determined in
\cite{Arutyunov:2000by} and for the fields of interest and
with our normalizations they read
\bea
(\Box - k (k-2)) \S^k_I &=& w_{Iij} S^1_i S^{1}_j; \label{S_eqn}\\
(\Box +1) S^1_i &=& w_{Iij} \S^k_I S^{1}_j; \nn \\
(\Box +1) S^{1}_j &=& w_{Iij} \S^k_I S^{1}_i; \nn
\eea
where
\be
w_{Iij} = \frac{k^3 (k+2) (k+4) (1-k/2)}{32 (k+1) \sqrt{k (k-1)}} a_{Iij}.
\ee
Notice that this coupling vanishes in the extremal case $k=2$.

The aim is to compute the 3-point
$\< \cao_{\S^k_I}(x_1) \cao_{S^1_i}(x_2) \cao_{S^{1}_j}(x_3)\>$,
but we start by
discussing 2-point functions. These are obtained by the first
variation of the 1-point functions
\bea
\< \cao_{\S^k_I}(x_1) \cao_{\S^k_J}(x_2) \> &=& -
\frac{\d \< \cao_{\S^k_I}(x_1) \>}{\d \S_{J(0)}^k(x_2)} =
-\left(\frac{n_1 n_5}{4 \pi} \right) (2 k- 2)
\frac{\d \S^k_{I(2k-2)}(x_1)}{\d \S_{J(0)}^k(x_2)}; \nn \\
\< \cao_{S^1_i}(x_1) \cao_{S^1_j}(x_2) \> &=& -
\frac{\d \< \cao_{S^1_i}(x_1) \>}{\d S_{j(0)}^1(x_2)} = -
\left(\frac{n_1 n_5}{4 \pi} \right) 2
\frac{\d \tilde{S}^1_{i(0)}(x_1)}{\d S_{j(0)}^1(x_2)}, \nn
\eea
where we used (\ref{mome}). It follows that in order to obtain
these 2-point functions we need to solve (\ref{S_eqn}) to linear order
in the sources (so the r.h.s is set equal to zero)
and then extract the appropriate coefficient. The details
of this computation can be found in section 6.3 of
\cite{Karch:2005ms} with the following result
\bea
\< \cao_{\S^k_I}(x_1) \cao_{\S^k_J}(x_2) \>
&=& \left(\frac{n_1 n_5}{4 \pi} \right)
\frac{(2 k-2) \Gamma(k)}{\pi \G(k-1)} \left(\frac{1}{x^{2k}}\right)_R \d_{IJ},
\qquad k \neq 1; \nn \\
\< \cao_{S^1_i}(x_1) \cao_{S^1_j}(x_2) \> &=& \left(\frac{n_1 n_5}{4 \pi} \right)
\frac{2}{\pi}
\left(\frac{1}{x^{2}}\right)_R \d_{ij},
\eea
where the subscript $R$ indicates that these are renormalized
correlators.

We now discuss the 3-point function with $k \neq 2$. We can can obtain
the 3-point function by
the second variation of the 1-point function of $O_{\S^k}$:
\bea
\< \cao_{\S^k_i}(x_1) \cao_{S^1_i}(x_2) \cao_{S^{1}_j}(x_3)\> &=&
\frac{\d^2 \< \cao_{\S^k_I}(x_1) \>}{\d S_{i(0)}^1(x_2) \d S_{j(0)}^{1}(x_3)} \nn \\
&=&\left(\frac{n_1 n_5}{4 \pi} \right) (2 k- 2)
\frac{\d^2 \S^k_{I(2k-2)}(x_1)}{\d S_{i(0)}^1(x_2) \d S_{j(0)}^{1}(x_3)}
\eea
It follows that we need to solve (\ref{S_eqn}) to quadratic order in the
sources and then extract the coefficient of order $z^k$. The steps
involved in this computation are spelled out in section 5.9 of
\cite{Skenderis:2002wp}. For the case at hand, the result is\footnote{
The normalization of the bulk-to-boundary
propagator in (5.52) when $\D=1$ is $C_1=1/\pi$.}
\bea
\< \cao_{\S^k_I}(x_1) \cao_{S^1_i}(x_2) O_{S^{1}_j}(x_3)\> &=&
- \left(\frac{n_1 n_5}{4 \pi} \right) w_{Iij}
\frac{2 \G(k)}{\pi^3 \G(k-1)} I_{k}(x_1,x_2,x_3)
\eea
where
\be
I_k(x_1,x_2,x_3) = \int \frac{d^2 x d z}{z^3}
\left(\frac{z}{z^2 + (\vec{x} - \vec{x}_1)^2}\right)^k
\left(\frac{z}{z^2 + (\vec{x} - \vec{x}_2)^2}\right)
\left(\frac{z}{z^2 + (\vec{x} - \vec{x}_3)^2}\right).
\ee
This integral was computed in \cite{Freedman:1998tz} with answer
\be
I_k(x_1,x_2,x_3) = \frac{\pi \G(1-k/2) (\G(k/2))^3}{2 \G(k)}
\frac{1}{|\vec{x}_1 - \vec{x}_2|^k  |\vec{x}_1 - \vec{x}_3|^k
|\vec{x}_2 - \vec{x}_3|^{2-k}}.
\ee
Notice that this integral diverges in the extremal case $k \to 2$.

The final answer for the correlator is thus
\be
\< \cao_{\S^k_I}(x_1) \cao_{S^1_i}(x_2) \cao_{S^{1}_j}(x_3)\> =
\frac{C^k_{Iij}}{|\vec{x}_1 - \vec{x}_2|^k  |\vec{x}_1 - \vec{x}_3|^k
|\vec{x}_2 - \vec{x}_3|^{2-k}}
\ee
where
\be
C^k_{Iij}
= - \left(\frac{n_1 n_5}{4 \pi} \right)
\frac{k^3 (k+2) (k+4) \G(k/2)^3 \G(2-k/2)}{32 \pi^2 (k+1) \G(k-1)
\sqrt{k (k-1)}} a_{Iij}\, .
\ee
This coefficient has a smooth limit as $k \to 2$; the zero in $w_{Iij}$ cancels
against the divergence in $I_2$, and we get
\be
C^2_{Iij} = - \left(\frac{n_1 n_5}{4 \pi} \right) \frac{1}{\sqrt{2}\pi^2} a_{Iij}
\, .
\ee
We will take this to be the correct extremal 3-point function, i.e.,
\be \label{ext}
\< \cao_{\S^2_I}(x_1) \cao_{S^1_i}(x_2) \cao_{S^{1}_j}(x_3)\> =
\frac{C^2_{Iij}}{|\vec{x}_1 - \vec{x}_2|^2  |\vec{x}_1 - \vec{x}_3|^2}\, ,
\ee
and use it to deduce the non-linear coupling in the 1-point function of
$\<\cao_{\S^2_I}\>$. As discussed in \cite{Skenderis:2006uy}, the form
of the 1-point function is uniquely fixed by general arguments to be
\be \la{cop1}
\<\cao_{\S^2_I}\> = \left(\frac{n_1 n_5}{4 \pi} \right)\left(\pi^{\S^2_I}_{(2)}
+ A_{Iij} \pi^{S^1_i}_{(1)} \pi_{(1)}^{S^{1}_j} \right)
\ee
The numerical coefficient $A_{Iij}$ should be determined by doing
holographic renormalization in 6 (rather than 3) dimensions.
We will fix it, however, such that the
the extremal correlator is correctly computed directly at $k=2$ (rather
than obtained as a limit from $k \neq 2$).
Since $w_{Iij}(k=2){=}0$ the only contribution comes
from the terms non-linear in momenta
\bea
\< \cao_{\S^k_I}(x_1) \cao_{S^1_i}(x_2) \cao_{S^{1}_j}(x_3)\> &=&
\left(\frac{n_1 n_5}{4 \pi} \right) 2 A_{Iij}
\left(\frac{\d \pi^{S^1_i}_{(1)}(x_1)}{\ S_{i(0)}^1(x_2)} \right)
\left(\frac{\d \pi^{S^{1}_j}_{(1)}(x_1)}{\ S_{j(0)}^{1}(x_3)} \right); \nn \\
&=& \left(\frac{n_1 n_5}{4 \pi} \right) A_{Iij}
\frac{8}{\pi^2} \frac{1}{|\vec{x}_1 - \vec{x}_2|^2  |\vec{x}_1 - \vec{x}_3|^2}
\eea
By comparing with (\ref{ext}) we find
\be \la{cop2}
A_{Iij} = - \frac{1}{4 \sqrt{2}} a_{Iij}.
\ee

\subsection{Non-extremal scalar three point functions}

We will also need other three-point functions for scalars due to
chiral primary operators. The relevant
cubic couplings in three dimensions were also computed in
\cite{Mihailescu:1999cj,Arutyunov:2000by} and are given by
\bea
&& - \frac{n_1 n_5}{4 \pi} \int d^3x \sqrt{-G} (T_{123}
S^1 S^2 \S^3 + U_{123} \S^1 \S^2 \S^3); \la{3pcou} \\
&& \equiv - \frac{n_1 n_5}{16 \pi} \int d^3x \sqrt{-G} V_{123}
\left ( \frac{S^1 S^2 \S^3}{ \sqrt{(k_1
    +1)(k_2 + 1)}} + \frac{(k_1^2 + k_2^2 + k_3^2 -2)}{(k_1 + 1) (k_2+1)}
\frac{\S^1 \S^2 \S^3}{6 \sqrt{(k_1 -1)(k_2 -
    1)}} \right ), \nn \\
&& \hsp V_{123} = \frac{\S (\S+2) (\S-2)
  \a_1 \a_2 \a_3 a_{123}}{(k_3 + 1) \sqrt{k_1 k_2 k_3 (k_3 - 1)}} \nn
\eea
where $k_a$ denotes the dimension of the operator dual to the field $\Psi^a$,
$\S = k_1 + k_2 + k_3$, $\a_1 = \half (k_2 + k_3 - k_1)$ etc and $a_{123}$
is shorthand for the spherical harmonic overlap. It is straightforward
to follow the same steps as before to compute the associated three
point functions:
\bea
\< \cao_{S^1}(x_1) \cao_{S^2}(x_2) \cao_{\S^3}(x_3)\> &=& \frac{N}{4
  \pi^3} \frac{W_{123} T_{123}}{|\vec{x}_1 - \vec{x}_2|^{2\a_3}
|\vec{x}_1 - \vec{x}_3|^{2 \a_2}
|\vec{x}_2 - \vec{x}_3|^{2 \a_1}}; \\
\< \cao_{\S^1}(x_1) \cao_{\S^2}(x_2) \cao_{\S^3}(x_3)\> &=& \frac{3 N}{4
  \pi^3} \frac{W_{123} U_{123}}{|\vec{x}_1 - \vec{x}_2|^{2\a_3}
|\vec{x}_1 - \vec{x}_3|^{2 \a_2}
|\vec{x}_2 - \vec{x}_3|^{2 \a_1}}; \nn \\
W_{123} &=& \frac{\G(\a_1 ) \G(\a_2    )\G(\a_3 )\G(\half (\S-2))}{
\G(k_1 -1)\G(k_2 -1)\G(k_3 -1)}. \nn
\eea
We will be interested in the case where $(S^1,S^2,\S^1,\S^2)$ have
dimension $k$ and $(S^2,\S^2)$ are chiral primary with $(S^1,\S^1)$
anti-chiral primary. Then charge conservation implies that the correlators
are only non-zero when $\S^3$ is neutral. In the case where ${\cal O}_{\S^3}$
has dimension two the explicit results for the correlators using the
spherical harmonic overlap of (\ref{sp-overlap}) are
\bea
\< (\cao_{S^p_k})^{\dagger}(x_1) \cao_{S^p_k}(x_2) \cao_{\S^2_0}(x_3)\> &=&
\frac{N \sqrt{3} }{2 \sqrt{2} \pi^3} \frac{k^3}{|\vec{x}_1 -
  \vec{x}_2|^{2 (k-1)}
|\vec{x}_1 - \vec{x}_3|^{2 }
|\vec{x}_2 - \vec{x}_3|^{2}}; \\
\< (\cao_{\S^p_k})^{\dagger} (x_1) \cao_{\S^p_k}(x_2)
\cao_{\S^2_0}(x_3)\> &=&
\frac{N \sqrt{3}}{2 \sqrt{2} \pi^3 (k+2)^3}
\frac{k(k-1) (k^4-1)}{|\vec{x}_1 - \vec{x}_2|^{2(k-1)}
|\vec{x}_1 - \vec{x}_3|^{2}
|\vec{x}_2 - \vec{x}_3|^{2}}. \nonumber
\eea
It will be useful to define normalized three point functions as
\bea
\< (\cao_{S^p_k})^{\dagger} \cao_{\S^2_0}(x) \cao_{S^p_k} \> & \equiv &
\frac{\< (\cao_{S^p_k})^{\dagger}(\infty )  \cao_{\S^2_0}(x)
  \cao_{S^p_k}(0) \>}{\< (\cao_{S^p_k})^{\dagger}(\infty)
  \cao_{S^p_k}(0) \>}
= \frac{\sqrt{3} k^3 }{\sqrt{2} \pi (k-1)^2} \frac{1}{| \vec{x}
  |^2}. \la{norm-tpf} \\
\< (\cao_{\S^p_k})^{\dagger} \cao_{\S^2_0}(x) \cao_{\S^p_k} \> & \equiv &
\frac{\< (\cao_{\S^p_k})^{\dagger}(\infty) \cao_{\S^2_0}(x) \cao_{\S^p_k}(0)
  \>}{\< (\cao_{\S^p_k})^{\dagger}(\infty)
  \cao_{\S^p_k}(0) \>}; \nn \\
&=& \hsp \frac{\sqrt{3} k (k+1) (k^2+1) }{\sqrt{2} \pi (k+2)^2} \frac{1}{| \vec{x}
  |^2}. \nn
\eea
(Implicitly we assume here that $k \neq 1$.)
Note that for $k \gg 1$ these expressions both tend to the same limit, $\sqrt{3}
k/\sqrt{2} \pi |\vec{x} |^2$.

\subsection{Two scalars and R symmetry current}

Finally we will need three point functions between two scalars (of the
same mass) and the R symmetry current. The relevant cubic couplings were again given in
\cite{Arutyunov:2000by}:
\be
- \frac{n_1 n_5}{8 \pi} \int d^3 x \sqrt{-G} A^{\pm \a}_{\mu} (S^k_I
D_{\m} S^{k}_J + \S^k_I D_{\m} \S^k_J) E^{\pm}_{\a I J},
\ee
where the triple overlap is defined in (\ref{gt2}). To compute the
corresponding three point functions one again follows the steps given
in \cite{Skenderis:2002wp}. This results in
\be \la{vec-3pt}
\< \cao_{S^k_I} (x_1) J^{\pm \a} (x) \cao_{S^k_J} (x_2) \> =
\< \cao_{\S^k_I} (x_1) J^{\pm \a} (x) \cao_{\S^k_J} (x_2) \>
= \mp i \frac{N}{8 \pi} E^{\pm}_{\a IJ} I_{\mp} (x,x_1,x_2),
\ee
where the AdS integral
\be
I_{\mp} (x,x_1,x_2) = \int \frac{d^3z}{z^3} K_{k}(z,\vec{x}_1) D^{\m}
K_{k}(z,\vec{x}_2)  {\cal{G}}_{\m \mp} (z,\vec{x}) =
\frac{(k-1)^2}{\pi^2} \frac{Z_{\mp}}{|\vec{x}_1 - \vec{x}_2 |^{2k}},
\ee
was computed in \cite{Freedman:1998tz}. In this integral $K_{k}(z,\vec{x})$ and
${\cal{G}}_{\m \mp} (z,\vec{x})$ are the standard AdS scalar and vector bulk to
boundary propagators respectively and
\be
Z_+ = \frac{1}{(w_1 - w)} - \frac{1}{(w_2 -w)}; \hsp
Z_- = \frac{1}{(\bar{w}_1 - \bar{w})} - \frac{1}{(\bar{w}_2 -\bar{w})}.
\ee
Here we have implicitly switched to Euclidean signature, $t = i \t$,
and introduced complex boundary coordinates $w = y + i \t$.

In deriving this result we use the standard vector
propagator, that following from the field equation $D_{\m} F^{\m \n} =
0$, although the (linearized) vector equation here is Chern-Simons, $F_{\m
  \n} = 0$. Whilst this step should be justified more rigorously, it
can be justified a posteriori by the fact that the three point
functions thus obtained are of the standard form for a two dimensional
CFT. To see this, consider the case where the scalar operators are
chiral primary. Using the specific values for the spherical harmonic
overlaps (\ref{ov-z1}) in (\ref{vec-3pt}) gives
\bea
\< (\cao_{S^{k}})^{\dagger} (x_1) J^{+3} (w) \cao_{S^k} (x_2) \>
&=& \frac{N}{8 \pi^3} k (k-1)^2 \left (
\frac{1}{(w_1 - w)} - \frac{1}{(w_2 -w)}
\right );
\\
&=& \< (\cao_{S^{k}})^{\dagger} (x_1) \cao_{S^k} (x_2) \>
\frac{k}{4 \pi} \left (
\frac{1}{(w_1 - w)} - \frac{1}{(w_2 -w)}
\right ), \nn
\eea
with the latter being the canonical form for the CFT three point
function between the (holomorphic) R current and operators charged
under it.
An analogous formula holds for the anti-holomorphic current,
$J^{-3}(\bar{w})$ and for
the correlators involving scalar operators dual to $\S^k$.
Again it is useful to define normalized three point functions such that
\bea
\< (\cao_{S^{k}})^{\dagger} J^{+3} (w) \cao_{S^k}  \>  \equiv
\frac{\< (\cao_{S^{k}})^{\dagger} (\infty) J^{+3} (w) \cao_{S^k} (0) \>
}{ \< (\cao_{S^{k}})^{\dagger} (\infty) \cao_{S^k} (0) \>} =
\frac{k}{4 \pi w}; \la{jj4} \\
\< (\cao_{\S^{k}})^{\dagger} J^{+3} (w)\cao_{\S^k}  \>  \equiv
\frac{\< (\cao_{\S^{k}})^{\dagger} (\infty) J^{+3} (w) \cao_{\S^k} (0) \>
}{ \< (\cao_{\S^{k}})^{\dagger} (\infty) \cao_{\S^k} (0) \>} =
\frac{k}{4 \pi w}, \nn
\eea
with analogous formulae holding for the anti-holomorphic currents. The
corresponding normalized three point functions for the spectrally
flowed operators in the R sector are then
\bea
\< (\cao_{S^{k}})_R^{\dagger} J^{+3} (w)(\cao_{S^k})_R \>  \equiv
\frac{\< (\cao_{S^{k}})_R^{\dagger} (\infty) J^{+3} (w) (\cao_{S^k})_R (0) \>
}{ \< (\cao_{S^{k}})_R^{\dagger} (\infty) (\cao_{S^k})_R (0) \>} =
\frac{k - N}{4 \pi w}; \la{jj5} \\
\< (\cao_{\S^{k}})_R^{\dagger} J^{+3} (w) (\cao_{\S^k})_R \>  \equiv
\frac{\< (\cao_{\S^{k}})_R^{\dagger} (\infty) J^{+3} (w) (\cao_{\S^k})_R (0) \>
}{ \< (\cao_{\S^{k}})^{\dagger} (\infty) \cao_{\S^k} (0) \>} =
\frac{k - N}{4 \pi w}, \nn
\eea
where ${\cal O}_R$ denotes the spectral flowed operator. Again
corresponding formulae hold for the anti-holomorphic currents.

\section{Holographic 1-point functions} \la{apc}

In this appendix we derive the 1-point function for the
stress energy tensor and the operators dual to $S^1_i$.
We omit the details of this computation since the
analysis is very similar to the Coulomb branch analysis in \cite{BFS1,BFS2}.
The asymptotic analysis of this system is also
presented (in a different coordinate system) in \cite{Henneaux:2004zi}
and the form of the counterterm was obtained in \cite{Karch:2005ms}.

The relevant action is given in (\ref{pp1}), retaining only the
graviton and scalar fields
$S^1_i$, and the most general asymptotic solution with Dirichlet boundary
conditions is given by the expansion in (\ref{as_exp}) with coefficients given by
\bea
\Tr\ g_{(2)} &=& -\half R - \frac{1}{2} \left(
2 (S_{i(0)}^1)^2 + (\td{S}^1_{i(0)})^2 \right) \nn \\
D^v g_{(2)uv} &=& = -D_{u} \left(\half R
+ \frac{1}{4} \left((\td{S}^1_{(0)i})^2 + 4 (S_{i(0)}^1)^2
- 2 S_{i(0)}^1 \td{S}^1_{i(0)} \right) \right) -
S_{i(0)}^1 D_u \td{S}^1_{i(0)} \nn \\
h_{(2)uv}&=& -\frac{1}{2} S_{i(0)}^1 \td{S}^1_{i(0)} g_{(0)uv} \nn \\
\tilde{h}_{(2)uv} &=& -\frac{1}{4} (S_{i(0)}^1)^2  g_{(0)uv}
\eea
The traceless transverse part of $g_{(2)}$ and
$\td{S}^1_{i(0)}$ (as well as the sources
$g_{(0)uv}$ and $S_{i(0)}^1$) are unconstrained. We will soon see
that these coefficients are related to the 1-point functions.

The counterterms needed to render the on-shell action finite
are
\be
S_{ct} = \frac{n_1 n_5}{ 4 \pi}
\int_{z=\e} d^2 x \sqrt{-\g} \left(2 - \log \e^2 \half R
+ \half (S^1_i)^2 \left(1 + \frac{2}{\log \e^2}\right) \right)
\ee
so the on-shell renormalized action consists of
(\ref{pp1}), the Gibbons-Hawking term and these counterterms
(along with additional counterterms
for the gauge fields, discussed in the main text).
The logarithmic terms determine the holographic conformal
anomalies \cite{Henningson:1998gx}.

The renormalized 1-point functions are \footnote{In comparing
with \cite{Karch:2005ms} one should note the factor of 2 difference
in the source.}
\bea
\< {\cal O}_{S^1_i} \> &=& \frac{n_1 n_5}{4 \pi} ( 2 \td{S}^1_{i(0)}); \la{1pts}\\
\< T_{uv} \> &=& \frac{n_1 n_5}{2 \pi} \left(
g_{(2)uv} + \half R g_{(0)uv}
\right .\nn
\\
&& \hsp \left .
+\frac{1}{4}\left( (\td{S}^1_{i(0)})^2 - 2 \td{S}^1_{i(0)} S^1_{i(0)}
+ 4 (S^1_{i(0)})^2 \right)g_{(0)uv} \right). \nn
\eea
Using the asymptotic solution one may verify that these
expressions satisfy the correct Ward identities
\bea
\<T^u_u \> &=& = - S_{i(0)}^1 \< {\cal O}_{S^1_i} \> + \cA \\
D^v \<T_{uv} \> &=& - \< {\cal O}_{S^1_i} \> D_u S^1_{i(0)}.
\eea
The first term on the r.h.s. is the standard term due to the
coupling of the source $S^1_{i(0)}$ to an operator of dimension one.
The conformal anomaly $\cA$ is given by
\be
\cA = \frac{c}{24 \pi} R + \frac{n_1 n_5}{2 \pi}
(S_{i(0)}^1)^2\, ; \qquad c= 6 n_1 n_5
\ee
The first term is the standard gravitational conformal anomaly and the
second the conformal anomaly induced by the short distance
singularities in the 2-point function of $\cao_{S^1_i}$ \cite{Petkou:1999fv}.

\section{Three point functions from the orbifold CFT} \la{apf}

In this appendix we discuss the relationship between three point
functions computed in the CFT on the symmetric product $S^N(T^4)$
with those in supergravity.
The chiral primary operators are summarized in (\ref{oper-2});
their detailed construction is not important here,
but note that they are $S_N$ invariant and orthonormal. The operators (\ref{oper-2})
manifestly have the correct dimensions and charges to correspond to
the fields $S_{k}^{(r )I}$ and $\S^I_k$ in supergravity. Moreover, as
discussed in section \ref{fti} the most natural
correspondence seems to be that given in (\ref{corr})
although this choice is not unique.

Extremal three point functions of these operators have the following
structure as $N \rightarrow \infty$ \cite{Jevicki:1998bm}
\bea
\left < {\cal O}_{n + k-1}^{(0,0) \dagger}  (\infty) {\cal
  O}_{k}^{(0,0)} (1) {\cal O} _{n}^{(0,0)} (0) \right > &=&
\frac{1}{\sqrt{N}} \left( (n+k-1)  nk \right )^{1/2};
\la{st-1} \\
\left < {\cal O}_{n + k-1}^{(i) \dagger}  (\infty) {\cal
  O}_{k}^{(0,0)} (1) {\cal O} _{n}^{(j)} (0) \right > &=&
\frac{1}{\sqrt{N}} \left ( (n+k-1) n k \right)^{1/2} \d^{ij}; \nn \\
\left < {\cal O}_{n + k- 1}^{(2,2) \dagger}  (\infty) {\cal
  O}_{k}^{(0,0)} (1) {\cal O} _{n}^{(2,2)} (0) \right > &=&
\frac{1}{\sqrt{N}} \left ( (n+k-1) n k   \right)^{1/2}; \nn \\
\left < {\cal O}_{n + k- 3}^{(2,2) \dagger}  (\infty) {\cal
  O}_{k}^{(0,0)} (1) {\cal O} _{n}^{(0,0)} (0) \right > &=&
\frac{2}{\sqrt{N}} \left ( (n+k-3) n k \right)^{1/2}; \nn \\
\left < {\cal O}_{n + k-1}^{(2,2) \dagger}  (\infty) {\cal
  O}_{k}^{(i)} (1) {\cal O} _{n}^{(j)} (0) \right > &=&
- \frac{1}{\sqrt{N}} \left ( (n+k-1) n k  \right)^{1/2} \w^i
  \ast \w^j; \nn \\
\left < {\cal O}_{n + k + 1}^{(2,2) \dagger}  (\infty) {\cal
  O}_{k}^{(2,2)} (1) {\cal O} _{n}^{(2,2)} (0) \right > &=& 0. \nn
\eea
(Here we use $\omega^i_{a \bar{a}}$ as a basis for
$H^{(1,1)}(T^4)$).
%There is a subtlety associated with the operator
%normalization - rescalings which preserve the two point functions affect these
%three point functions - but this ambiguity will not affect the
%subsequent discussion.

The cubic couplings between scalars in supergravity were
determined in \cite{Mihailescu:1999cj,Arutyunov:2000by}. From
(\ref{3pcou}) one sees that the couplings
$\S \S \S$ and $\S S S $ are generically non-zero whereas the
couplings $S S S $ and $S \S \S$ are always zero. This implies that
the corresponding extremal three point functions between chiral
primaries determined in supergravity have the following structures
\bea
\left < {\cal O}^{\dagger}_{\S^p_{\D}} {\cal O}_{\S^p_{\D_1}} {\cal O}_{\S^p_{\D_2}}
\right > \neq 0; \hsp
\left < {\cal O}^{\dagger}_{\S^p_{\D}} {\cal O}_{S^p_{\D_1}} {\cal O}_{S^p_{\D_2}}
\right > \neq 0; \hsp
\left < {\cal O}^{\dagger}_{S^p_{\D}} {\cal O}_{\S^p_{\D_1}} {\cal O}_{S^p_{\D_2}}
\right > \neq 0; \la{st-2} \\
\left < {\cal O}^{\dagger}_{S^p_{\D}} {\cal O}_{S^p_{\D_1}} {\cal O}_{S^p_{\D_2}}
\right > = 0; \hsp
\left < {\cal O}^{\dagger}_{\S^p_{\D}} {\cal O}_{\S^p_{\D_1}} {\cal O}_{S^p_{\D_2}}
\right > = 0; \hsp
\left < {\cal O}^{\dagger}_{S^p_{\D}} {\cal O}_{\S^p_{\D_1}} {\cal O}_{\S^p_{\D_2}}
\right > = 0, \nn
\eea
where $\D = \D_1 + \D_2$. Note that such correlators would be
determined in supergravity either by a careful limiting procedure of
non-extremal correlators (which uses directly the cubic couplings
mentioned above) or by reducing the six-dimensional action including
all boundary terms. In the latter case given that there are no bulk
couplings $S S S $ and $S \S \S$ it seems that there would be no
boundary couplings between such fields, and hence no non-zero extremal
correlators.

The correlators (\ref{st-1}) and (\ref{st-2}) clearly disagree if one makes
the identification proposed in (\ref{corr}). Given that this
identification was not unique, one might wonder whether there is a
different linear map between supergravity and orbifold CFT operators
such that the correlators agree. Whilst we have not proved in full
generality that this is impossible, the following argument suggests
that it is unlikely. Let ${\cal O}^a_1 = ({\cal O}_{2}^{(0,0)},
{\cal O}_{1}^{(i=1)})$ denote two of the dimension one CFT operators and
${\cal O}^\a_2 = ({\cal O}_{3}^{(0,0)}, {\cal O}_{2}^{(i=1)}, {\cal
  O}_{1}^{(2,2)})$ denote three of the dimension two CFT
operators. Let
 $\hat{{\cal O}}^a_1 = {\cal O}_{S_1^a}$
denote two dimension one operators dual to sugra scalar fields and
$\hat{{\cal O}}^\a_2 = ({\cal O}_{S^a_2}, {\cal O}_{\S_2})$ denote
three of the dimension two
operators dual to sugra fields.
Next write the fusion coefficients in
the corresponding extremal three
point functions in the orbifold CFT and supergravity as $C_{\a a b}$
and $\hat{C}_{\a a b}$ respectively. Since these are symmetric on the
last two indices, rewrite them as (square) matrices $D_{\a \b}$ and
$\hat{D}_{\a \b}$. Now the key point is that (\ref{st-1}) and
(\ref{st-2}) imply that  $D_{\a \b}$ has non-zero determinant, but
$\hat{D}_{\a \b}$ has zero determinant. Any linear maps between ${\cal
  O}^a_1$ and $\hat{{\cal O}}^a_1$, and between ${\cal O}^\a_2$ and
$\hat{{\cal O}}^\a_2$ which preserve the two point functions will not
map $D_{\a \b}$ to a zero determinant matrix and therefore one cannot
get agreement between (\ref{st-1}) and (\ref{st-2}) by making a
different identification between operators.

Whilst we have not extended this argument to higher dimension operators,
it seems more likely that the
supergravity and orbifold CFT correlators disagree because of
renormalization; there is no known non-renormalization theorem for
these correlators.


\begin{thebibliography}{99}

\bibitem{Lunin:2001jy}
  O.~Lunin and S.~D.~Mathur,
  ``AdS/CFT duality and the black hole information paradox,''
  Nucl.\ Phys.\ B {\bf 623}, 342 (2002)
  [arXiv:hep-th/0109154].

\bibitem{Lunin:2002bj}
  O.~Lunin, S.~D.~Mathur and A.~Saxena,
  ``What is the gravity dual of a chiral primary?,''
  Nucl.\ Phys.\ B {\bf 655} (2003) 185
  [arXiv:hep-th/0211292].

\bibitem{Lunin:2002iz}
  O.~Lunin, J.~Maldacena and L.~Maoz,
  ``Gravity solutions for the D1-D5 system with angular momentum,''
  arXiv:hep-th/0212210.



\bibitem{Lunin:2001fv}
  O.~Lunin and S.~D.~Mathur,
  ``Metric of the multiply wound rotating string,''
  Nucl.\ Phys.\ B {\bf 610}, 49 (2001)
  [arXiv:hep-th/0105136].

\bibitem{Balasubramanian:2000rt}
  V.~Balasubramanian, J.~de Boer, E.~Keski-Vakkuri and S.~F.~Ross,
  ``Supersymmetric conical defects: Towards a string theoretic description  of
  black hole formation,''
  Phys.\ Rev.\ D {\bf 64}, 064011 (2001)
  [arXiv:hep-th/0011217].

\bibitem{Maldacena:2000dr}
  J.~M.~Maldacena and L.~Maoz,
  ``De-singularization by rotation,''
  JHEP {\bf 0212} (2002) 055
  [arXiv:hep-th/0012025].


\bibitem{Mathur:2005zp}
  S.~D.~Mathur,
  ``The fuzzball proposal for black holes: An elementary review,''
  arXiv:hep-th/0502050.

%\cite{Callan:1995hn}
\bibitem{Callan:1995hn}
  C.~G.~Callan, J.~M.~Maldacena and A.~W.~Peet,
  ``Extremal Black Holes As Fundamental Strings,''
  Nucl.\ Phys.\ B {\bf 475}, 645 (1996)
  [arXiv:hep-th/9510134].
  %%CITATION = HEP-TH 9510134;%%

%\cite{Dabholkar:1995nc}
\bibitem{Dabholkar:1995nc}
  A.~Dabholkar, J.~P.~Gauntlett, J.~A.~Harvey and D.~Waldram,
  ``Strings as Solitons and Black Holes as Strings,''
  Nucl.\ Phys.\ B {\bf 474}, 85 (1996)
  [arXiv:hep-th/9511053].
  %%CITATION = HEP-TH 9511053;%%

\bibitem{Taylor:2005db}
  M.~Taylor,
  ``General 2 charge geometries,''
  JHEP {\bf 0603} (2006) 009
  [arXiv:hep-th/0507223].
  %%CITATION = HEP-TH 0507223;%%


\bibitem{KST} I.~Kanitscheider, K.~Skenderis and M.~Taylor,
in preparation.

%\cite{Rychkov:2005ji}
\bibitem{Rychkov:2005ji}
  V.~S.~Rychkov,
  ``D1-D5 black hole microstate counting from supergravity,''
  JHEP {\bf 0601}, 063 (2006)
  [arXiv:hep-th/0512053].
  %%CITATION = HEP-TH 0512053;%%

%\cite{Palmer:2004gu}
\bibitem{Palmer:2004gu}
  B.~C.~Palmer and D.~Marolf,
  ``Counting supertubes,''
  JHEP {\bf 0406}, 028 (2004)
  [arXiv:hep-th/0403025].
  %%CITATION = HEP-TH 0403025;%%

%\cite{Bak:2004rj}
\bibitem{Bak:2004rj}
  D.~Bak, Y.~Hyakutake and N.~Ohta,
  ``Phase moduli space of supertubes,''
  Nucl.\ Phys.\ B {\bf 696}, 251 (2004)
  [arXiv:hep-th/0404104]; D. Bak, Y. Hyakutake, S. Kim and N. Ohta,
``A Geometric Look on the Microstates of Supertubes'',
Nucl. Phys. B 712 (2005) 115 [hep-th/0407253].
  %%CITATION = HEP-TH 0404104;%%

%\cite{Skenderis:2006ah}
\bibitem{Skenderis:2006ah}
  K.~Skenderis and M.~Taylor,
  ``Fuzzball solutions for black holes
and D1-brane---D5-brane microstates,''
Phys. Rev. Lett. {\bf 98}, 071601 (2007) [arXiv:hep-th/0609154].
  %%CITATION = HEP-TH 0609154;%%

%\cite{Alday:2006nd}
\bibitem{Alday:2006nd}
  L.~F.~Alday, J.~de Boer and I.~Messamah,
  ``The gravitational description of coarse grained microstates,''
  arXiv:hep-th/0607222.




%\cite{Witten:1997yu}
\bibitem{Witten:1997yu}
  E.~Witten,
  ``On the conformal field theory of the Higgs branch,''
  JHEP {\bf 9707}, 003 (1997)
  [arXiv:hep-th/9707093].
  %%CITATION = HEP-TH 9707093;%%

%\cite{Dijkgraaf:1998gf}
\bibitem{Dijkgraaf:1998gf}
  R.~Dijkgraaf,
  ``Instanton strings and hyperKaehler geometry,''
  Nucl.\ Phys.\ B {\bf 543}, 545 (1999)
  [arXiv:hep-th/9810210].
  %%CITATION = HEP-TH 9810210;%%

%\cite{Seiberg:1999xz}
\bibitem{Seiberg:1999xz}
  N.~Seiberg and E.~Witten,
  ``The D1/D5 system and singular CFT,''
  JHEP {\bf 9904}, 017 (1999)
  [arXiv:hep-th/9903224].
  %%CITATION = HEP-TH 9903224;%%



%\cite{Skenderis:2006uy}
\bibitem{Skenderis:2006uy}
  K.~Skenderis and M.~Taylor,
  ``Kaluza-Klein holography,'' JHEP {\bf 0605}, 057 (2006)
  [arXiv:hep-th/0603016].

%\cite{Skenderis:2002wp}
\bibitem{Skenderis:2002wp}
  K.~Skenderis,
  ``Lecture notes on holographic renormalization,''
  Class.\ Quant.\ Grav.\  {\bf 19} (2002) 5849
  [arXiv:hep-th/0209067].
  %%CITATION = HEP-TH 0209067;%%

%\cite{Papadimitriou:2004ap}
\bibitem{Papadimitriou:2004ap}
  I.~Papadimitriou and K.~Skenderis,
  ``AdS / CFT correspondence and geometry,''
  arXiv:hep-th/0404176.

\bibitem{Papadimitriou:2004rz}
  I.~Papadimitriou and K.~Skenderis,
  ``Correlation functions in holographic RG flows,''
  JHEP {\bf 0410}, 075 (2004)
  [arXiv:hep-th/0407071].
  %%CITATION = HEP-TH 0407071;%%

%\cite{deHaro:2000xn}
\bibitem{deHaro:2000xn}
  S.~de Haro, S.~N.~Solodukhin and K.~Skenderis,
  ``Holographic reconstruction of spacetime and renormalization in the  AdS/CFT
  correspondence,''
  Commun.\ Math.\ Phys.\  {\bf 217}, 595 (2001)
  [arXiv:hep-th/0002230].
  %%CITATION = HEP-TH 0002230;%%

\bibitem{D'Hoker:1999ea}
  E.~D'Hoker, D.~Z.~Freedman, S.~D.~Mathur, A.~Matusis and L.~Rastelli,
  ``Extremal correlators in the AdS/CFT correspondence,''
  arXiv:hep-th/9908160.

\bibitem{Arutyunov:2000by}
  G.~Arutyunov, A.~Pankiewicz and S.~Theisen,
  ``Cubic couplings in D = 6 N = 4b supergravity on AdS(3) x S(3),''
  Phys.\ Rev.\ D {\bf 63} (2001) 044024
  [arXiv:hep-th/0007061].


\bibitem{Pank}  A.~Pankiewicz, ``Six-dimensional supergravities and the
AdS/CFT correspondence,'' Diploma Thesis, University of Munich,
October 2000.


%\cite{Alday:2005xj}
\bibitem{Alday:2005xj}
  L.~F.~Alday, J.~de Boer and I.~Messamah,
  ``What is the dual of a dipole?,''
  Nucl.\ Phys.\ B {\bf 746}, 29 (2006)
  [arXiv:hep-th/0511246].
  %%CITATION = HEP-TH 0511246;%%




\bibitem{Bena} I.~Bena and P.~Kraus,
``Three charge supertubes and black hole hair,''
Phys.\ Rev.\ D {\bf 70}, 046003 (2004)
[arXiv:hep-th/0402144];
O.~Lunin,  ``Adding momentum to D1-D5 system,''
  JHEP {\bf 0404}, 054 (2004)
  [arXiv:hep-th/0404006];
I.~Bena,
  ``Splitting hairs of the three charge black hole,''
  Phys.\ Rev.\ D {\bf 70}, 105018 (2004)
  [arXiv:hep-th/0404073];
 S.~Giusto, S.~D.~Mathur and A.~Saxena,
  ``Dual geometries for a set of 3-charge microstates,''
  Nucl.\ Phys.\ B {\bf 701}, 357 (2004) [arXiv:hep-th/0405017].
S.~Giusto, S.~D.~Mathur and A.~Saxena,
 ``3-charge geometries and their CFT duals,''
  Nucl.\ Phys.\ B {\bf 710}, 425 (2005)
  [arXiv:hep-th/0406103];
I.~Bena and N.~P.~Warner,
``One ring to rule them all ... and in the darkness bind them?,''
Adv. Theor. Math. Phys. 9 (2006) 1-35
[arXiv:hep-th/0408106];
S.~Giusto and S.~D.~Mathur,
   ``Geometry of D1-D5-P bound states,'' Nucl.\ Phys.\ B {\bf 729}, 203 (2005)
  [arXiv:hep-th/0409067];
 I.~Bena and P.~Kraus,
``Microstates of the D1-D5-KK system,''
  Phys.\ Rev.\ D {\bf 72}, 025007 (2005)
  [arXiv:hep-th/0503053];
P.~Berglund, E.~G.~Gimon and T.~S.~Levi,
``Supergravity microstates for BPS black holes and black rings,'' JHEP {\bf 0606}, 007 (2006)
  [arXiv:hep-th/0505167];
  A.~Saxena, G.~Potvin, S.~Giusto and A.~W.~Peet,
``Smooth geometries with four charges in four dimensions,'' JHEP {\bf 0604}, 010 (2006)
  [arXiv:hep-th/0509214];
  I.~Bena, C.~W.~Wang and N.~P.~Warner,
``Sliding rings and spinning holes,'' JHEP {\bf 0605}, 075 (2006)
  [arXiv:hep-th/0512157];
S.~Giusto, S.~D.~Mathur and Y.~K.~Srivastava,
``A microstate for the 3-charge black ring,''
  arXiv:hep-th/0601193;
 I.~Bena, C.~W.~Wang and N.~P.~Warner,
  ``The foaming three-charge black hole,''
  arXiv:hep-th/0604110;
V.~Balasubramanian, E.~G.~Gimon and T.~S.~Levi,
  ``Four dimensional black hole microstates: From D-branes to spacetime foam,''
  arXiv:hep-th/0606118;
  S.~Ferrara, E.~G.~Gimon and R.~Kallosh,
  `Magic supergravities, N = 8 and black hole composites,''
  arXiv:hep-th/0606211;
I.~Bena, C.~W.~Wang and N.~P.~Warner,
  ``Mergers and typical black hole microstates,''
  arXiv:hep-th/0608217;
 Y.~K.~Srivastava,
  ``Bound states of KK monopole and momentum,''
  arXiv:hep-th/0611124.

\bibitem{Balasubramanian:2005qu}
  V.~Balasubramanian, P.~Kraus and M.~Shigemori,
  ``Massless black holes and black rings as effective geometries of the D1-D5
  system,''
  Class.\ Quant.\ Grav.\  {\bf 22}, 4803 (2005)
  [arXiv:hep-th/0508110].

%\cite{Skenderis:2003da}
\bibitem{Skenderis:2003da}
  K.~Skenderis and M.~Taylor,
  ``Properties of branes in curved spacetimes,''
  JHEP {\bf 0402}, 030 (2004)
  [arXiv:hep-th/0311079].
  %%CITATION = HEP-TH 0311079;%%

\bibitem{Romans} L.~J.~Romans, ``Self-duality for interacting fields:
  covariant field equations for six dimensional chiral
  supergravities,'' Nucl. Phys. {\bf B276} (1986) 71.

\bibitem{Sez}
  S.~Deger, A.~Kaya, E.~Sezgin and P.~Sundell,
  ``Spectrum of D = 6, N = 4b supergravity on AdS(3) x S(3),''
  Nucl.\ Phys.\ B {\bf 536} (1998) 110
  [arXiv:hep-th/9804166].

\bibitem{Jackson} J.D. Jackson, ``Classical Electrodynamics'',  2nd Edition,
Wiley, see the discussion in section 3.6.

\bibitem{Skenderis:2006di}
  K.~Skenderis and M.~Taylor,
  ``Holographic Coulomb branch vevs,''
JHEP {\bf 0608}, 001 (2006)
  [arXiv:hep-th/0604169].






\bibitem{Mihailescu:1999cj}
  M.~Mihailescu,
  ``Correlation functions for chiral primaries in D = 6 supergravity on  AdS(3)
  x S(3),''
  JHEP {\bf 0002}, 007 (2000)
  [arXiv:hep-th/9910111].






%\cite{Kim:1985ez}
%\bibitem{Kim:1985ez}
%  H.~J.~Kim, L.~J.~Romans and P.~van Nieuwenhuizen,
%  ``The Mass Spectrum Of Chiral N=2 D = 10 Supergravity On $S^5$,''
%  Phys.\ Rev.\ D {\bf 32}, 389 (1985).
%  %%CITATION = PHRVA,D32,389;%%


\bibitem{Ban} M.~Banados, O.~Miskovic and S.~Theisen,
   ``Holographic Currents In First Order Gravity And Finite Fefferman-Graham
  Expansions,'' arXiv:hep-th/0604148.

\bibitem{Kraus}J.~Hansen and P.~Kraus,
  ``Generating charge from diffeomorphisms,''
arXiv:hep-th/0606230.


\bibitem{Jevicki:1998bm}
  A.~Jevicki, M.~Mihailescu and S.~Ramgoolam,
  ``Gravity from CFT on S**N(X): Symmetries and interactions,''
  Nucl.\ Phys.\ B {\bf 577}, 47 (2000)
  [arXiv:hep-th/9907144].


\bibitem{Lunin:2001pw}
  O.~Lunin and S.~D.~Mathur,
  ``Three-point functions for M(N)/S(N) orbifolds with N = 4 supersymmetry,''
  Commun.\ Math.\ Phys.\  {\bf 227} (2002) 385
  [arXiv:hep-th/0103169]; O.~Lunin and S.~D.~Mathur,
  ``Correlation functions for M(N)/S(N) orbifolds,''
  Commun.\ Math.\ Phys.\  {\bf 219}, 399 (2001)
  [arXiv:hep-th/0006196].


\bibitem{BFS1}
  M.~Bianchi, D.~Z.~Freedman and K.~Skenderis,
  ``How to go with an RG flow,''
  JHEP {\bf 0108}, 041 (2001)
  [arXiv:hep-th/0105276].
  %%CITATION = HEP-TH 0105276;%%

\bibitem{BFS2}
  M.~Bianchi, D.~Z.~Freedman and K.~Skenderis,
  ``Holographic renormalization,''
  Nucl.\ Phys.\ B {\bf 631}, 159 (2002)
  [arXiv:hep-th/0112119].
  %%CITATION = HEP-TH 0112119;%%


%\cite{deBoer:1998ip}
\bibitem{deBoer:1998ip}
  J.~de Boer,
   ``Six-dimensional supergravity on S(3) x AdS(3) and 2d conformal field
  theory,''
  Nucl.\ Phys.\ B {\bf 548}, 139 (1999)
  [arXiv:hep-th/9806104].
  %%CITATION = HEP-TH 9806104;%%


%\cite{David:2002wn}
\bibitem{David:2002wn}
  J.~R.~David, G.~Mandal and S.~R.~Wadia,
  ``Microscopic formulation of black holes in string theory,''
  Phys.\ Rept.\  {\bf 369}, 549 (2002)
  [arXiv:hep-th/0203048].




%\bibitem{Izquierdo:1994jz}
%  J.~M.~Izquierdo and P.~K.~Townsend,
%  ``Supersymmetric space-times in (2+1) adS supergravity models,''
%  Class.\ Quant.\ Grav.\  {\bf 12}, 895 (1995)
%  [arXiv:gr-qc/9501018].



%\cite{Freedman:1998tz}
\bibitem{Freedman:1998tz}
  D.~Z.~Freedman, S.~D.~Mathur, A.~Matusis and L.~Rastelli,
  ``Correlation functions in the CFT($d$)/AdS($d+1$) correspondence,''
  Nucl.\ Phys.\ B {\bf 546}, 96 (1999)
  [arXiv:hep-th/9804058].
  %%CITATION = HEP-TH 9804058;%%


%\cite{Henneaux:2004zi}
\bibitem{Henneaux:2004zi}
  M.~Henneaux, C.~Martinez, R.~Troncoso and J.~Zanelli,
  ``Asymptotically anti-de Sitter spacetimes and scalar fields with a
  logarithmic branch,''
  Phys.\ Rev.\ D {\bf 70} (2004) 044034
  [arXiv:hep-th/0404236].
  %%CITATION = HEP-TH 0404236;%%

%\cite{Karch:2005ms}
\bibitem{Karch:2005ms}
  A.~Karch, A.~O'Bannon and K.~Skenderis,
  ``Holographic renormalization of probe D-branes in AdS/CFT,''
 JHEP {\bf 0604}, 015 (2006)
  [arXiv:hep-th/0512125].
  %%CITATION = HEP-TH 0512125;%%

\bibitem{Henningson:1998gx}
  M.~Henningson and K.~Skenderis,
  ``The holographic Weyl anomaly,''
  JHEP {\bf 9807}, 023 (1998)
  [arXiv:hep-th/9806087];
  %%CITATION = HEP-TH 9806087;%%
``Holography and the Weyl anomaly,''
  Fortsch.\ Phys.\  {\bf 48}, 125 (2000)
  [arXiv:hep-th/9812032].
  %%CITATION = HEP-TH 9812032;%%

%\cite{Petkou:1999fv}
\bibitem{Petkou:1999fv}
  A.~Petkou and K.~Skenderis,
  ``A non-renormalization theorem for conformal anomalies,''
  Nucl.\ Phys.\ B {\bf 561}, 100 (1999)
  [arXiv:hep-th/9906030].
  %%CITATION = HEP-TH 9906030;%%






\end{thebibliography}
\end{document}